\documentclass{dune}
\pdfoutput=1

\input{common/preamble}

\begin{document}

\pagestyle{titlepage}
%\cleardoublepage

% This should be \input first thing after \begin{document}

\pagestyle{titlepage}
\date{}
\title{
\normalsize \snowmasstitle 

\vspace{5mm}
\scshape\Large Report of the Topical Group on Artificial Neutrino Sources for Snowmass 2021\\

\vspace{5mm}
\normalsize Neutrino Frontier Topical Group 09 \\
\vspace{20mm}
}

\renewcommand\Authfont{\scshape\small}
\renewcommand\Affilfont{\itshape\footnotesize}

%Authors go here with affiliation indexed by [X]
%\author[1]{L.~Fields$^{\thanks{NF09 Conveners}}$}
\author[1]{\textbf{NF09 Topical Group Conveners:}\\ L.~Fields}
\author[2]{A.~D.~Marino}
\author[3]{J.~P.~Ochoa-Ricoux}
\author[4]{J.~Spitz}
%\author[5]{Other Authors}

\vspace{-0.5cm}
%Affiliations go here
\affil[1]{University of Notre Dame, Notre Dame, Indiana, USA }
\affil[2]{University of Colorado Boulder, Boulder, Colorado, USA}
\affil[3]{University of California, Irvine, California, USA}
\affil[4]{University of Michigan, Ann Arbor, Michigan, USA}

%\affil[5]{Other Affilliations}

\maketitle

\renewcommand{\familydefault}{\sfdefault}
\renewcommand{\thepage}{\roman{page}}
\setcounter{page}{0}

\pagestyle{plain} 
\clearpage
\textsf{\tableofcontents}

%\textsf{\listoffigures}
%\clearpage

%\textsf{\listoftables}
%\clearpage

%For acronym list to appear just after TOC, TOF, TOT
%\printnomenclature
%\clearpage

%\iffinal\else
%\textsf{\listoftodos}
%\clearpage
%\fi

\renewcommand{\thepage}{\arabic{page}}
\setcounter{page}{1}

\pagestyle{fancy}

% Set how header/footers look
%\newcommand{\chaptermark}[1]{%
%\markboth{Chapter \thechapter:\# 1}{}}
%\renewcommand{\chaptermark}[1]{%
%\markboth{Chapter \thechapter:\ #1}{}}
\fancyhead{}
%\fancyhead[RO,LE]{\textsf{\footnotesize \thepage}}
\fancyhead[RO]{\textsf{\footnotesize \thepage}}
%\fancyhead[LO,RE]{\textsf{\footnotesize \rightmark}}
\fancyhead[LO]{\textsf{\footnotesize \rightmark}}

\fancyfoot{}
\fancyfoot[RO]{\textsf{\footnotesize Snowmass 2021}}
\fancyfoot[LO]{\textsf{\footnotesize NF09 Topical Group Report}}
\fancypagestyle{plain}{}

\renewcommand{\headrule}{\vspace{-4mm}\color[gray]{0.5}{\rule{\headwidth}{0.5pt}}}

% Not all main documents have any citations.
% When not built in "final" mode, add in one citation just to let the
% document build.
% If, after substantial editing a main document still lacks any
% citations then it should have its whole bibliography removed.
%\ifdefined\isfinal\nocite{}\else\nocite{CD0}\fi
%\nocite{nothing}

% see also preamble.tex
%\input{common/acronyms}

\clearpage

\section*{Executive Summary}
\label{sec:summary}

The NF09 topical group was charged with soliciting input to Snowmass on the topic of artificial neutrino sources. In this report, we attempt to catalogue all new or upgraded artificial neutrino sources that are being considered by the global neutrino physics community over the next decade. This report also highlights projects that can improve our knowledge of the fluxes from these sources, which is important to maximize their use. 
%To get the most out of these sources it is necessary for them to have well-understood fluxes, and this report also highlights projects that can improve our knowledge of the fluxes from these sources.

The current landscape for neutrino beams that result from focused hadrons includes the long- and short-baseline program at Fermilab and the J-PARC neutrino beam to T2K.  Both of these laboratories will be a major international focus for long-baseline experiments over the next decade with the upgrade of the J-PARC beamline for Hyper-Kamiokande in Japan and the construction of the initial phase of the LBNF neutrino beamline for DUNE at Fermilab.  A beam power upgrade of the LBNF beamline from 1.2 to 2.4 MW in the second phase of DUNE will be crucial to maximizing its physics reach in a timely manner. Other upcoming global possibilities for accelerator-generated beams include an intense new beam using the protons at the new European Spallation Source.  The NA61/SHINE, EMPHATIC and DsTau/NA65 experiments are in the midst of program of measurements aimed at improving flux determination at these beamlines.  

Neutrino experiments at spallation neutron sources use neutrinos produced with a well understood energy spectrum and generally rely on neutrino interactions with a precisely predicted interaction cross section. This combination is particularly advantageous for a number of physics measurements including those involving oscillations, neutrino interactions, coherent neutrino-nucleus scattering, related new physics searches involving non-standard neutrino interactions, and probes of dark matter and other exotica. Both new sources and future upgrades to existing facilities, domestically and internationally, promise a rich physics program for many decades to come.

Nuclear reactors continue to be widely used as MeV-scale neutrino sources and a next-generation reactor neutrino program spanning a wide range of physics goals is under preparation. These experiments will benefit from the precise characterization of reactor neutrino fluxes made by experiments in the last decade, which uncovered discrepancies with the models while also providing important clues about their origin. Direct neutrino measurements with greater precision, as well as in the uncharted region below the $1.8$~MeV IBD threshold, are anticipated in the next decade that will further enhance existing constraints on reactor neutrino emission. The community is also interested in pursuing ancillary nuclear physics measurements that will allow full resolution of existing data vs. model discrepancies and enable improved reactor neutrino flux modeling. 
%to make precision measurements of neutrino oscillations, search for sterile neutrinos, make new tests of the Standard Model, and demonstrate the use of neutrino detectors for nuclear safeguards

Several new varieties of neutrino sources are also being considered for the future.  Last year, the FASER$\nu$ experiment observed the first neutrino interactions at the LHC and will collect additional neutrino data in the upcoming LHC Run-3, as will the SND@LHC experiment.  There is a proposal for a new Forward Physics Facility at the LHC, which would dramatically expand the physics possibilities with LHC neutrinos.  Another novel source is IsoDAR, a planned experiment at Yemilab in South Korea that will use a 60 MeV proton beam and subsequent neutron capture to produce an intense antineutrino source.  There are several proposals for use of stored muon beams as neutrino sources, ranging from the relatively modest NuStorm facility, which would make precise measurements of both muon and electron neutrino cross sections, to long-baseline neutrino factories to a muon collider.  All of these new potential sources would benefit from R\&D over the coming decade.

%%%%%%%%%%%%%%%%%%%%%%%%%%%%%%%%%%%%%%%%%%%%%%%%%%%%%%%%%%%

\cleardoublepage

\section{Introduction and Motivation}
\label{sec:introduction}

During the Snowmass 2021 process, the NF09 topical group was charged with soliciting input to Snowmass on the topic of artificial neutrino sources.  The topics under the purview of NF09 included, but were not limited to,
\begin{itemize}
    \item Focused-Hadron neutrino beams (both new and upgrades)
\item Novel neutrino sources (including beta beams, neutrino factories and IsoDAR)
\item Nuclear Reactors
\item Radioactive sources
\item Neutrinos from Spallation Neutron Sources (pi/mu/K decay-at-rest)
\item Neutrino beam instrumentation
\item Neutrino flux determination from artificial sources (including hadron production measurements)
\end{itemize}
    
In this report, we attempt to catalogue all new or upgraded artificial neutrino sources that are currently being considered by the global neutrino physics community.  We also consider efforts to characterize the neutrino flux from these sources.    

%%%%%%%%%%%%%%%%%%%%%%%%%%%%%%%%%%%%%%%%%%%%%%%%%%%%%%%%%%%
%\cleardoublepage

\section{Focused-Hadron Neutrino Beams}
\label{sec:convb}

Conventional neutrino beams use high-energy protons striking a target to generate short-lived hadrons (mainly $\pi^\pm$ and $K^{\pm}$) that decay into neutrinos.  This type of source has a long history, going back to the beam used to discover the muon neutrino~\cite{Danby:1962nd}.   These beams can produce neutrinos that either decay-at-rest, as was done for the LSND experiment~\cite{LSND:1996jxj}, or decay-in-flight, like the current high-energy long-baseline beams discussed in Section~\ref{subsec:currentconv}.  Decay-at-rest beams produce neutrinos primarily through $\pi^+ \rightarrow \mu^+ + \nu_{\mu}$ and the subsequent $\mu^+ \rightarrow \bar{\nu}_{\mu} + \nu_{e} + e^{+}$, resulting in an isotropic flux of neutrinos.  Current and future decay-at-rest sources using spallation neutron sources and isotopes will be discussed in Sections~\ref{sec:spall} and~\ref{subsec:isodar} respectively.

Decay-in-flight beams rely on the decays of pions and kaons, notably $\pi^{\pm} \rightarrow \mu^{\pm} +  \nu_{\mu} / \bar{\nu}_{\mu}$, $K^{\pm} \rightarrow \mu^{\pm} +  \nu_{\mu} / \bar{\nu}_{\mu}$, $K_{L} \rightarrow \pi^{\pm} + \mu^{\mp}  + \bar{\nu}_{\mu}/ \nu_{\mu}$, and $K_{L} \rightarrow \pi^{\pm} + e^{\mp} +  \bar{\nu}_{e}/\nu_{e}$, resulting in fluxes that are primarily muon neutrinos (or antineutrinos).  Typically one or more magnetic horns are used to focus the charged hadrons produced from the target into a region where they decay to neutrinos.  Selecting positively charged hadrons results in a beam that is primarily $\nu_{\mu}$, while selecting negatively charged hadrons results in a beam that is primarily $\bar{\nu}_{\mu}$.  Near detectors  and other beamline instrumentation help to constrain the flux uncertainties.  Measurements of the production of hadrons in the interactions from dedicated hadron production experiments also improve our knowledge of the neutrino fluxes in accelerator-generated beams.

\begin{figure}[ht]
    \centering
  \includegraphics[width=0.95\textwidth]{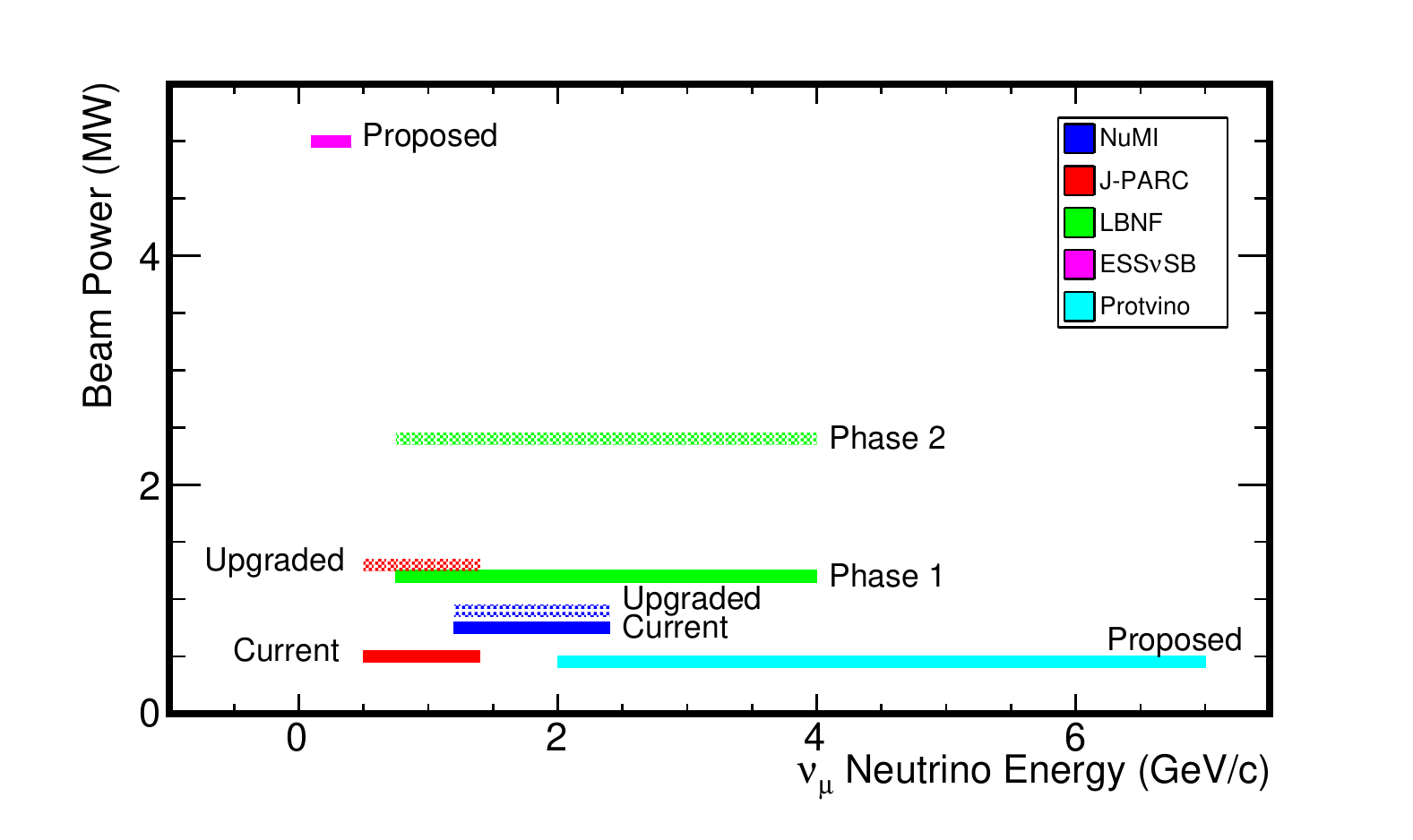}
  \caption{Neutrino beam power vs. energy for current and future long-baseline neutrino beams.  The energy range shown here is the approximate width to half-max around the beam peak energy.}
    \label{fig:focusedbeamsummary}
\end{figure}

In this section the current landscape of conventional focused-hadron neutrino beams will be briefly described as well as future projects and complementary measurements that will enhance our understanding of these beams.  A summary of the current long-baseline neutrino beams is shown in Figure~\ref{fig:focusedbeamsummary}.

% too hard to show the initial phase of P2O or the booster beam on this plot

\subsection{Current Landscape}
\label{subsec:currentconv}

The current landscape for conventional focused-hadron neutrino beams includes both short-baseline and long-baseline neutrino experiments.  

\subsubsection{Short-Baseline Beams}
A Short-Baseline Neutrino (SBN) physics program~\cite{MicroBooNE:2015bmn} is currently operating at Fermilab, with  a suite of three liquid-argon detectors located along the Booster Neutrino Beam (BNB)~\cite{BNB_TDR}, at distances from 110 m to 600 m from the target.  The BNB uses 8 GeV/$c$ protons from Fermilab's booster on a 71-cm long beryllium target with a single focusing horn. The resulting beam peaks around a neutrino energy of 1 GeV. (The physics program will be discussed in more detail in the NF02 Topical Group Report.) At present there are no plans to continue the operation of this beamline beyond the current SBN program, as the Fermilab booster will likely need to be replaced in the next decade or so (See Section~\ref{sec:lbnf}).

\subsubsection{Long-Baseline Beams}

Two long-baseline neutrino beamlines are currently operating with baselines of hundreds of kilometers: the Neutrinos at the Main Injector (NuMI) beam~\cite{Crane:1995ky} at Fermi National Accelerator Laboratory (Fermilab) in Illinois and the neutrino beam~\cite{T2K:2011qtm} at the Japan Proton Accelerator Research Complex (J-PARC) in Japan.  The NuMI beamline accelerates protons up to a momentum of 120 {GeV/$c$} to produce neutrinos and has achieved an hourly power record of over 800 kW, while the J-PARC beam uses protons with a momentum of 30~{GeV/$c$} and has achieved a power of 500 kW. Near term upgrade plans for both of these beamlines are discussed in the next subsections.

\subsection{NOvA Future Program}

The NuMI beamline is scheduled to provide beam to the NuMI Off-Axis Neutrino (NOvA) long-baseline  experiment~\cite{NOvA:2007rmc} until approximately 2026, 
% updated date based on email from P. Vahle
when the Fermilab accelerator complex is scheduled to shut down
to allow for final construction of Proton Improvement Plan II (PIP-II)~\cite{Lebedev:2017vnu} linac replacement and completion of the Long-Baseline Neutrino Facility(LBNF) beamline~\cite{DUNE:2016evb} in preparation for the start of the Deep Underground Neutrino Experiment (DUNE).  Fermilab is
now pursuing several modest improvements that stand to deliver power to NuMI in excess of 900 kW in the
next several years.  Improvements to the NuMI beam will require reducing losses primarily in the \SI{8}{\GeV} booster, relying
in part on new collimators and dampers included in the scope of the PIP-II project. Depending on the schedule,  NOvA could accumulate a total of  $63 \times 10^{20}$ protons-on-target (POT) by the end of its run~\cite{NOvALOI}.

\subsection{J-PARC Upgrade}

The J-PARC neutrino beam provides neutrinos to the Tokai-to-Kamioka (T2K) long-baseline neutrino oscillation experiment~\cite{T2K:2011qtm}, which is 295 km from J-PARC.  The beam is aimed 2.5$^\circ$ off-axis from the Super-Kamiokande detector~\cite{Super-Kamiokande:2002weg}.
The current phase of the J-PARC beam will continue for several more years, aiming to deliver $10\times10^{21}$ POT to the T2K long-baseline experiment~\cite{T2KLOI}.  A magnet power supply upgrade in 2021 will increase the beam power by decreasing the time between beam pulses from 2.48 s to 1.32 s. Additional machine development and RF upgrades should result in a beam power of approximately 1 MW by 2025-2026~\cite{T2KLOI}.

In 2017 the Japan Association of High Energy Physicists (JAHEP) updated the its strategy
for future particle physics, including a recommendation to pursue the Hyper-Kamiokande project~\cite{Hyper-Kamiokande:2018ofw}, which will deploy new detectors on the J-PARC neutrino beamline (at a very similar baseline  to T2K).  Hyper-Kamiokande was approved in 2020 and is expected to begin operation in 2027. The Hyper-Kamiokande physics potential will be discussed in the NF01 Topical Group Report and is also discussed in a contributed White Paper~\cite{Hyper-Kamiokande:2022smq}. To achieve its physics goals the J-PARC neutrino beamline will be upgraded to a power of 1.3 MW~\cite{JapanLOI}.

\begin{figure}[htb]
    \centering
  \includegraphics[width=0.48\textwidth]{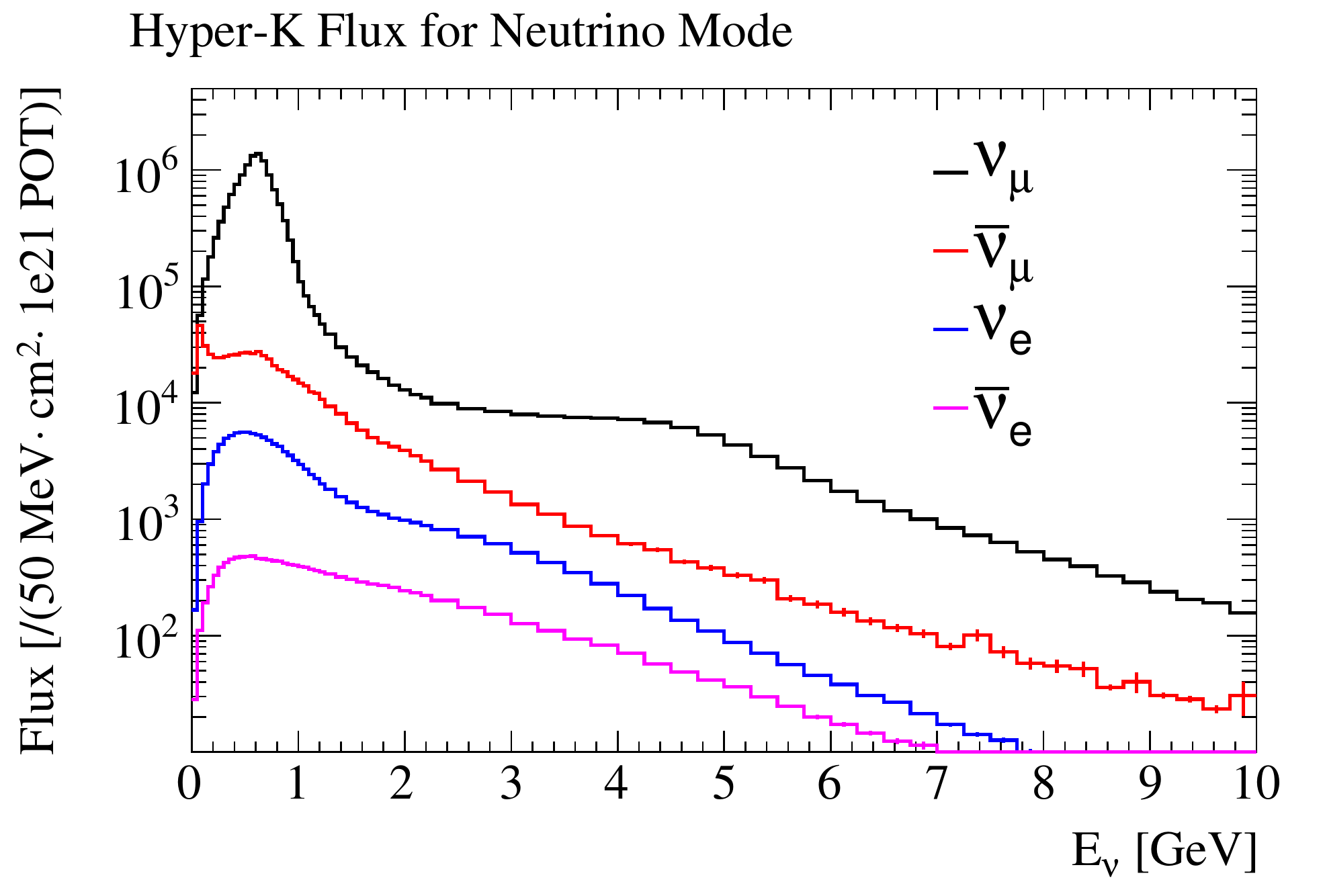}
  \includegraphics[width=0.48\textwidth]{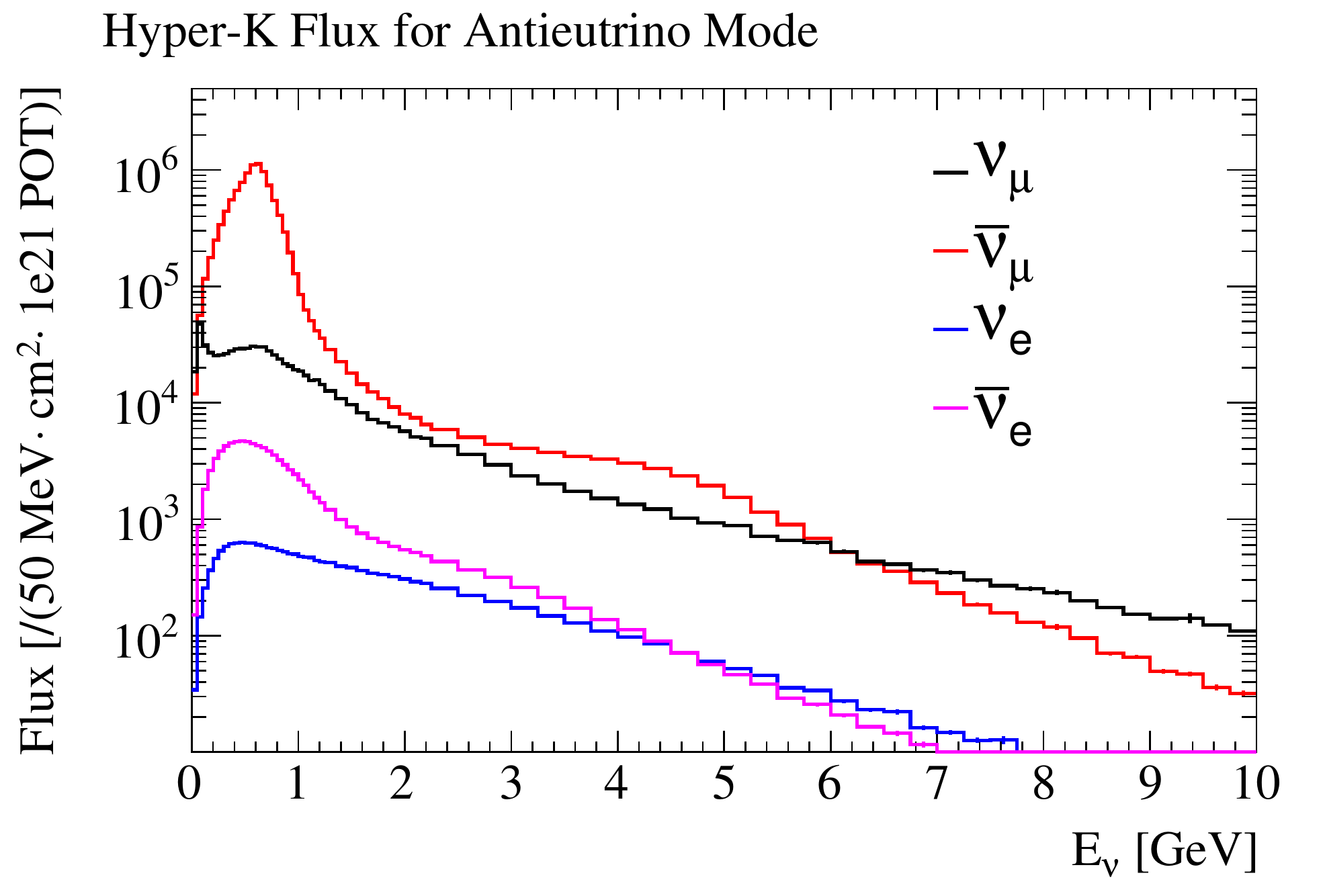}
\caption{The neutrino spectra at Hyper-K for the neutrino-enhanced (left) and antineutrino-enhanced (right) horn current polarities  with the absolute horn current set to 320\,kA, from ~\cite{Hyper-Kamiokande:2018ofw}.}
    \label{fig:hkflux}
\end{figure}

\subsection{LBNF Beamline and Upgrade}
\label{sec:lbnf}

\begin{figure}[ht]
    \centering
    \includegraphics[width=0.45\textwidth]{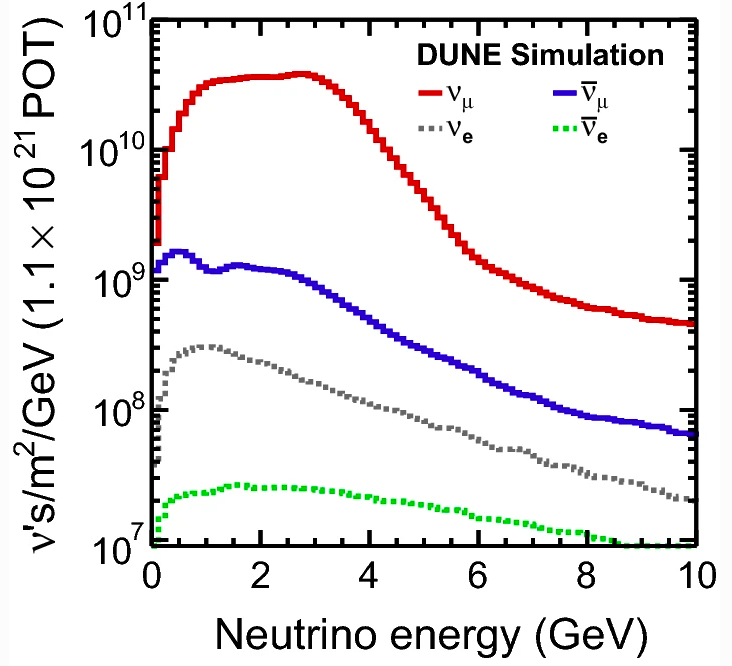}
    \includegraphics[width=0.45\textwidth]{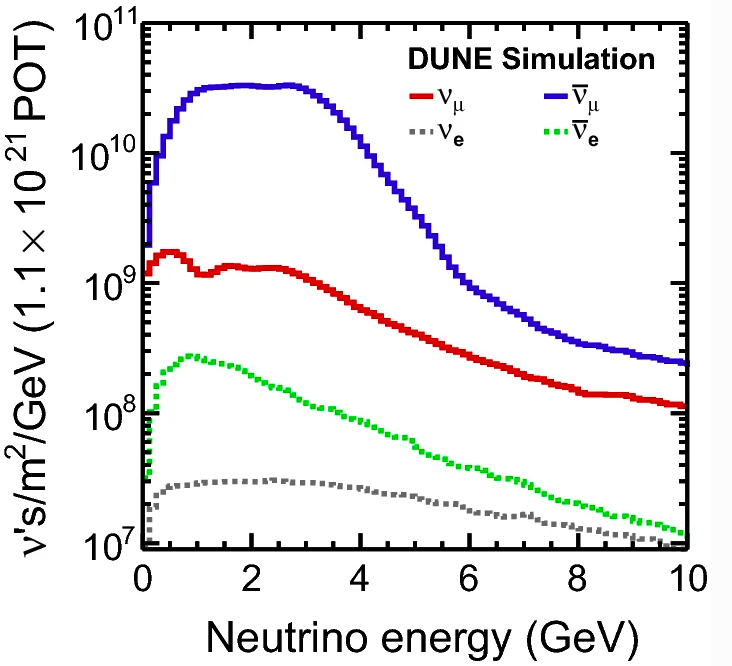}
    \caption{LBNF neutrino fluxes at the DUNE far detector for neutrino-enhanced, FHC, beam running (left) and antineutrino-enhanced, RHC, beam running (right ) from~\cite{DUNE:2020jqi}}
    \label{fig:duneflux}
\end{figure}

The centerpiece of the US neutrino program over the next decade will be the DUNE long-baseline experiment, currently under construction, which will use the LBNF beamline to be constructed at Fermilab. The DUNE far detectors in South Dakota are approximately 1300 km from the source. The physics goals of DUNE are described in several of the topical group reports.  The neutrino beamline in the Long-Baseline Neutrino Facility is described in detail in~\cite{DUNE:2016evb}. The predicted fluxes at the far detector are shown in Figure~\ref{fig:duneflux}.  

The LBNF/DUNE construction schedule is funding-limited,
and the experiment construction is currently divided into two phases. In Phase I, the LBNF beam will have a power of 1.2 MW, two far detector modules, and a more limited near detector.  These items will allow DUNE to do world-class physics, but they will not be sufficient to achieve the long-term physics goals of DUNE. Phase II will include upgrading the LBNF beam to a power of 2.4 MW, constructing far detector modules 3 and 4, and adding a more capable near detector~\cite{DUNE:2022aul}.   

\begin{figure}[ht]
    \centering
 \includegraphics[width=0.32\linewidth]{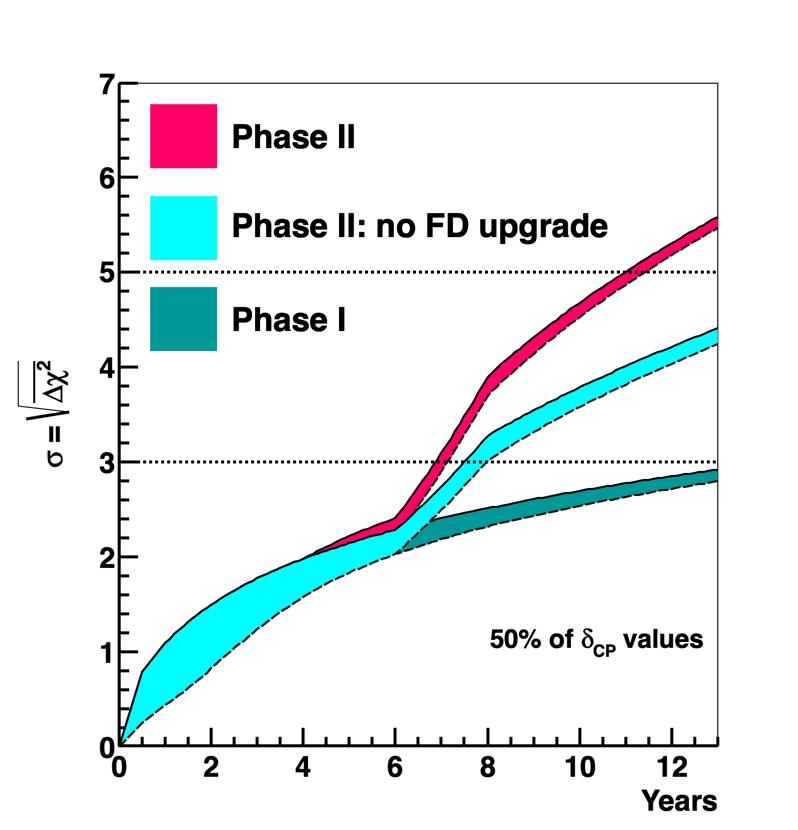}
  \includegraphics[width=0.32\linewidth]{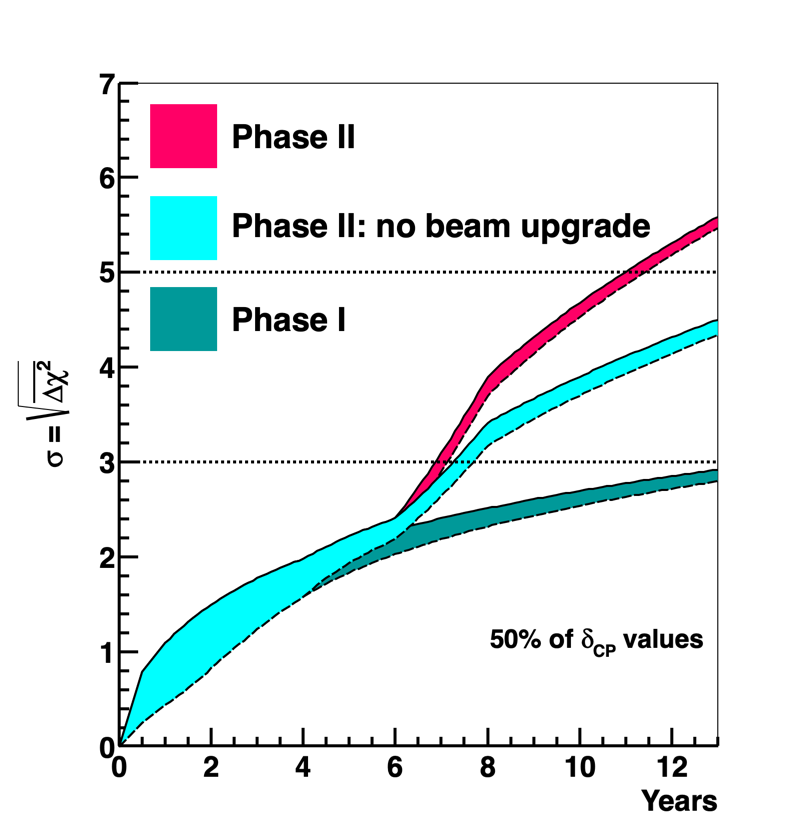}
  \includegraphics[width=0.32\linewidth]{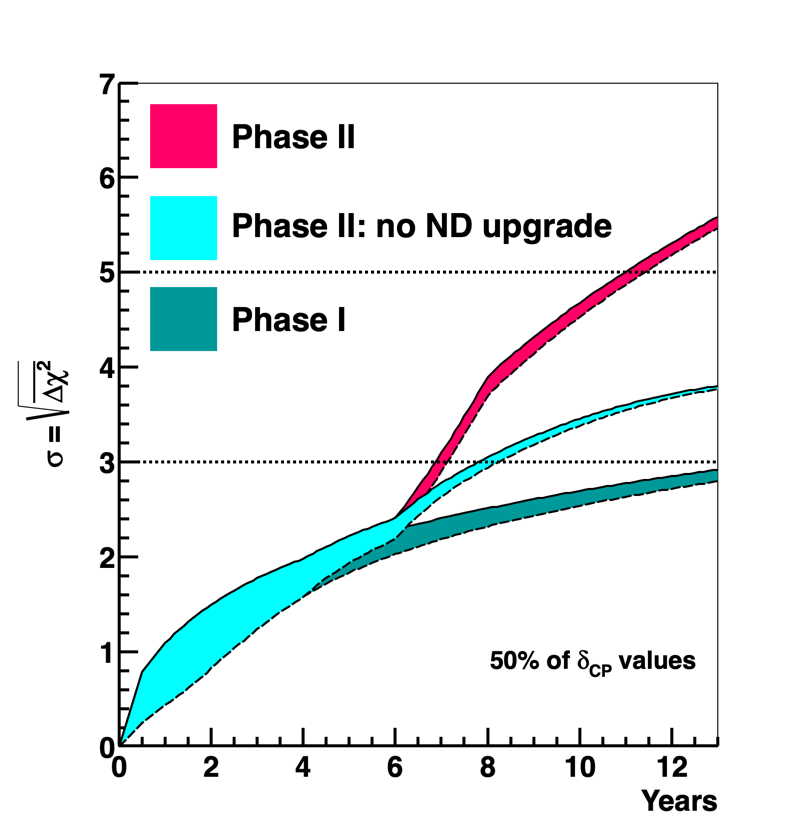}
  \caption{DUNE Sensitivity to CP violation for 50\% of $\deltacp$ values, as a function of time in calendar years, reproduced from ~\cite{DUNE:2022aul}. The width of the bands shows the impact of potential beam power ramp up; the solid upper curve is the sensitivity if data collection begins with 1.2 MW beam power and the lower dashed curve shows a conservative beam ramp scenario where the full power is achieved after 4 years. The green bands show the Phase I sensitivity and the red bands shows the Phase II sensitivity. In each plot the cyan band shows the Phase II sensitivity if one of the three upgrades does not occur.
  The left plot shows the sensitivity without the FD upgrade, the middle plot shows the sensitivity without the beam upgrade, and the right plot shows the sensitivity without the ND upgrade, illustrating that each is necessary to achieve DUNE's physics goals.}
    \label{fig:lbnf24}
\end{figure}

DUNE will need all four far detector modules, the more capable near detector, and the upgraded 2.4~MW beam to achieve its physics goals  on a ~10-year timescale. As shown in Fig.~\ref{fig:lbnf24}, Phase I alone will not reach 3$\sigma$ sensitivity for 50\% of the possible $\delta_{CP}$ values, while having the full Phase II ready in year 6 will allow 5$\sigma$ sensitivity for 50\% of the possible $\delta_{CP}$ values in $\sim$10 years. The  $\delta_{CP}$ sensitivity without the beam upgrade is shown in the middle panel of Fig.~\ref{fig:lbnf24}, and while this will allow DUNE to reach 3$\sigma$ sensitivity for 50\% of the possible $\delta_{CP}$ values, it will not reach 5$\sigma$ within 12 years.

 The previous P5 panel recommended that the LBNF beamline be more than 1 MW and upgradable to multi-megawatt power~\cite{HEPAPSubcommittee:2014bsm}. The PIP-II project will replace the current 400 MeV linac at Fermilab with a new 800 MeV proton linac capable of delivering the 1.2 MW power for Phase I of LBNF. A proposal to add an 800 MeV PIP-II Accumulator Ring (PAR) can maximize the physics potential with the PIP-II beam and help the DUNE Phase I power ramp up more quickly~\cite{Pellico:2022dju}.  However, after PIP-II the amount of beam that can be transmitted to the Main Injector will be limited by the capacity of the 8 GeV Booster beam. To reach the higher power needs of LBNF Phase II, the Booster will need to be replaced.  The Booster replacement could be based on a continuation of
the 800 MeV linac up to 2-3 GeV, which can then be followed by either a new rapid cycling synchrotron~\cite{Eldred:2022vxi,Nagaitsev:2021xzy,Ainsworth:2021ahm} or continuing the linac to all the way up 8 GeV~\cite{Belomestnykh:2022kal}. All of these options can provide a 2.4 MW beam to LBNF, and some can potentially go up to 4 MW. Some of the additional physics opportunities with these two upgrade paths are discussed in~\cite{Arrington:2022pon}.  

In addition to the upgrades to the accelerator complex, additional upgrades will need to be made to the LBNF beamline to handle the higher beam power~\cite{HurhCSS}. Some components that are difficult to access after beam begins are being designed now for 2.4~MW beam power, such as the shielding, absorber, and decay pipe. However others, such as the target, first horn, and the upstream decay pipe window will need to be redesigned for Phase II. Significant R\& D on target materials will be essential in the near term to maintain target reliability in Phase II and allow for accurate prediction of the component lifetimes~\cite{PellemoineCSS}.  As an example of this, the Radiation Damage In Accelerator Target Environments (RaDIATE) collaboration, an international effort between the accelerator and fusion/fission communities, aims to generate new data on the material properties of target materials~\cite{radiate,Pellemoine:2022wmo}. Novel target materials must also be investigated~\cite{Ammigan:2022ogd}.

%%%%%%%%%%%%%%%%%%%%%%%%%%%%%%%%%%%%%%%%%%%%%%%%%%%%%%%%%%%

\subsection{ESSnu Super Beam}
\label{essnusb}
\begin{figure}[!htbp]
    \centering
       \includegraphics[width=0.49\linewidth]{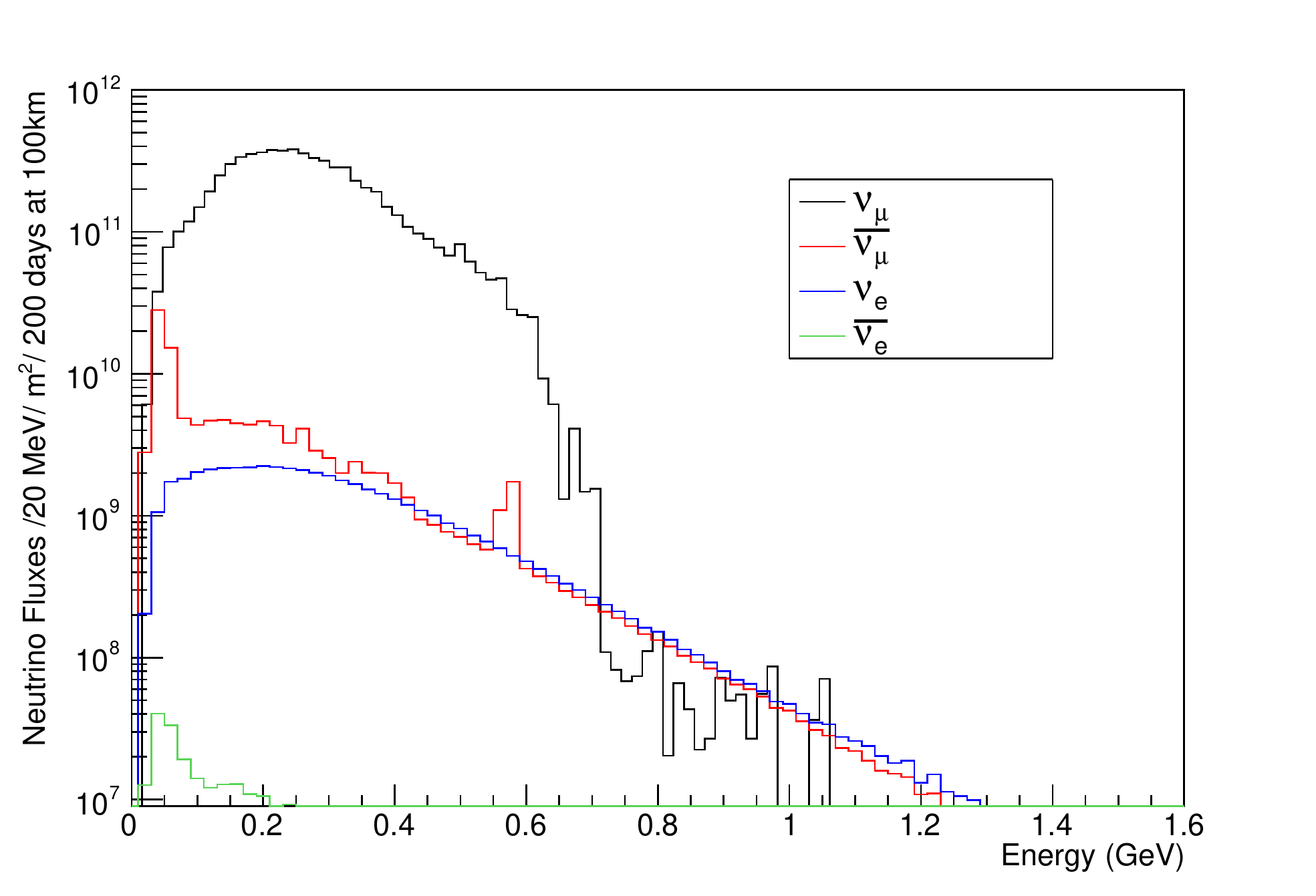}
       \includegraphics[width=0.49\linewidth]{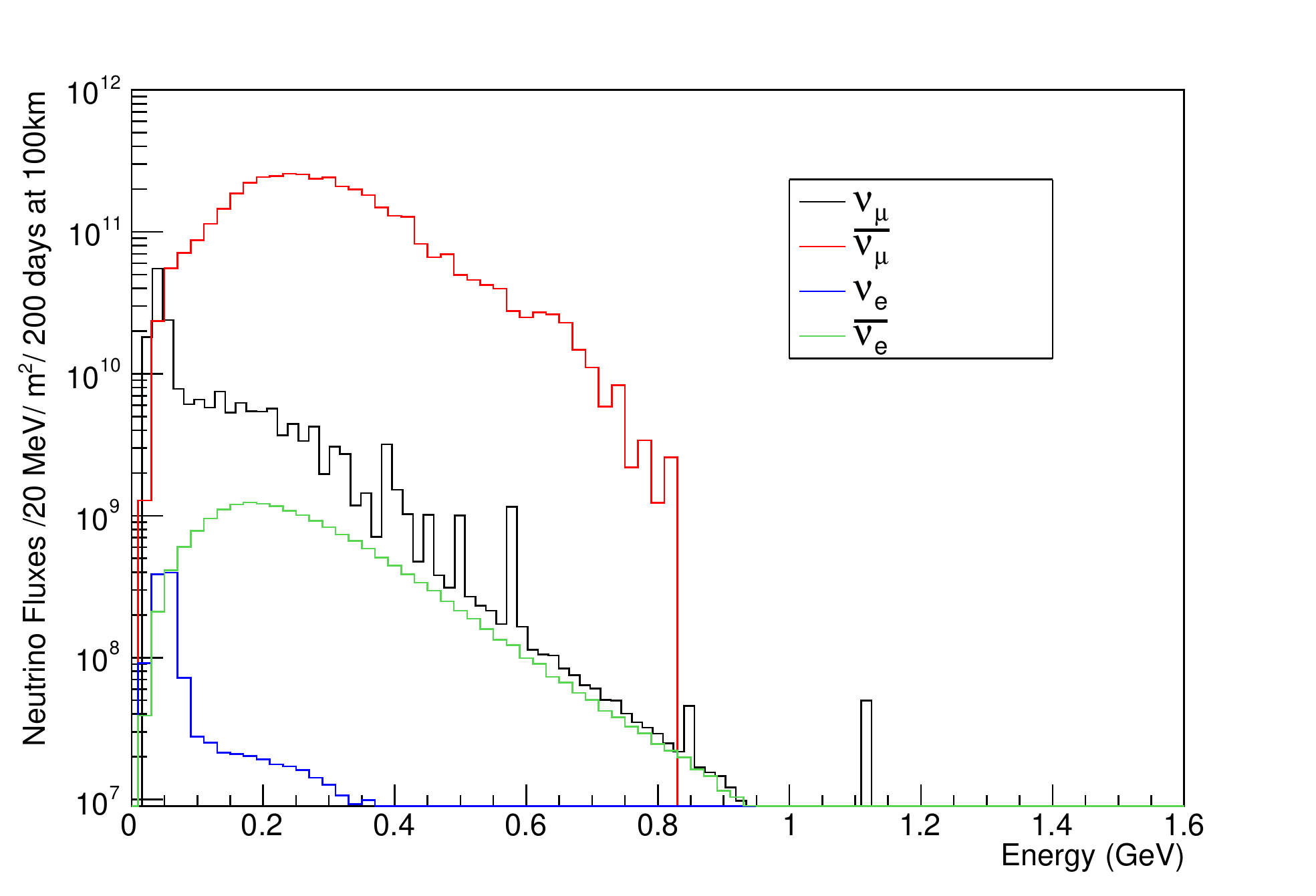} 
    \caption{ESS$\nu$SB neutrino (left) and antineutrino (right) energy spectrum at \SI{100}{\kilo\meter} from the neutrino source, from~\cite{Alekou:2022mav}.}
    \label{fig:ESSNuFlux}
\end{figure}

The European Spallation Source (ESS)~\cite{Garoby:2017vew} is currently under construction near Lund, Sweden, and is expected to deliver its first physics in 2023~\cite{ESSnuLOI,Alekou:2022mav}.  Its linac will produce an intense 5 MW beam of 2 GeV (kinetic energy)  protons to produce spallation neutrons.  (The neutrinos produced from this spallation source will be discussed in Section~\ref{sec:spall}.) The high power of the ESS linac presents an opportunity to produce an intense second-generation neutrino beam. A design study began in 2018 to evaluate the feasibility of also using this proton beam to create a neutrino super beam, the European Spallation Source Neutrino Super Beam (ESS$\nu$SB), and the study has recently produced a conceptual design report~\cite{Alekou:2022emd}. The goals of this study include specifying the necessary upgrades to the current ESS linear accelerator in order to
raise the average beam power from 5 MW to 10 MW (so that a 5 MW beam of H$^{-}$ ions can go the to ESS$\nu$SB accumulator ring while concurrently providing the 5 MW beam of protons for the spallation neutron source), increasing the beam energy to 2.5 GeV, 
increasing the repetition rate to 28 Hz (with 14 pulses per second going into the neutrino beam),  
designing an intermediate proton-beam accumulator ring needed for the generation of a time-compressed and well-focused neutrino
beam, and designing a neutrino-production target station.
The high intensity of the beam will be a challenge for the beam target.  The current plan features four solid targets, each embedded in a magnetic horn with a horn current of 350 kA.  A beam switching system will be used to hit the targets sequentially with proton pulses, reducing the power hitting each target to 1.25 MW~\cite{Ekelof:2019hqi}. The predicted fluxes are shown in Figure~\ref{fig:ESSNuFlux}.  The physics opportunities for such a beam with a megaton-scale water Cherenkov detector placed near the second neutrino oscillation maximum include measuring leptonic CP violation, and ESS$\nu$SB can achieve 5$\sigma$ discovery of CP violation for approximately 70\% of $\delta$ values in about eight years~\cite{Alekou:2022emd}.

%%%%%%%%%%%%%%%%%%%%%%%%%%%%%%%%%%%%%%%%%%%

\subsection{Protvino to ORCA Beamline}

There is also interest in using the accelerator complex in Protvino to generate a neutrino beam aimed towards the KM3NeT/ORCA detector in the Mediterranean Sea as a far detector~\cite{Akindinov:2019flp}. Protvino-to-ORCA (P2O) would have a very long 2595 km baseline and the beam would need to be angled down at 11.7$^\circ$ below the horizon.  The accelerator chain normally operates with a beam energy of 50 GeV to 70 GeV, and the neutrino beamline would include a target station, horns, and a $\sim$180 m decay pipe. Some simulations have been performed resulting in neutrino beam with a neutrino energy plateau between 2 and 7 GeV.  Modest beam improvements could be made to achieve a beam power of 90 kW. A new chain of injection accelerators could allow for an upgrade to 450 kW.  The NF01 report contains more details on the physics capabilities of P2O.

%%%%%

\subsection{Additional Opportunities with Focused Beams}

\subsubsection{Time-bunched sources}

\begin{figure}[htb]
\centering
\includegraphics[]{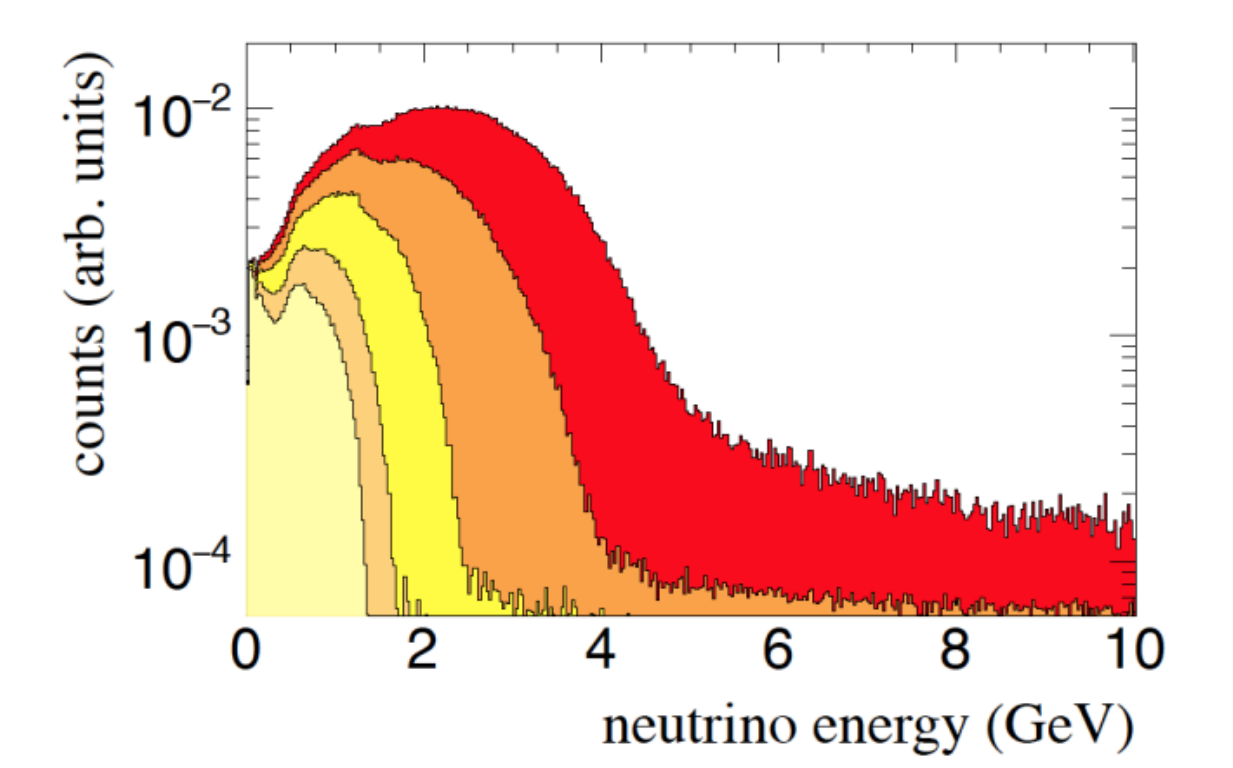}
\caption{The LBNF neutrino flux at DUNE (red), and subsets of that flux corresponding to different arrival times at the DUNE far detector: 250 ps after the start of the neutrino bunch
(orange), 500 ps after (yellow), 750 ps (dark beige), and 1 ns
(light beige).  No time spread of the protons is assumed.  Figure taken from~\cite{Angelico:2019gyi}.}.  
\label{fig:time_bunched}
\end{figure}

One potential new capability of focused-hadron beams would be a modest variation that is proposed as an upgrade to LBNF/DUNE~\cite{StroboscopicLOI,Angelico:2019gyi}.  By combining a tightly-bunched proton beam with precision timing detectors, a wide-band neutrino beam can be separated into narrower components by selecting events based on their time of arrival at the detector.  The energy spectra of neutrinos with various arrival times at the DUNE far detector are shown in Figure~\ref{fig:time_bunched}.  This technique would require $\sim$100~ps bunches separated by $\sim$2~ns.  Initial studies indicate that the modification to the proton beam bunch structure could be accomplished with a single RF cavity and minimal losses to the number of protons on target. Studies have also indicated that detector timing requirements are achievable in Liquid Argon Time Projection Chamber (TPC) detectors (through the measurement of Cherenkov light)~\cite{Angelico:2020gyn} as well as water-based detectors such as the Theia concept~\cite{TheiaLOI}.

\subsubsection{Tau Neutrinos at LBNF}  
\label{sec:nutausrc}

\begin{figure}[htb]
\centering
\includegraphics[width=.6\linewidth]{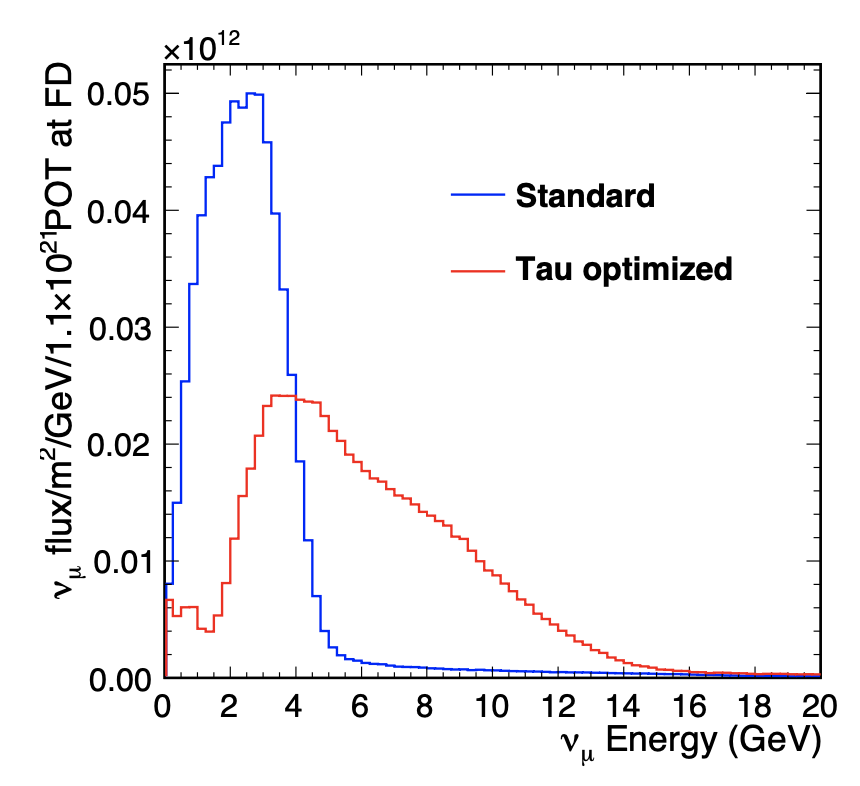}
\caption{Comparison between the standard CP-optimized $\nu_{\mu}$ flux and a potential tau-optimized neutrino fluxes for the LBNF beamline, from~\cite{DUNE:2020ypp}.}
\label{fig:lbnfnutau}
\end{figure}

There is less direct experimental knowledge of tau neutrinos than any other Standard Model particle, and there is great interest in improving this over the next two decades~\cite{tauphysics,Abraham:2022jse}. Long-baseline experiments like DUNE cam make more precise measurements of $\nu_{\mu} \rightarrow \nu_{\tau}$ appearance, probing the unitarity of the PMNS matrix and searching for signs of non-standard interactions.  The nominal LBNF neutrino beam configuration is optimized for the search for CP-violation, and much of the $\nu_{\tau}$ component in the far detector is below the 3.4 GeV kinematic threshold for $\nu_{\tau}$ CC interactions. However, the
beamline could potentially be modified to generate a higher energy $\nu_{\mu}$ beam where a significant portion of the neutrinos are above this threshold. This could
be done by replacing the three focusing horns with two parabolic horns, separated by a distance of 17.5 m, and by modifying the target~\cite{Abraham:2022jse,DUNE:2020ypp}. The neutrino flux from this tau-optimized beam is compared to the standard CP-optimized neutrino flux in Figure~\ref{fig:lbnfnutau}.

%%%%%%%%%%%%%%%%%%%%%%%%%%%%%%%%%%%%%%%%%%%%%%%%%%%%%%%%%%%

\subsection{Measurements to Characterize Neutrino Beams}

Precision measurements of neutrino oscillations requires well-understood neutrino beams.   Current conventional beams have neutrino flux uncertainties approaching 5\%, future precision oscillation measurements may require flux uncertainties at the 2-3\% level.   \emph{In-situ} measurements can be made with beamline instrumentation, such as hadron or muon monitors.  Near detectors are also essential to constraining neutrino fluxes and monitoring the stability of the beam.   \emph{Ex-situ} measurements can also be made with fixed-target beam experiments to better constrain the production of hadrons that decay into neutrinos in the beam target.

\subsubsection{Beam Instrumentation}
\label{Sec:beam_instrumentation}

As neutrino beams reach multi-megawatt power, there are new challenges for beam instrumentation.  R\&D for radiation hardened beam instruments to monitor and characterize the primary beam and target is essential~\cite{Yonehara:2022lmk}.

A novel approach to neutrino beam monitoring is being pursued by the ENUBET (Enhanced NeUtrino BEams from kaon Tagging) experiment~\cite{Brizzolari:2022seb}\cite{Meregaglia:2016vxf}. ENUBET features an instrumented
decay tunnel, allowing for the monitoring of kaon decays in the beamline. 
Counting the large-angle positrons from $K_{e3}$ decays provides a direct $\nu_{e}$ flux determination, while the $\nu_{\mu}$ flux can be determined by monitoring the muons from $K \rightarrow \mu + \nu_{\mu}$  and $\pi  \rightarrow \mu + \nu_{\mu}$.  These particles are measured in a calorimeter that encloses the decay pipe, and an active R\&D program is progressing in Europe with testing at CERN.  The
physics goal of ENUBET is to constrain both the the $\nu_{e}$ and $\nu_{\mu}$ fluxes and beam flavor composition to the 1\% level~\cite{Brizzolari:2022seb}. 

\subsubsection{Hadron Production Measurements}

Several fixed-target experiments such as NA56/SPY~\cite{SPY:1998czp,SPY:1998jku,NA56SPY:1999zez}, HARP~\cite{HARP:2005clh,HARP:2007dqt,HARP:2008sqs},  BNL E910~\cite{E910:2007puw}, NA49~\cite{NA49:2006oyk,NA49:2012jna}, MIPP~\cite{MIPP:2010vnr,MIPP:2014shj}, and earlier experiments have made measurements of the production of hadrons (primarily charged pions) resulting from the interactions of protons on (mostly) thin targets, with proton beam momenta from approximately 9 GeV/$c$ to 450 GeV/$c$.  Several current and future experiments are improving on these measurements, especially for the momenta and targets that are used in current and near future neutrino beams.

\textbf{NA61/SHINE}  

A series of measurements were made in the NA61/SHINE experiment with 31 GeV/c protons on carbon targets~\cite{NA61SHINE:2015bad}, which are crucial for predicting the flux in the J-PARC neutrino beam~\cite{T2K:2012bge}. Starting in 2021, oscillation analyses in the T2K experiment incorporate the charged pion yields measured by NA61/SHINE from a T2K-replica target~\cite{NA61SHINE:2018rhe} to constrain the beam flux
prediction and this will expand to include more species in the future. These measurements have reduced the flux uncertainty  from about 10\% to around 5\% near the flux peak. Future hadron production data at lower energies and
high-statistics replica-target data will reduce these uncertainties further~\cite{T2KLOI}.

In the future, measurements for LBNF/DUNE will also be a high priority for NA61/SHINE, including a high-statistics measurement of particle yields from a replica of the
LBNF/DUNE target~\cite{NA61LOI}.  The feasibility of future low-momentum beams in NA61/SHINE (below 31 GeV/$c$ down to a few GeV/$c$) is being explored, which will useful for flux predictions for lower-energy beamlines and also for understanding secondary interactions in the beamline materials of the J-PARC neutrino beamline.

\textbf{EMPHATIC}

EMPHATIC~\cite{EMPHATIC:2019xmc} is a hadron production detector located at the Fermilab Test Beam Facility. It will be used primarily for thin target measurements, and its spectrometer has a 400 mrad solid angle acceptance for out-going hadrons. EMPHATIC plans a series of upgrades that will ultimately provide particle
identification capabilities up to 15 GeV/$c$, allowing measurements of hadron scattering and production that
will reduce accelerator-based neutrino beam flux uncertainties by a factor of two, to approximately 5\%~\cite{EMPHATICLOI}.  The experiment also plans to collect data downstream of a replica NuMI target and a pulsed focusing horn.  This will be the first direct measurements of hadrons exiting a focusing horn.    

\textbf{DsTau/NA65}

The properties of $\nu_{\tau}$ are still poorly measured and future experiments can probe $\nu_{\tau}$ cross sections. (See Section~\ref{sec:nutausrc} for example.)  The overall accuracy of the these cross section measurements will depend on the $\nu_{\tau}$ flux uncertainty.  The DsTau experiment at CERN (also known as NA65) aims to study $\nu_{\tau}$ production by making a measurement of  $D_{s} \rightarrow  \tau  \rightarrow  \nu_{\tau} + X $ decays following high-energy proton-nucleus interactions ~\cite{DsTau:2019wjb}.   It is expected to provide an independent $\nu_{\tau}$ flux prediction for future neutrino beams with an accuracy under 10\%, which would be a substantial improvement over the current $\sim50\%$ uncertainties on this process.

\subsubsection{Near Detectors}

Near neutrino detectors in long-baseline neutrino experiments are also an essential tool to understanding neutrino beams. They measure the initial unoscillated neutrino energy spectra, provide essential
input for the neutrino interaction model, and monitor the beam.

One difficulty for predicting the unoscillated event rate is the fact that the near and far detectors do not have identical fluxes due to differences in the geometrical acceptances at the two locations and the presence of oscillations at the far detector. So extrapolating from the near detector to the far detector requires well-constrained neutrino interaction models as a function of neutrino energy. As detectors move to larger off-axis angles
relative to the beam direction, the peak energy of the
neutrino energy spectrum is lowered and the beam energy becomes narrower as the high-energy tail is reduced (a property that is exploited by the off-axis beams used in the T2K and NOvA experiments).  The novel  Precision Reaction-Independent Spectrum Measurement or ``PRISM"  concept will be employed by the next generation of long-baseline experiments, collecting data in a near detector at several angles and then taking linear combinations of the measurements at several locations to effectively reproduce data with different neutrino energy spectra.  These new spectra can be narrow ``pseudo-monoenergetic" Gaussians to measure cross sections or even the oscillated energy spectrum at the far detector.

\begin{figure}[ht]
    \centering
    \includegraphics[width=0.45\textwidth]{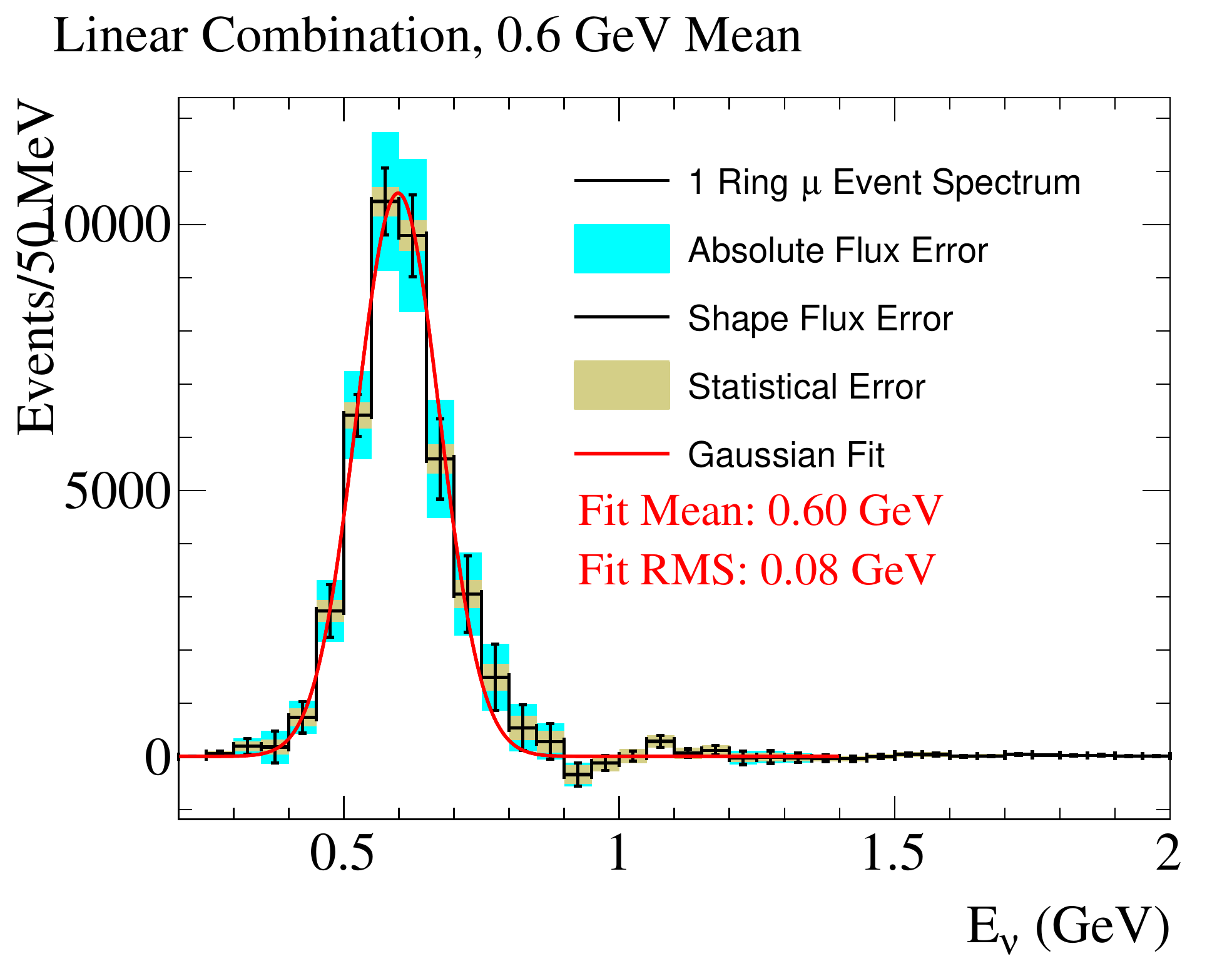}
    \includegraphics[width=0.45\textwidth]{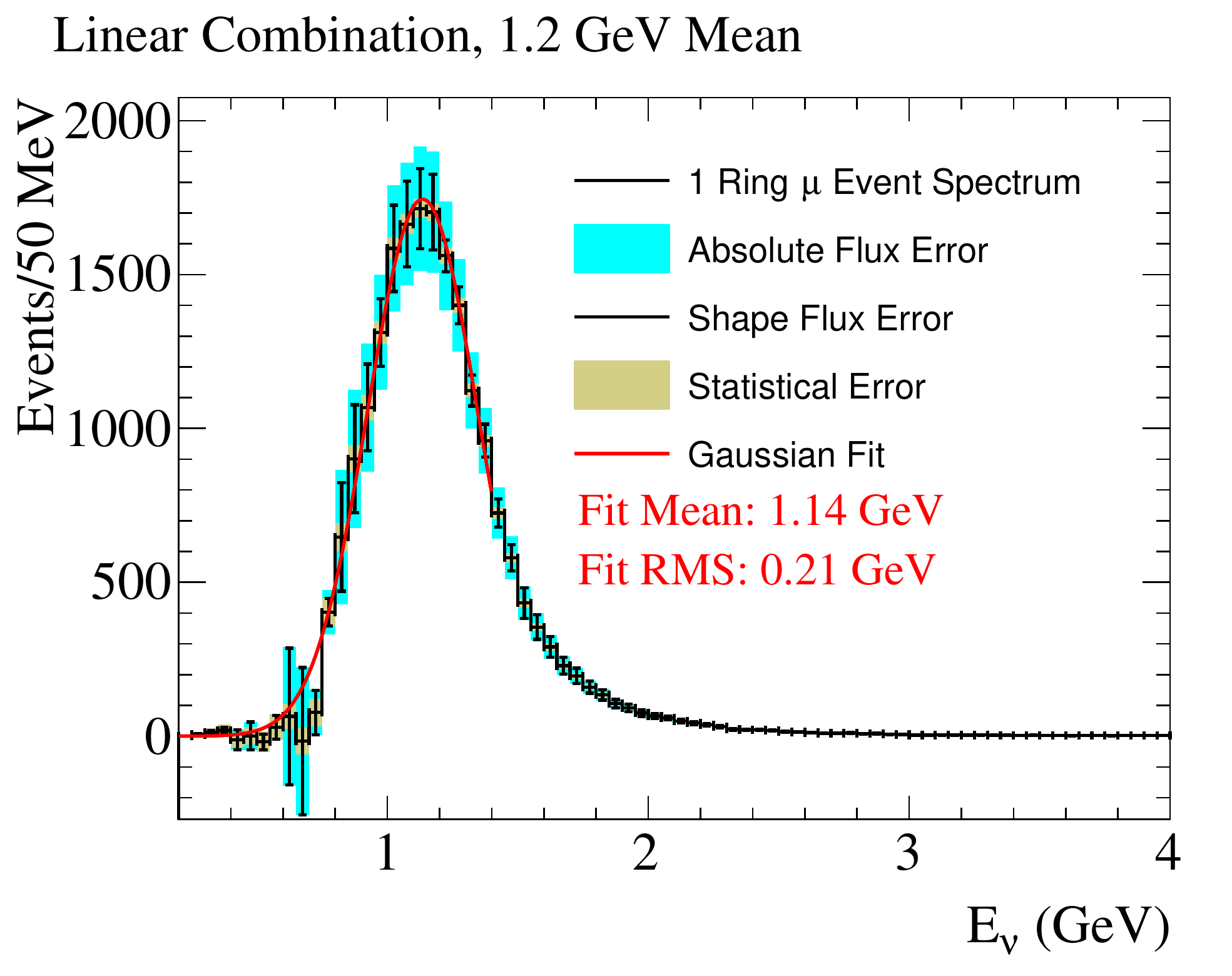}
    \caption{Two NuPRISM “pseudo-monochromatic” spectra centered at 0.6 (left)and 1.2 (right) GeV, built of linear combinations of off-axis fluxes,  from~\cite{nuPRISM:2014mzw}.}
    \label{fig:numprismflux}
\end{figure}

Proposed for use in the J-PARC neutrino beam, nuPRISM~\cite{nuPRISM:2014mzw} will deploy a new near water Cherenkov detector that spans the range of 1$^{\circ}$-4$^{\circ}$ off-axis. Some examples of “pseudo-monochromatic” energy spectra that can be built from linear combinations of the data taken at several off-axis locations are shown in Figure~\ref{fig:numprismflux}.

The DUNE near detector comprises three subsystems ~\cite{DUNE:2021tad}.  The first component encountered by the beam is a liquid argon TPC, ND-LAr.  Just downstream of ND-LAr is a high-pressure gaseous argon TPC surrounded by a calorimeter and a magnet, ND-GAr.  ND-LAr and ND-GAr can take data at distances of up to 33 m off-axis of the beam center.
This capability to probe different neutrino spectra in off-axis positions is part of the DUNE-PRISM concept~\cite{DUNE:2021tad}. The third component of the DUNE near detector, the
SAND detector~\cite{SANDLOI}, consisting of of a large magnet and calorimeter surrounding a tracker, will remain on-axis to monitor the stability of the beam spectrum. SAND can also determine the absolute flux of the beam by measuring $\nu e \rightarrow \nu e $ elastic scattering~\cite{SANDLOI}.  For DUNE to achieve its ultimate physics goals, the full near detector suite will be essential, in addition to having the full 2.4 MW beam~\cite{DUNE:2022aul,DUNE:2022yni}.

%\cleardoublepage

\section{Decay-At-Rest Sources}
\label{sec:spall}

\subsection{Spallation Neutron Sources}
Spallation neutron sources utilize intense $\sim$1-3~GeV proton beams on metal (usually liquid mercury) targets to produce controlled sources of neutrons, mainly for the study of materials science. Remarkably, these sources double as the most powerful accelerator-based sources of neutrinos in the world, and a number of relevant neutrino experiments at these sites are presented in this section. With the target usually surrounded by various forms of concrete and steel/iron shielding, the pions, muons, and kaons produced from the proton-on-target interactions tend to come to rest before decaying, producing a well understood flux of isotropic neutrinos [$\pi^+$ ($\rightarrow \mu^+ + \nu_\mu$; 29.8~MeV), $\mu^+$ ($\rightarrow e^+ + \overline{\nu}_\mu + \nu_e$; 0-52.8~MeV), and K$^+$ ($\rightarrow \mu^+ + \nu_\mu$; 236~MeV)]. The neutrinos from spallation sources can be employed for a wide variety of neutrino physics measurements, including short-baseline oscillations, neutrino interaction studies relevant for oscillation searches at both short- and long-baseline and astrophysics, coherent neutrino-nucleus scattering and related new physics searches involving non-standard neutrino interactions, and probes of dark matter using nearby neutrino detectors. 

\begin{figure}
\centering
\includegraphics[width=4.5in]{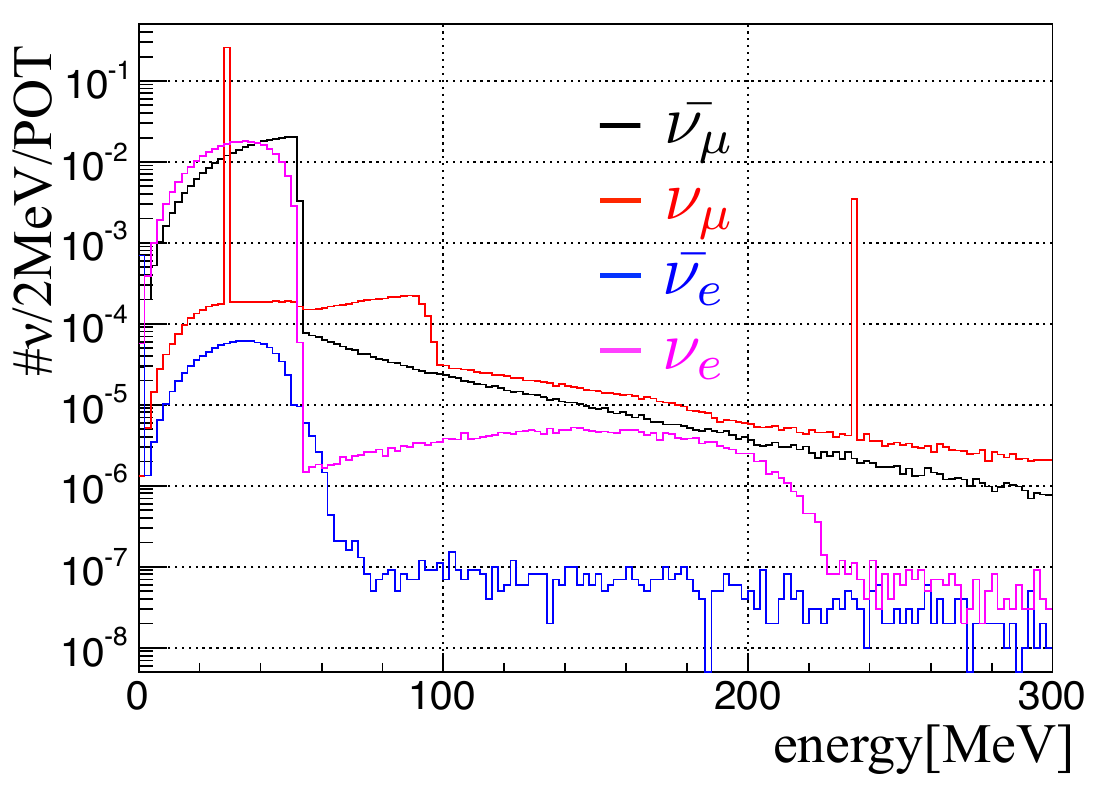}
\vspace{-.0cm}
\caption{The expected J-PARC spallation neutron source flux, originating from 3~GeV protons on a mercury target, at the JSNS$^2$ detector~\cite{Ajimura:2017fld}.}
\label{dar_flux}
\end{figure}

As compared to other facilities, spallation neutron (and non-spallation beam dump, discussed in Section~\ref{beamdumpsources}) sources  offer a number of advantages that make them highly attractive for probing the properties of the neutrino: (1) \textbf{Power.} First and foremost, these sources are incredibly powerful and produce unmatched fluxes of neutrinos. For example, the 1.4~MW spallation neutron source at ORNL (1~GeV) and 830~kW (1~MW planned) spallation neutron source at J-PARC (3~GeV) are both well over an order of magnitude more powerful than the oft-used Booster Neutrino Beam (32 kW; 8~GeV) at Fermilab, whilst being used for highly-related and complementary short-baseline oscillation physics. (2) \textbf{Timing.} Spallation neutron sources also tend to have impressive timing capabilities, with very low duty factors, which can be used by surface-based neutrino experiments to mitigate steady-state backgrounds, especially cosmics. For example, without considering neutrino parent lifetimes, the J-PARC MLF source duty factor is $\sim5\cdot10^{-6}$, with two $\sim$100~ns bunches separated by 540~ns at 25~Hz. (3) \textbf{Well understood flux.} Understanding the properties of the neutrinos when they are created remains of tantamount importance to almost all measurements in the field of neutrino physics. While normalization can be somewhat uncertain, the energy dependence and flavor composition of the pion, muon, and kaon decay-at-rest sources produced by these spallation neutron complexes is extremely well understood. Carefully understanding the energy dependence of the neutrinos at creation is incredibly powerful for probing neutrino oscillations and interactions, in particular. These sources also tend to have only a small fraction (sub-\%-level) of decay-in-flight neutrinos amongst the dominant decay-at-rest components. For example, Figure~\ref{dar_flux} shows the J-PARC MLF spallation neutron source neutrino flux. The $\pi^+$ ($\rightarrow \mu^+ + \nu_\mu$; 29.8~MeV), $\mu^+$ ($\rightarrow e^+ + \overline{\nu}_\mu + \nu_e$; 0-52.8~MeV), and K$^+$ ($\rightarrow \mu^+ + \nu_\mu$; 236~MeV) decay-at-rest components can easily be seen, dominating (by orders of magnitude) the rest of the (decay-in-flight) flux.

Although spallation neutron sources for neutrino experiments have a number of attractive features, somewhat unfortunately, at least for neutrino physics, they are primarly designed with an eye for neutron physics and materials science. The J-PARC Spallation Neutron Source is inside the ``Materials and Life Science Facility'' (MLF) building, after all. This overarching issue with spallation neutron sources is a problem for neutrino physics measurements for three reasons, in particular: (1) \textbf{Shielding.} In consideration of background needs for neutrino experiments, the shielding near these sources can be inadequate and unpredictable. Beam-based neutrons and gammas, for which accelerator timing cannot always mitigate effectively since they can arrive with a similar time structure as signal neutrinos, can present an issue for neutrino measurements, especially those involving rare event searches. In general, it can be difficult to find a low-background space close to the source to accommodate a (large enough) neutrino detector and peripherals. (2) \textbf{Facilities issues.} Just because neutrinos go right through the materials surrounding the source does not mean that placing a neutrino detector proximal to a spallation neutron source is easy, even in the case that shielding is adequate. The 1/$r^2$ dependence of an isotropic decay-at-rest neutrino source tends to mean that detector placement close to the source is optimal, although maximizing sensitivity to particular $\Delta m^2$ values in the context of a short-baseline oscillation search can change this view (e.g. maximizing sensitivity to $\Delta m^2\sim1$~eV$^2$ oscillations for 40~MeV neutrinos, chacteristic of a muon decay-at-rest source, for example, is accomplished with a baseline of 50~m). As an example of these facilities complications, the JSNS$^2$ collaboration~\cite{JSNS2:2021hyk} is required to completely remove their 24~m baseline detector, including both the 50~tons of liquid scintillator inside the vessel and the vessel itself, separately, every year so that the MLF target and target area can be accessed for maintenance. (3) \textbf{Source purity.} The neutron beamlines and variety of different materials close to the target complicate understanding, and can also detract from the purity, of the source. In general, the target and surrounding geometry at these sources is/has not been optimized for producing a pure flux of neutrinos. In particular, the intrinsic $\overline{\nu}_e$ component of a spallation neutron source is highly dependent on the amount of low-A materials close to the target. For neutron beamlines, these kinds of materials may be heavily utilized, depending on the facility. Such materials reduce the probability that produced $\pi^-$, and subsequent $\mu^-$, capture on a nucleus before decaying, which can lead to significant intrinsic $\overline{\nu}_e$ production and reduces the overall purity of the $\pi^+$ and $\mu^+$ decay-at-rest source. A significant flux of background $\overline{\nu}_e$ can seriously compromise a search for $\overline{\nu}_\mu \rightarrow \overline{\nu}_e$ appearance.

%BNB in dump mode, similar properties as a spallation source (decay at rest source)

%Modest detectors, but impactful physics. JSNS2-II competitive for very modest price.

\begin{figure}
\centering
\includegraphics[width=6.in]{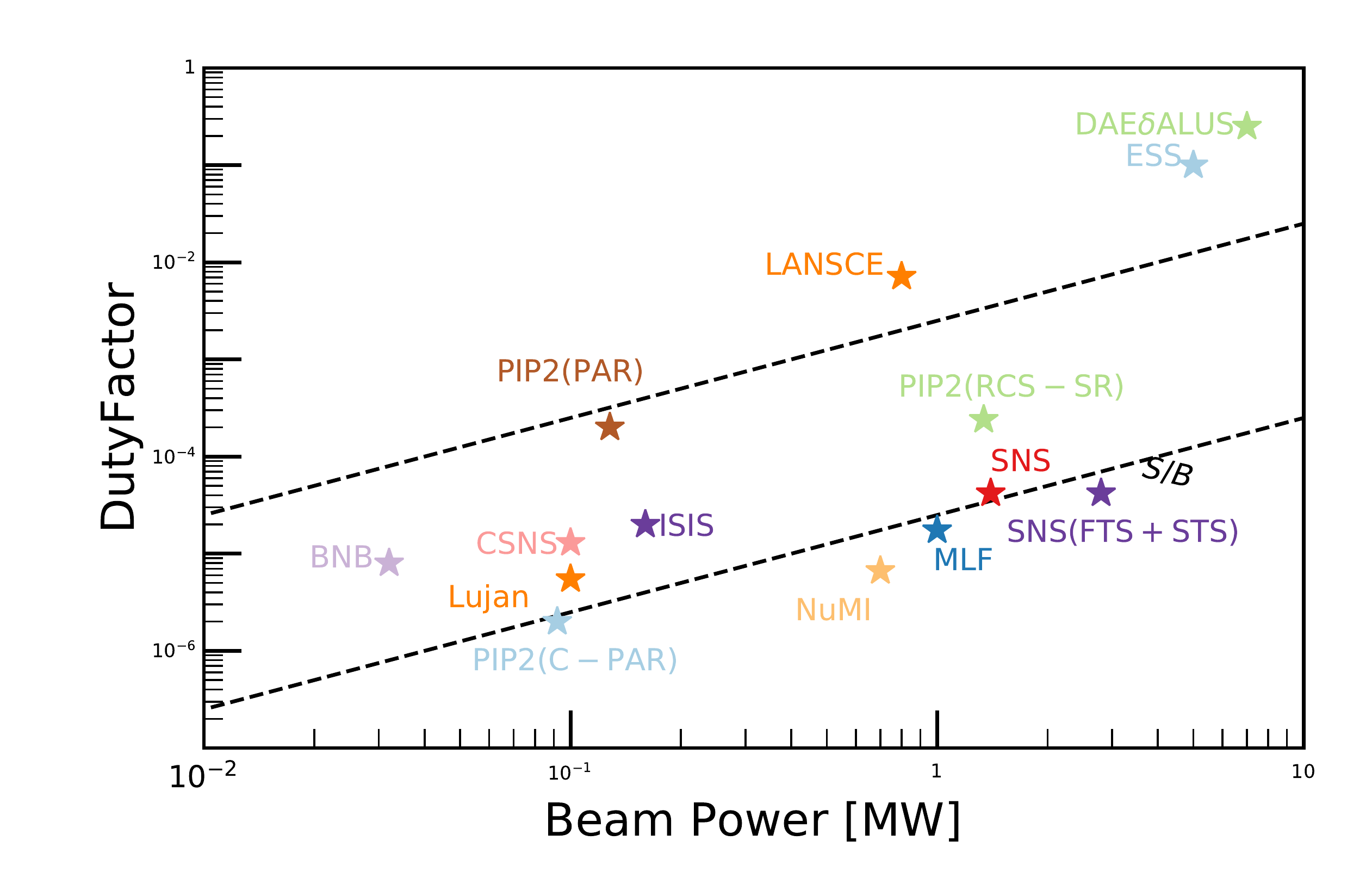}
\caption{Duty factor vs. beam power for a number of proton accelerator facilities around the world, including the spallation neutron sources discussed in the text~\cite{Kelly:2021jgj,Toups:2022yxs,Zettlemoyer}. The over-simplistic iso-lines of signal-to-background shown are in consideration of power as a proxy for signal and beam duty factor as a proxy for steady-state background (e.g. cosmics) rejection factor.}
\label{landscape_sns}
\end{figure}

Despite the overall attractiveness of spallation neutron sources for neutrino physics, there are, at present, very few neutrino experiments making measurements at these sources: CCM~\cite{CCM:2021leg}, COHERENT~\cite{COHERENT:2017ipa}, and JSNS$^2$~\cite{JSNS2:2021hyk} are the only experiments presently taking data. In fact, for many years before COHERENT started taking data at ORNL in 2015 there were no neutrino detectors taking advantage of these existing, powerful, and pure sources. Fortunately, this lack of neutrino experiments at spallation neutron sources paradigm is slowly changing as the importance of these facilities to neutrino physics becomes more clear to the wider scientific community and funding agencies. Indeed, there are a wide variety of proposals, featuring both physics measurements and technological development, that can take advantage of these \textit{existing} and \textit{highly upgradeable} sources in the future. The current landscape of accelerator-based sources (including spallation neutron sources), near-past, present, and future, in terms of primary beam energy, accelerator duty factor, and beam power is shown in Figure~\ref{landscape_sns}.

Notably, these spallation neutron sources (and, usually, the associated neutrino detectors) can be utilized for sensitive dark matter searches as well (see, e.g., Refs.~\cite{Kahn:2014sra,Jordan:2018gcd}). While outside of the scope of this Neutrino Frontier focused document, it bears mentioning that, for certain classes/masses of dark matter, high-power, GeV-scale proton-on-dump configurations, coupled with a downstream, nominally-neutrino (or, more dedicated, dark matter) detector, may represent an optimal setup for achieving sensitivity. Put simply, there is a non-negligible possibility that the spallation neutron/neutrino sources discussed here double as dark matter sources. 

%cite miniboone dark matter search, DAEdALUS toups idea paper. pseudo-dirac Jordan paper

%flux plot showing pion muon kaon DAR.

%Scholberg-style flux plot, with and without muon timing.

\subsubsection{Future and Far-Future COHERENT Program, including ORNL Second Target Station and Flux Normalization Measurements}
\label{coherent}
The COHERENT experiment relies on the 1~GeV, 1.4~MW Spallation Neutron Source (SNS) at Oak Ridge National Lab for measuring coherent elastic neutrino-nucleus scattering (CE$\nu$NS) with an array of detector technologies and varying nuclear targets~\cite{Akimov:2022oyb}. Following on from COHERENT's impactful first observation of CE$\nu$NS~\cite{COHERENT:2017ipa}, the collaboration now seeks to measure this process across a wide range of nuclear targets, while improving statistics, kinematic resolutions, and thresholds. Given the well predicted Standard Model cross section, including as a function of nuclear target, these measurements are highly sensitive to a range of new physics scenarios that would cause a deviation from the precise and well understood prediction (see, e.g., Ref.~\cite{Lindner:2016wff}).
Substantial sensitivity increases are expected with more neutrino events and bigger and better detectors featuring an array of nuclear targets.

Along with the detector size and technological improvements themselves, such physics is enabled by the duty factor and world-leading power of the Spallation Neutron Source at ORNL. However, there is significant room for improvement, especially in terms of the facilities issues described above. Aside from higher power, and therefore more neutrinos, perhaps driven by a second target station, dedicated space in low background areas for COHERENT's neutrino detectors, increasing in number and size, is paramount to the success of the experiment moving forward. Further, knowledge of the flux normalization, which is currently at the 10\% level, can provide another handle on the characterization of coherent neutrino-nucleus scattering, relevant for all future measurements, on any number of nuclear targets~\cite{Akimov:2022oyb}. COHERENT is also expected to take advantage of the properties of the neutrino source as a testbed for detector technology R\&D using a pulsed, low-background neutrino source.

Aside from coherent scattering, there are also plans to study neutrino inelastic interactions on a number of targets within the COHERENT collaboration~\cite{Akimov:2022oyb}. Such measurements are highly relevant for understanding supernova evolution and, eventually, being able to characterize and fully take advantage of the neutrinos collected from the next supernova neutrino burst. In particular, many of the most relevant neutrino-inelastic cross sections are poorly understood theoretically and either unmeasured or poorly measured.

\subsubsection{JSNS$^2$ at JPARC}
\label{jsns2}
\begin{figure}
\centering
\includegraphics[width=6.in]{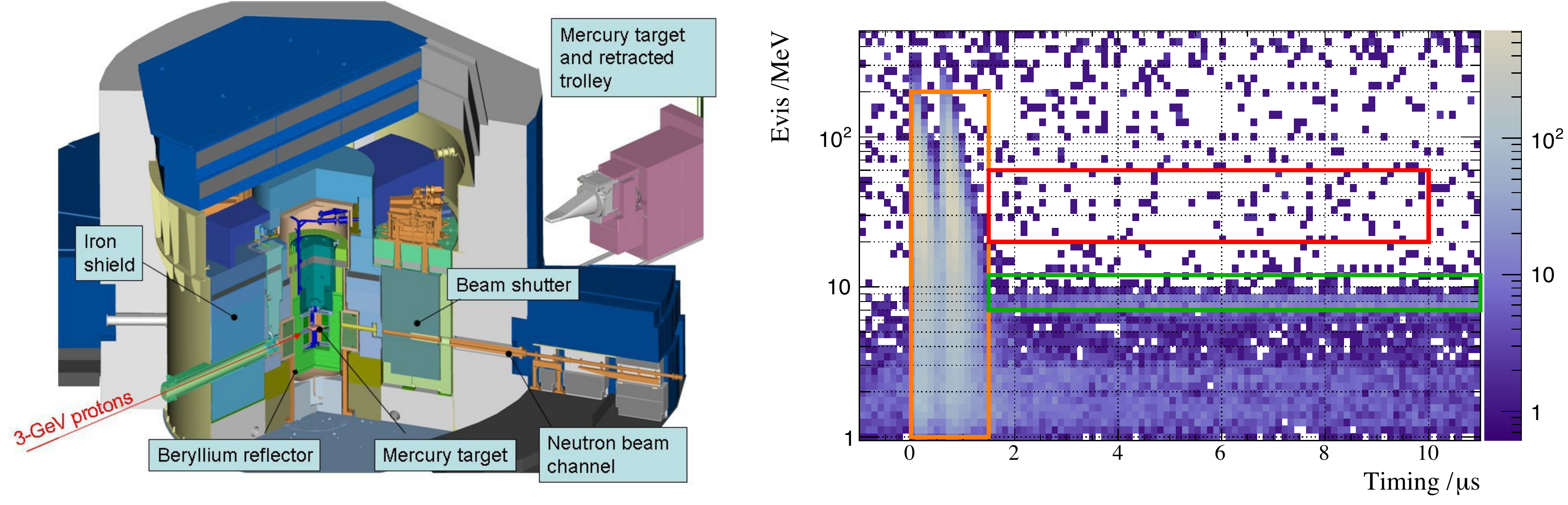}
\caption{(Left) A schematic of the J-PARC spallation neutron (neutrino) source.  (Right) The timing characteristics of the beam (two 100~ns bunches separated by 540~ns at 25~Hz), as measured by JSNS$^2$~\cite{Hino:2021uwz}. The orange represents the on-bunch events, while the red and green show the prompt and delayed signal event selection regions relevant for their muon-induced-$\overline{\nu}_\mu \rightarrow \overline{\nu}_e$ IBD search, respectively.}
\label{fig:jsns}
\end{figure}

The J-PARC Sterile Neutrino Search at the J-PARC Spallation Neutron Source (JSNS$^2$) uses liquid scintillator loaded with Gadolinium detector technology combined with a MW-scale spallation neutron source to provide sensitivity to short-baseline oscillations~\cite{JSNS2:2021hyk}, as a $\textit{direct}$ test of the LSND anomaly~\cite{LSND:1996ubh}, and a set of neutrino interaction measurements relevant for understanding oscillations, nuclear physics, and supernova neutrinos~\cite{Ajimura:2017fld}. Like other decay-at-rest experiments at spallation neutron sources, JSNS$^2$ takes advantage of the tight accelerator timing window (two~100~ns bunches separated by 540~ns at 25~Hz) to mitigate steady-state background. A drawing of the source and a plot displaying the measured timing characteristics of the beam are shown in Fig.~\ref{fig:jsns}. The ongoing physics data taking run began in summer 2020~\cite{Hino:2021uwz} with a 50~ton (17~ton fiducial volume) detector 24~m away from the 3~GeV proton and currently 830~kW (with 1~MW expected in the next couple years) pion, muon, and kaon decay-at-rest neutrino source. Oscillation sensitivity comes from studying the $\mu^+$ decay-at-rest ($\mu^+\rightarrow e^+ + \nu_e + \overline{\nu}_\mu$) component of the source and searching for $\overline{\nu}_\mu\rightarrow \overline{\nu}_e$ appearance using the inverse beta decay channel ($\overline{\nu}_e+p\rightarrow e^+ + n$) with 96~PMTs inside of the detector (and an additional 24~PMTs in the veto surrounding the target volume). Such events will produce a double-coincidence light signal from the initial $e^+$, which provides an antineutrino energy determination, followed by a characteristic $\sim$8~MeV neutron capture on Gadolinum signal $\sim30$~$\mu$s later. In addition, a 32~ton fiducial volume far detector (``JSNS$^2$-II''), which will sit at a baseline of 48~m, is now fully funded, approved, and at an advanced stage of construction~\cite{Ajimura:2020qni}. First data with JSNS$^2$-II is expected in 2023/2024. 

\subsubsection{ESS Dump/spallation neutrino source}
\label{ess}
The European Spallation Source Neutrino Super Beam (ESS$\nu$SB) will service a long-baseline neutrino oscillation experiment (see Section~\ref{essnusb}), featuring a proposed megaton water Cherenkov far detector placed at the second neutrino oscillation maximum in order to extract $\delta_{CP}$~\cite{Alekou:2022mav}. In addition, there are plans to create a MW-class fixed-target decay-at-rest source with the accelerator facility. Similar to ORNL's SNS and J-PARC's MLF, such a source would provide a powerful means to accomplish decay-at-rest physics (see, e.g., Ref.~\cite{Baxter:2019mcx}). Construction and commissioning of the beam facility is expected in the 2030s~\cite{Alekou:2022mav}.

\subsubsection{LANSCE (Los Alamos) and CCM}
\label{lansce}
The Lujan center at the Los Alamos Neutron Science Center (LANSCE) provides 800 MeV (100~kW) protons onto a tungsten target with $\sim300$~ns pulses at 20~Hz, which produces an intense source of decay-at-rest neutrinos. The 10~ton Coherent CAPTAIN Mills (CCM) liquid argon detector there is sampling these neutrinos in a search for coherent neutrino-nucleus scattering, with an eye towards new physics sensitivity, including short-baseline neutral-current-only oscillations~\cite{CCM:2021leg}. The latter possibility is particularly exciting because a disappearance signature in a neutral current channel would be both smoking gun evidence for oscillations involving a sterile flavor, and would also provide a uniquely direct measurement of the sterile flavor composition of the fourth (or more) sterile mass eigenstate~\cite{Anderson:2012pn}.

Upgrades to the Lujan source may allow the beam spill width to be compressed to 30~ns with little intensity loss~\cite{VandeWater:2022qot}. With a focus on the fast $\pi^+$ decay-at-rest component signal $\nu_\mu$ at 29.8~MeV, rather than the much slower $\mu^+$ decay-at-rest neutrinos, such a timing improvement would increase the signal-to-background by a factor of $\sim$100, with resulting sensitivity increase by an order of magnitude for sterile neutrino and dark matter searches.

% note that the above line throws an error.  Apprarently one cannot use math mode in headings.  Will need to fin a way to fix this.

\subsection{Beam Dump Sources}
\label{beamdumpsources}
The physics measurements discussed above in the context of spallation neutron sources are also possible at dedicated, non-spallation-neutron facilities, again featuring GeV-scale proton interactions on a beam dump (target with surrounding shielding, and no decay pipe) and there are a number of future ideas for pursuing this fixed-target physics and more, including with a 100s~of~GeV beam. Beam-dump facilities can be thought of as conventional neutrino beams without a decay volume.  Neutrinos produced via decay-in-flight are suppressed, creating beams enriched in decay-at-rest neutrinos, tau neutrinos, and, potentially, beyond-the-standard model particles.  A number of such facilities are being considered for the future.
\subsubsection{$\mathcal{O}(1)$ GeV Fixed Target Facility at Fermilab}
The PIP-II project is constructing a new superconducting LINAC at Fermilab that will be the proton driver for LBNF/DUNE.  This new LINAC opens the possibility of additional facilities at Fermilab beyond LBNF, including an $\mathcal{O}(1$ GeV) beam dump~\cite{O1GeVBeamDumpLOI}.  This would be a completely new facility and is envisioned to include a variety of detectors located at different distances from the target and angles with respect to the primary proton beam.  Being designed from the ground up with HEP in mind, it would have some advantages over e.g. spallation sources in that it could be designed to supress neutron production and with lighter targets that enhance proton production.  This facility would require a new storage ring (possibly located within the existing Booster enclosure) that would produce $\sim$320~ns pulses, initially operating at 800 GeV beam energy, 100~kW beam power and $\mathcal{O}(10^{-5})$ duty factor, but upgradable to $\mathcal{O}(1$ GeV) beam energy and higher power.  Detectors located at this facility would be sensitive to dark sector particles produced directly in hadronic interactions or through the decay of light mesons, and would likely make use of the CE$\nu$NS channel. Notably as well, a CE$\nu$NS-based disappearance search provides sensitivity to oscillations involviong a sterile neutrino~\cite{Anderson:2012pn}.   

\subsubsection{$\mathcal{O}(10)$ GeV Fixed Target Facility at Fermilab}
Another possible beam dump facility at Fermilab would be a new dedicated target station in the Booster Neutrino Beam (BNB)~\cite{O10GeVBeamDumpLOI}.  With the increases in power to the BNB made possible by the PIP-II Linac upgrade, such a facility could receive $6\times10^{21}$ POT in five years of operation while 35~kW of protons are also delivered to the BNB neutrino beam.  The BNB neutrino detectors would be used for both physics programs.  The  primary physics driver of this facility (search for sub-GeV Dark Matter) is not neutrino physics, and the facility would in fact be optimized to minimize the neutrino flux.  However, it would be a source for mono-energetic neutrinos from kaon decay-at-rest that could have several uses, including a search for sterile neutrinos~\cite{Spitz:2012gp,Spitz:2014hwa,Axani:2015dha,MiniBooNE:2018dus}.   

\subsubsection{SHiP}

\begin{figure}
\centering
\includegraphics[width=\linewidth]{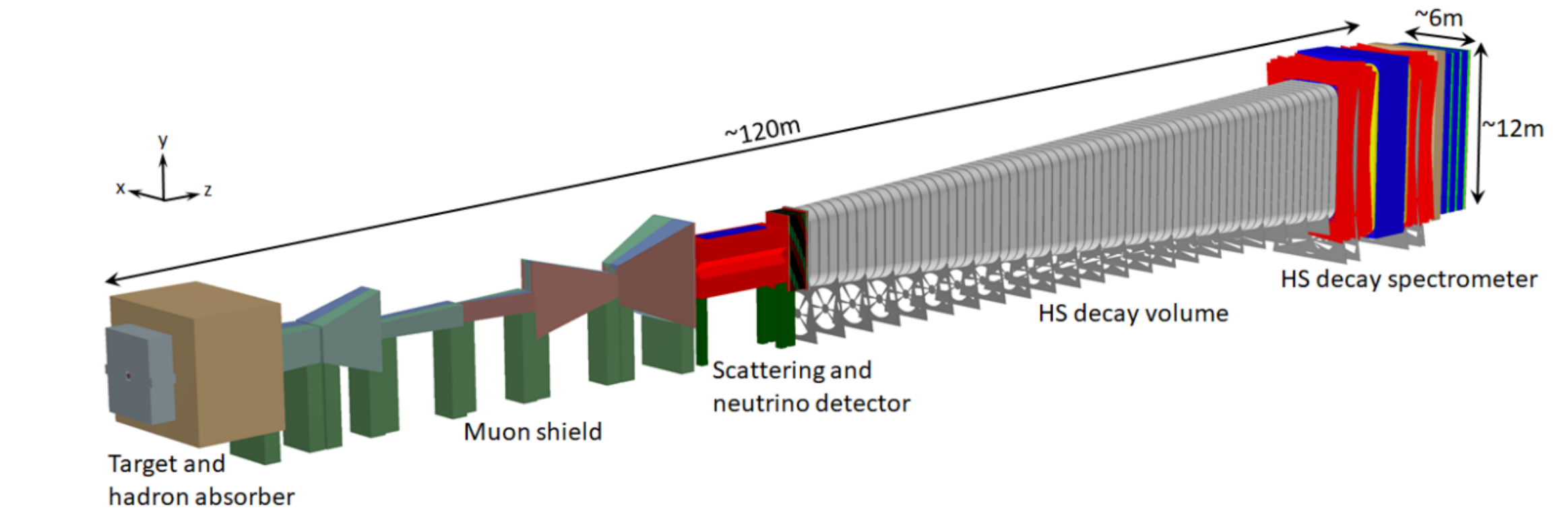}
\caption{A schematic of the SHiP experiment at the CERN SPS.  Figure taken from~\cite{SHIP1}}.  
\label{fig:ship}
\end{figure}
The Search for HIdden Particles (SHiP) is a proposed beam dump experiment at the CERN SPS with the goals of searching for dark-sector particles, measuring tau neutrinos, and studying charm mesons~\cite{SHIP1}.  A schematic of the beam dump and detector is shown in Figure~\ref{fig:ship}.  A slow-extracted beam of 400 GeV protons with collide with a 12 interaction length molybdenum target, with an expected exposure of $4\times10^{19}$ POT per year (or $2\times10^{20}$ over the planned 5-year run).  In five years, the experiment would collect $\mathcal{O}(10^4)$ $\nu_{\tau}$ and $\bar{\nu_\tau}$ candidates with a mean $\nu_\tau$ ($\bar{\nu_\tau}$) energy of 52 GeV (70 GeV).  This would be by far the largest sample of tau neutrinos ever produced and would enable precision measurements of $\nu_\tau$ and $\bar{\nu_\tau}$ deep inelastic scattering cross sections.  In particular, the experiment would be sensitive to the $F_4$ and $F_5$ structure functions, which are suppressed in $\nu_\mu$ and $\nu_e$ interactions and have never been measured.  

%\cleardoublepage

\section{Nuclear Reactors}
\label{sec:reac}

\subsection{Nuclear Reactors as Neutrino Sources}\label{sec:nrintro}
%\subsection{Importance of Precise Reactor / Nuclear Knowledge}

Nuclear reactors have had a central role in experimental neutrino physics ever since the discovery of these elusive particles in 1956~\cite{Cowan:1956rrn}. As neutrino sources, they have the following advantages: 
\begin{itemize} 
\item {\bf Intensity}: About $2\times10^{20}$ electron antineutrinos ($\bar\nu_e$'s) are produced every second from a 1~GW$_\textrm{th}$ commercial reactor core. They are in fact the most intense man-made source of neutrinos. 
\item {\bf Availability}: There are currently over 400 nuclear reactors in operation in more than 30 countries worldwide. The world's nuclear power capacity continues to increase steadily, with about 55 new reactors currently under construction~\cite{wnassoc}. 
\item {\bf Flavor-purity}: Antineutrinos are produced from the beta decays of nuclear fission products through the $n \rightarrow p + e^- + \bar{\nu}_e$ reaction. Consequently, only $\bar{\nu}_e$'s are emitted. 
\item {\bf Cost-effectiveness}: Researchers do not typically have to bear the costs involved in designing, building, and operating nuclear reactors, whose primary goals include power generation (commercial reactors) and neutron production (research reactors). This in turn allows for small experiments with a lower barrier to entry and faster timescales from design to data-taking compared to large international projects, providing important opportunities for workforce development~\cite{reactorwp}. 
%\item Predictability: knowing the reactor's power and fission fractions (fraction of the total amount of fissions occurring due to a certain isotope .... Physics behind are well-understood. There are models that allow to predict rate and shape to a few percent, and also rich data acquired by several experiments above the $1.8$~MeV IBD threshold.(to be continued) 
\end{itemize} 

%% Note to self: should talk about some of the disadvantages? Continuous flux, high neutron flux at research reactors, difficult site access. 

The above list includes at least one more advantage, which is {\bf predictability}. The physics behind reactor antineutrino emission is well understood. Commercial reactors, which typically operate in the GW$_\mathrm{th}$ regime and are the most commonly used type of reactor for physics purposes, use fuel with a relatively low amount of $^{235}$U enrichment, commonly denoted as low-enriched uranium (LEU). Over 99.7\% of reactor antineutrinos emitted from these reactors originate in the fission of four isotopes, $^{235}$U, $^{239}$Pu, $^{241}$Pu, and $^{238}$U, with the first two accounting for $\gtrsim 80\%$ of all $\bar{\nu}_e$'s. Research reactors, which typically operate in the $\sim$10-100~MW$_{\textrm{th}}$ regime, are smaller in spatial extent and use fuel with a higher level of $^{235}$U enrichment, commonly denoted as highly-enriched uranium (HEU). In this case, the great majority of $\bar{\nu}_e$'s are emitted from the fission of $^{235}$U. For both types of reactors, the resulting $\bar{\nu}_e$ spectrum decreases rapidly with neutrino energy, reflecting the endpoint energies of the decaying isotopes that range from a fraction of an MeV until about 10~MeV. Knowing a reactor's power and its fission fractions, namely the amount of fissions occurring with a certain isotope over the total, allows the prediction of the reactor antineutrino rate and spectral shape to roughly within $5\%$. Several measurements have been carried out in the last decade that benchmark these models and even allow, in some circumstances, making data-driven predictions with better precision. These measurements have revealed several inconsistencies with the models that are still not fully resolved but which do not take away from the fact that reactor antineutrino fluxes can already be predicted with very good precision.  

\begin{figure}[!h]
  \centering
  \includegraphics[width=0.95\textwidth]{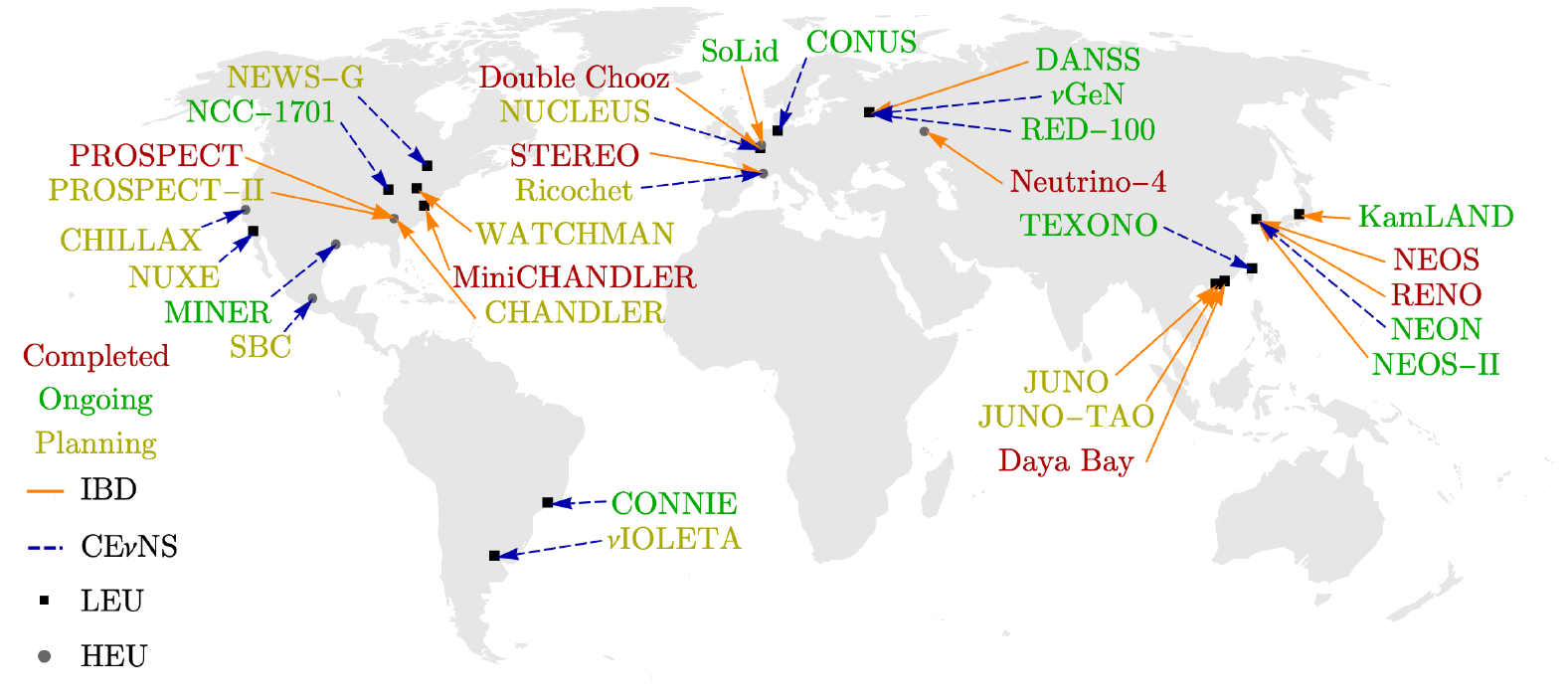}
  \caption{Map of planned, current, and recently completed reactor antineutrino experiments relying on the IBD or CE$\nu$NS detection channels. The text color indicates experimental status, while the arrow color indicates the interaction channel used by the experiment. Only completed experiments taking data after 2010 are included. Image from Ref.~\cite{reactorwp}.}
  \label{fig:RxMap}
\end{figure}

Because of all these reasons, nuclear reactors have been and continue to be widely used in experimental neutrino physics. 
%Breakthroughs achieved with reactor antineutrinos include the discovery of these particles~\cite{Cowan:1956rrn}, the first observation of neutrino oscillation using terrestrial neutrinos~\cite{KamLAND_rate}, and the first unambiguous observation of a non-zero value of the $\theta_{13}$ mixing angle~\cite{DayaBay:2012fng,bib:dc,RENO:2012mkc}. 
%Reactor neutrino experiments also continue to provide some of the most precise measurements of neutrino oscillation parameters to date~\cite{ParticleDataGroup:2020ssz} and some of the most stringent tests for the existence of sterile neutrinos~\cite{lightsterilewp}. 
%In the coming decade, nuclear reactors will continue to be one of the best tools at our disposal for experimental neutrino physics. 
As shown in Fig.~\ref{fig:RxMap}, a large number of reactor neutrino experiments are currently in operation or in the planning stage throughout the world. The goals of these experiments include making precision measurements of neutrino oscillations, searching for sterile neutrinos, making new or more stringent tests of the Standard Model, and demonstrating the use of neutrino detectors for nuclear safeguards~\cite{reactorwp}. The great majority of reactor experiments to date have relied on the Inverse Beta Decay (IBD) reaction, given by $\bar{\nu}_e + p \rightarrow e^+ + n$, but a new generation of experiments relying on coherent elastic neutrino-nucleus scattering (CE$\nu$NS) is in the early stages. The IBD reaction has the advantage that both products can be detected, which greatly suppresses the backgrounds, but has the disadvantage that $\bar{\nu}_e$'s must have a minimum energy of 1.8~MeV to initiate it. In contrast, the CE$\nu$NS reaction is virtually threshold-less and has a much larger cross section, several orders of magnitude higher~\cite{Cogswell:2016aog}, but has a nuclear recoil as its only signal, requiring a combination of extremely low and well-understood background contamination, very low energy detection thresholds, and specialized materials for its observation. Both types of reactor experiments, as well as those relying on neutrino-elastic scattering and possibly on neutral current inelastic nuclear scattering, will continue to play a central role in advancing our knowledge of the neutrino and its role in the universe.

%The direct neutrino measurements carried out by these experiments 
%One of the outcomes of this vigorous experimental program is 
%This vigorous experimental program, in combination with inputs from the nuclear and theoretical physics communities, 
%Some of the neutrino measurements produced in the next decade will play an important role in furthering our knowledge of reactor antineutrino fluxes. 
Precise and self-consistent knowledge of reactor antineutrino emission is essential to the community for several reasons. Very importantly, such knowledge bolsters the global experimental effort by affording reactor experiments a greater physics reach without the need for ancillary measurements. Resolving the discrepancies that have arisen between the data and the models will result either in important discoveries or in more efficient uses of the community's resources. The effort to better understand and model reactor antineutrino fluxes benefits other areas beyond basic neutrino research, such as by spurring investments and improvements in non-neutrino nuclear physics measurements and nuclear data. Reactor antineutrino measurements are in fact a powerful probe of nuclear data evaluations that underlie many areas of nuclear physics~\cite{wondram2021}.
%These and other measurements will further our knowledge of reactor antineutrino fluxes. %Experiments using the IBD reaction will produce improved estimates of isotopic antineutrino emissions, including a measurement of the reactor antineutrino spectral shape with unprecedented energy resolution. Very importantly, CE$\nu$NS experiments will be able to make direct measurements of the reactor antineutrino flux below the 1.8~MeV IBD threshold. 
%These improvements will greatly benefit the global experimental effort, by affording reactor experiments a greater physics reach without the need for ancillary measurements. By helping resolve the data/model discrepancies described in Section~\ref{sec:datavsmodel}, they will also play an important role in resolving some of the anomalies that have arisen in the field. Moreover, the impact will extend beyond basic neutrino research. Traditionally, the effort to better understand and model reactor antineutrino fluxes has spurred investments and improvements in non-neutrino nuclear physics measurements and nuclear data. Reactor antineutrino measurements are a powerful probe of nuclear data evaluations that underlie many areas of nuclear physics~\cite{wondram2021}. 
%For example,  the total $\bar{\nu}_e$ spectrum includes contributions from short-lived, high Q-value isotopes that are difficult or impossible to measure by other means (this sentence almost verbatim from white paper). 
Finally, precise knowledge of reactor antineutrino emission, together with the technology development that enables it, undergirds the use of neutrino detectors for nuclear safeguards applications. For a discussion on the wider impact and applications of reactor antineutrinos, please refer to the report from the NF07 topical group in Ref.~\cite{nf07report}. 
%% Point out that monitoring requires predicted spectral difference in neutrino emission between uranium and plutonium isotopes?

The rest of this section is organized as follows: Sec.~\ref{sec:currentknowledge} describes the state-of-the-art knowledge on reactor antineutrino emission from models as well as from direct neutrino measurements, and describes how well they match against each other. Sec.~\ref{sec:futureimprovements} discusses the improvements anticipated from future neutrino and non-neutrino measurements, including a list of recommendations for the latter. Both of these subsections rely heavily on the community's input provided in Ref.~\cite{reactorwp}, as well as on reports summarizing the outcomes of recent workshops~\cite{wondram2021,bib:IAEA}.

\subsection{Current Knowledge of Reactor Neutrino Fluxes}\label{sec:currentknowledge}
\subsubsection{Models}\label{sec:reactormodels}

As mentioned in the previous section, each fission isotope produces a different $\bar{\nu}_e$ spectrum through its fission and the subsequent beta decay of the products. In order to predict the total (aggregate) antineutrino flux and energy spectrum, the isotopic spectra must be combined using the knowledge of the reactor's power and fuel composition at any one time. The normalization of each spectrum requires knowledge of the isotopic yield per fission, which is a measure of the total number of antineutrinos emitted per fission of a given isotope. As a reactor evolves along its fuel cycle, so do the fission fractions and consequently the total spectrum and flux. 

There currently exist two complementary methods for modeling the isotopic reactor antineutrino yields and spectra. The first one is the `summation' or `\textit{ab-initio}' method. As the name suggests, here the antineutrino spectrum is calculated from the bottom up by using tabulated information on cumulative fission yields and beta decays for each fission product, a process that requires summing over $\sim$1000 isotopes and thousands of beta branches. The fission yields are normally extracted from nuclear databases such as ENDF~\cite{BROWN20181} and JEFF~\cite{bib:JEFF33}, while data or theoretical predictions of beta decay spectra are available from ENSDF databases~\cite{bib:ENSDF}. Unfortunately, the beta decay information available is often inaccurate, incomplete, or entirely missing~\cite{Huber:2016fkt}. This, compounded with the fact that tabulations did not account for correlations in fission yield and decay uncertainties between isotopes and branches~\cite{reactorwp}, means that it is extremely difficult to estimate a reliable uncertainty for this method. The inclusion of improved beta decay data collected using total absorption gamma-ray spectroscopy (TAGS)~\cite{PhysRevLett.105.202501,tas_lots,tas_few,tas_br,tas_nb,tas_nb2,tas_rb,tas_rbbr,tas_rbi}, as well as recent efforts to estimate the correlations between independent and cumulative fission yields~\cite{Matthews2021}, holds the promise of better understood summation uncertainties~\cite{reactorwp}. Despite these challenges, this method performs quite well, as shown in Section~\ref{sec:datavsmodel}. 
% from White paper
%Recently, cataloguing of fission yield correlations~\cite{Matthews2021} and addition of improved decay data using total absorption spectroscopy (TAGS) techniques~\cite{PhysRevLett.105.202501,tas_lots,tas_few,tas_br,tas_nb,tas_nb2,tas_rb,tas_rbbr,tas_rbi} has provided the promise of reducing and better understanding summation uncertainties.  

The second one is the `conversion' method and relies on the measurement of the integral electron (beta) spectra from the fissions of $^{235}$U, $^{239}$Pu and $^{241}$Pu. These measurements were obtained in the 1980s by exposing foils of these elements to a thermal neutron flux in the Institut
% no final e in Institut in French
Laue-Langevin (ILL) research reactor~\cite{SCHRECKENBACH1985325,VONFEILITZSCH1982162,HAHN1989365} and using a magnetic spectrometer with electron/gamma separation capabilities~\cite{Huber:2016fkt}. Given that every electron antineutrino is simultaneously produced with an electron during beta decay, it is possible to convert the latter's spectrum into the former, hence the method's name. 
%this method converts the electron spectra into the corresponding antineutrino spectra for each of these isotopes, hence the method's name. 
This is done by fitting the electron spectra with virtual beta decay branches until their sum is consistent with the measurements. Compared with the summation method, the conversion method has the advantage that it is more robust against uncertainties from unknown or unmeasured data of individual beta decay branches. However, the virtual branches do not fully represent the $\sim$1000 beta branches actually present in the spectrum. Corrections must be applied to account for forbidden transitions~\cite{hayes_first,Hayen:2019eop} and finite-size effects~\cite{Wang:2016rqh}, among others, which depend on the nuclear structure and thus introduce uncertainties that are likely larger than originally anticipated. The measurements at ILL did not involve $^{238}$U, which only fissions with fast neutrons and contributes $<10\%$ of the total flux from an LEU reactor. An electron spectrum measurement and the corresponding conversion was carried out recently~\cite{Haag:2013raa} and found to be in good agreement with an ab-initio calculation from 2011~\cite{Mueller:2011nm}. The latter, together with the latest conversion of the ILL electron spectra for $^{235}$U, $^{239}$Pu and $^{241}$Pu, constitutes the so-called Huber-Mueller (HM) model~\cite{Huber:2011wv,Mueller:2011nm} that has recently served as a reference for many experiments. 

The uncertainties of both methods are largely uncorrelated, with those of the conversion method typically estimated to be smaller and on the order of a few percent. The isotopic spectral shapes from the latest iterations of the summation method~\cite{bib:fallot2} are in good agreement with those of the conversion method. However, the isotopic yields are in tension, particularly for $^{235}$U. More on this is said in Section~\ref{sec:datavsmodel}. 

%Predicting the $\bar{\nu}_e$ flux from~$^{238}$U, which only fissions with fast neutrons,  relies on theoretical calculations~\cite{Vogel:1980bk}, experimental data~\cite{Haag:2013raa}, or nuclear database summation~\cite{Mueller:2011nm}. 
%Effects such as forbidden transitions~\cite{hayes_first,Hayen:2019eop}, which can make up a large fraction of the antineutrinos, or weak magnetism~\cite{bib:HayesMag}, require additional additional corrections and introduce additional uncertainties. 

%% From white paper 
%This data-driven approach has the advantage of being immune to uncertainties from unknown or unmeasured beta decay spectra. However, the virtual branches do not fully represent the $\sim$1000 fission-produced beta branches actually present in the spectrum. Theoretical corrections, including forbidden transitions~\cite{hayes_first,Hayen:2019eop} and weak magnetism corrections~\cite{bib:HayesMag}, add additional uncertainties. Flux prediction of neutrinos from~$^{238}$U, and other non-fissile isotopes in reactor facilities, still rely on other experimental data or nuclear database summation.  

\subsubsection{Direct Neutrino Measurements}

Reactor antineutrino fluxes have been measured by a wide variety of experiments operating at different baselines from both LEU and HEU reactors. Great strides were made in the last decade thanks primarily to experiments focused on making a precise measurement of the $\theta_{13}$ neutrino mixing angle, namely Daya Bay, RENO and Double Chooz, and those searching for sterile neutrino oscillations at baselines of a few meters, such as PROSPECT and STEREO. Generally speaking, the improvements brought about by these experiments owe to their larger size and superior energy resolution ($\lesssim 10\%$ at 1~MeV) compared to their predecessors. These measurements have resulted in an increasingly complete picture of $\bar{\nu}_e$ production largely independent of theoretical models. 

Table~\ref{tab:exphighlights} highlights a subset of IBD experiments that have provided us with some of the most precise measurements to date. These measurements are divided into three categories: flux, spectrum, and evolution. The first refers to the measurement of the total (aggregate) flux from all the isotopes, which can be converted to a time-averaged yield per fission using the rate of reactor fission in the core and the absolute detection efficiency. The uncertainty in the latter typically dominates these measurements, which have been done as well as $1\%$ as shown in Table~\ref{tab:exphighlights}. For HEU experiments, whose flux largely originates from $^{235}$U fissions, this is equivalent to a measurement of this isotope's yield per fission. The STEREO experiment recently reported such a measurement with a precision of about $2.5\%$~\cite{STEREO:2020fvd}.  %Fig.~\ref{fig:raagiunti} in Section~\ref{sec:datavsmodel} shows a summary of the flux measurements done by many reactor experiments and their comparison to the HM prediction. 
%with the $^{235}$ measurement done by STEREO represented by the green band.  

The second category refers to the measurement of the time-integrated reactor antineutrino spectral shape. This is commonly done by reporting the yield per fission, or a quantity proportional to it, such as the number of detected events, as a function of positron detected energy. Since experiments have different detector responses, most notably detector nonlinearities and energy resolutions, these spectra are not directly comparable. However, it is possible to take out the detector response and express the spectra in terms of antineutrino energy, a process commonly referred to as unfolding. Among the LEU experiments, this has been done by Daya Bay~\cite{DayaBay:2016ssb,DayaBay:2021dqj} and RENO~\cite{RENO:2020dxd}, achieving a remarkable precision of about 2\% for the $2-5$~MeV energy region. Once again, for HEU experiments this is essentially a measurement of the antineutrino spectrum from $^{235}$U fissions. The STEREO collaboration has unfolded their measured spectrum and achieved a precision of about $3-4$\% in the $2-5$~MeV region~\cite{STEREO:2020hup}. 

\begin{table}[t]
    \centering
    \begin{tabular}{c|c|c|c|c}
    \toprule\toprule
    & Experiment  &   Flux & Spectrum & Evolution \\
    \midrule
    \multirow{5}{*}{LEU} & Bugey-3 &  & \checkmark & \\
                         & Bugey-4 & 1.4\% &  & \\
                         & Daya Bay & 1.5\% & \checkmark & \checkmark \\
                         & RENO & 2.1\% & \checkmark & \checkmark \\
                         & Double Chooz & 1.0\% & \checkmark & \\
    \midrule
    \multirow{4}{*}{HEU} & ILL & 9.1\% & & \\%\multirow{2}{*}{Not} \\
                         & Savannah River & 2.9\% & & Not \\
                         & STEREO & 2.5\% & \checkmark & applicable \\%\multirow{2}{*}{applicable} \\
                         & PROSPECT & & \checkmark & \\
    %ILL    & 1  & 0  & 0  & 0  &   flux    \\
    %Savannah River & 1  & 0  & 0  & 0  &   flux  \\
    %STEREO    & 1  & 0  & 0  & 0  &   flux/spect    \\
    %PROSPECT    & 1  & 0  & 0  & 0  &   spect    \\
    \bottomrule
    \end{tabular}
    \caption{Examples of IBD experiments' measurement of reactor antineutrino flux, spectrum, and evolution. For the flux, the number indicated is the precision with which the time-averaged IBD yield has been determined. Table adapted from Ref.~\cite{reactorwp}.}
    \label{tab:exphighlights}
\end{table}

\begin{figure}[!h]
    \centering
    \includegraphics[width=0.65\textwidth]{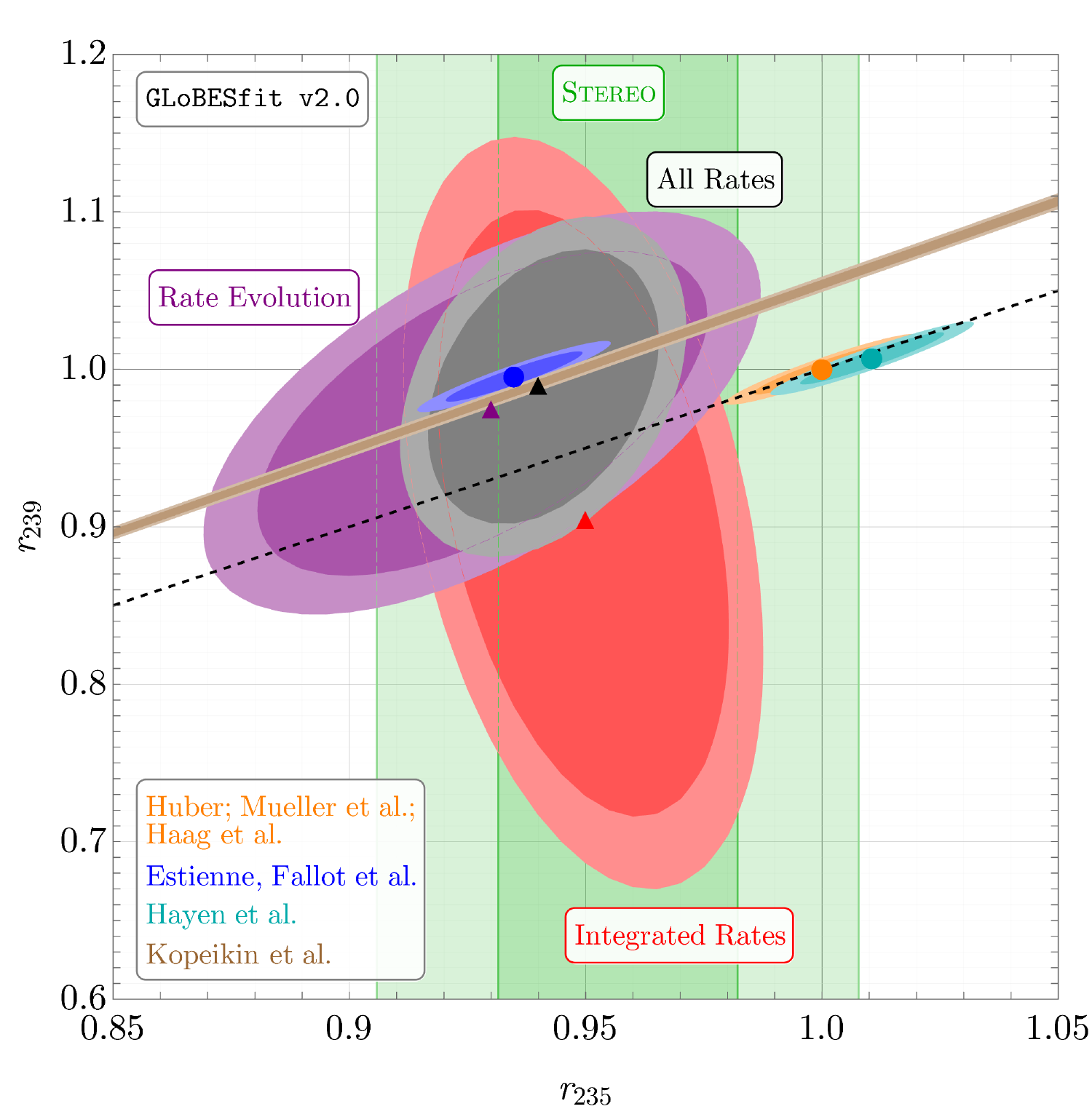}
    \caption{The 95\% C.L. (dark) and 99\% C.L. (light) contours in $r_{235}$--$r_{239}$ plane for integrated rate (red), fuel evolution (purple) and all reactor experiments (black), where $r_{X}$ is the ratio of the flux predicted/measured for isotope $X$ over its HM prediction. The result from STEREO \cite{STEREO:2020fvd} is shown in green; the bands represent the $1\sigma$ (dark) and $2\sigma$ (light) regions for one degree of freedom. The orange, blue and cyan ellipses represent the expectations from the HM~\cite{Huber:2011wv,Mueller:2011nm}, SM2018~\cite{bib:fallot2} and HKSS~\cite{Hayen:2019eop} flux models, respectively; $1\sigma$ ($2\sigma$) is shown in dark (light) shades. The brown bands represent the $1\sigma$ (dark) and $2\sigma$ (light) determination of the $^{239}$Pu/$^{235}$U ratio from the Kurchatov Institute~\cite{Kopeikin:2021rnb,Kopeikin:2021ugh}. The black, dashed line represents the line along which $r_{235}=r_{239}$. The triangles represent the best-fit values for the three fits, and the circles show the central values for the flux models. Figure and caption obtained from Ref.~\cite{reactorwp}, adapted from Ref.~\cite{huber_berryman}.}
    \label{fig:r235r239}
\end{figure}

The third category refers to the measurement of the IBD yield at various stages in the fuel cycle of LEU reactors. If done with enough precision, these measurements can disentangle the contributions from individual isotopes. Such data have so far been reported by Daya Bay~\cite{DayaBay:2017jkb} and RENO~\cite{RENO:2018pwo}, and have made it possible to place important constraints on the individual $^{235}$U, $^{239}$Pu and $^{238}$U yields~\cite{Giunti:2017nww,Gebre:2017vmm,Giunti:2019qlt}, which are shown for the two dominant isotopes $^{235}$U and $^{239}$Pu as the purple regions in Fig.~\ref{fig:r235r239}. Combining these constraints with time-integrated flux measurements, shown in red, yields the regions shown in grey. Moreover, Daya Bay's high statistics have enabled performing this measurement as a function of energy, resulting in unfolded spectra for $^{235}$U and for the combination of the $^{239}$Pu and $^{241}$Pu isotopes. These spectra can be used to make a data-driven prediction of the reactor antineutrino spectral shape at any other reactor complex with a precision of $\sim$2\%~\cite{DayaBay:2016ssb}. HEU experiments have additionally produced measurements of the $^{235}$U spectral shape that have been found to be consistent between them~\cite{Stereo:2021wfd} and with LEU experiments~\cite{DayaBay:2021owf}. A joint analysis between Daya Bay and PROSPECT resulted in a $^{235}$U spectrum extracted to a precision of about $3\%$ in most of the $2-5$~MeV region~\cite{DayaBay:2021owf}, as shown on the right of Fig.~\ref{fig:speccomp}. 
%The $^{235}$U spectra from the STEREO, PROSPECT and Daya Bay experiments are generally found to be consistent. 
%A joint unfolding analysis of Daya Bay and PROSPECT yields the $^{235}$U spectrum to about $3\%$ in most of the $2-5$~MeV region. 
%with a precision of about $\sim$4\% and $5-6$\% in the 2-5~MeV region~\cite{DayaBay:2016ssb}, respectively. The $^{235}$U spectrum obtained in this fashion is consistent with the one measured by PROSPECT and STEREO~\cite{DayaBay:2021owf,Stereo:2021wfd}. By applying small corrections on the time-integrated spectrum, which is known very precisely, these unfolded spectra also allow to predict the spectrum at any other reactor complex to $\sim$2\%~\cite{DayaBay:2016ssb}. 

%\begin{figure}[!h]
%  \centering
%  \includegraphics[width=0.5\textwidth]{./graphics/Global_IBD_Xsection.png}
%  \caption{Allowed regions for isotopic IBD yields of $^{235}$U, $^{239}$Pu, and $^{238}$U~provided by a fit of time-integrated and `flux evolution' IBD yield datasets.  Sterile neutrino oscillations are assumed to be negligible.  Image from Ref~\cite{white paper}. }
%  \label{fig:yieldssummary}
%\end{figure}

% from white paper 
%Measured \uFive~isotopic spectra have been demonstrated to be generally consistent between Daya Bay, PROSPECT, and STEREO~\cite{bib:prosDBjoint,bib:prosSTEREOjoint}.

%Recent neutrino experiments have been very successful in advancing the state of knowledge of reactor antineutrino emissions, most notably by uncovering the reactor flux and spectrum anomalies.(from white paper; need to modify) 

% determine the extent to which the data prefer a rescaling of the HM fluxes

%Measurements via the IBD channel will  Experiments requiring greater precision than what was afforded by reactor antineutrino flux models were forced to perform relative measurements, most notably through the use of near detectors in . These experiments produced some of the measurements that the community now benefits from. 

%It is worth mentioning that it was the need for greater precision beyond what was afforded by models of reactor antineutrino emission what resulted in a series of measurements, most notably in the near detectors of oscillation experiments of experiments like Daya Bay~\cite{}.  pushed oscillation experiments 

\subsubsection{Data vs. Model Comparisons}\label{sec:datavsmodel}

\begin{figure}[!h]
    \centering
    \includegraphics[width=0.9\textwidth]{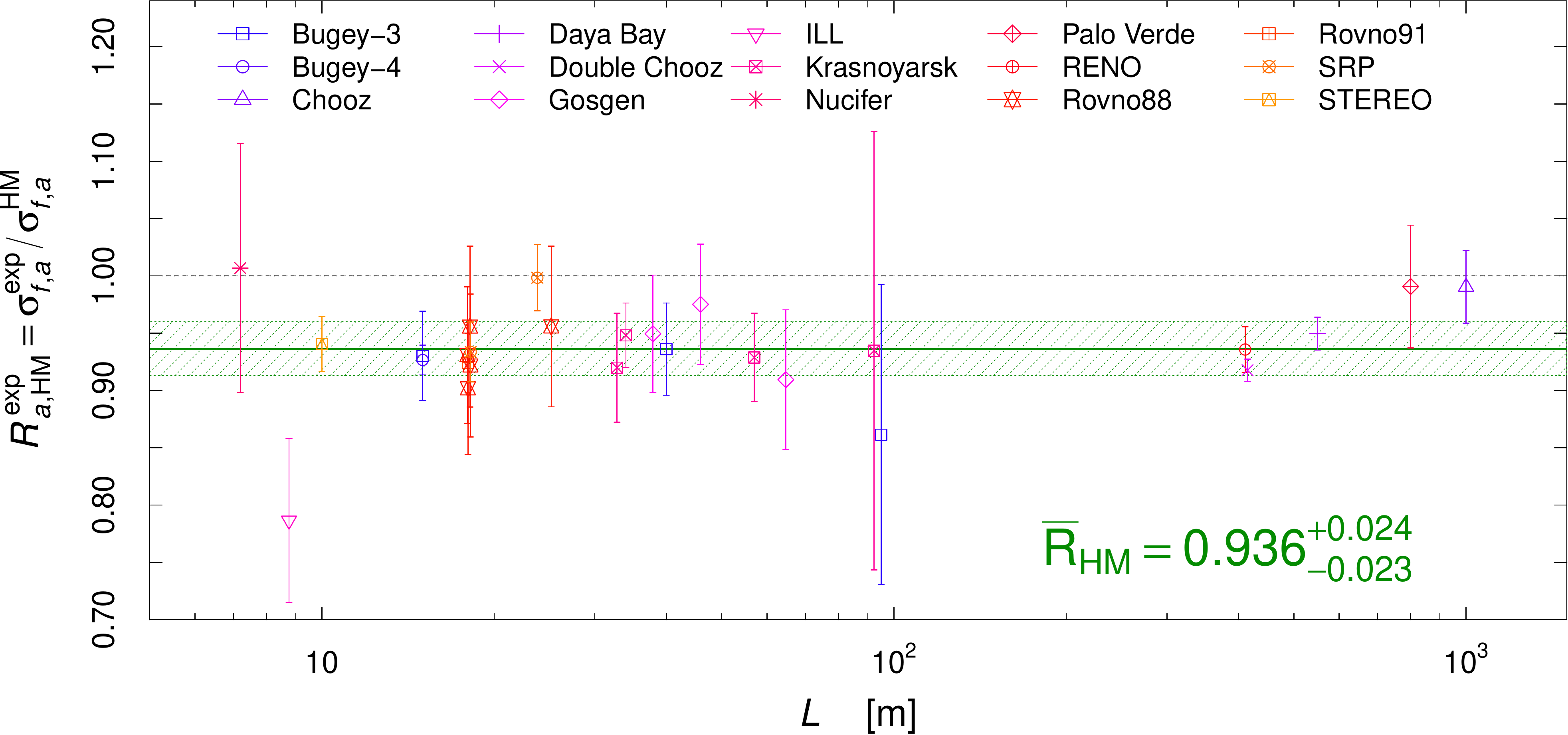}
    \caption{Ratio of measured over predicted time-averaged IBD yields per fission as a function of baseline from the reactor. The prediction is obtained with the HM model. The horizontal green band shows the average ratio and its uncertainty, highlighting the so-called reactor antineutrino anomaly. Image from Ref.~\cite{Giunti:2021kab}.}
    \label{fig:raagiunti}
\end{figure}

The improved precision in IBD yield and spectrum measurements in the last decade has uncovered several discrepancies with the models. First came the realization in 2011 that most time-averaged yield measurements are consistently below the expectation from the latest iteration of the conversion method~\cite{Mueller:2011nm,Huber:2011wv}, a discrepancy commonly referred to as the reactor antineutrino anomaly. This effect is illustrated in Fig.~\ref{fig:raagiunti}, which includes the most recent measurements available, and can also be seen in Fig.~\ref{fig:r235r239} as the incompatibility between the red and orange ellipses. The flux evolution data from Daya Bay and RENO, represented by the purple ellipses in that same figure, as well as STEREO's measurement of $^{235}$U's yield in green, suggest that the discrepancy originates primarily from a mis-modelling of the $^{235}$U yield per fission, while the $^{239}$Pu yield is consistent with the HM model. This is seen by the fact that both of these regions are offset to the left of the vertical line at $r_{235}=1$. This figure also shows how the flux data, both time-averaged and evolution, are in good agreement with one of the latest iterations of the summation model~\cite{bib:fallot2}, represented by the blue ellipses, but in disagreement with another method that used conversion techniques~\cite{Hayen:2019eop}, shown in light blue. Finally, the figure also includes a new re-evaluation of the ratio between the cumulative fission beta spectra of $^{239}$Pu and $^{235}$U done at the Kurchatov Institute. This measurement revealed an offset of about $5.4\%$ with respect to the measurements done at ILL, represented by the brown band, but notably found no significant difference in the spectral shape~\cite{Kopeikin:2021rnb,Kopeikin:2021ugh}. The confluence of all these data sets strongly suggest that the ILL-measured beta spectrum inputs to conversion methods, particularly for $^{235}$U, are to blame for flux data/model discrepancies. If this is confirmed, the conversion and summation approaches would finally converge. 

\begin{figure}[!h]
    \centering
    \includegraphics[width=0.98\textwidth]{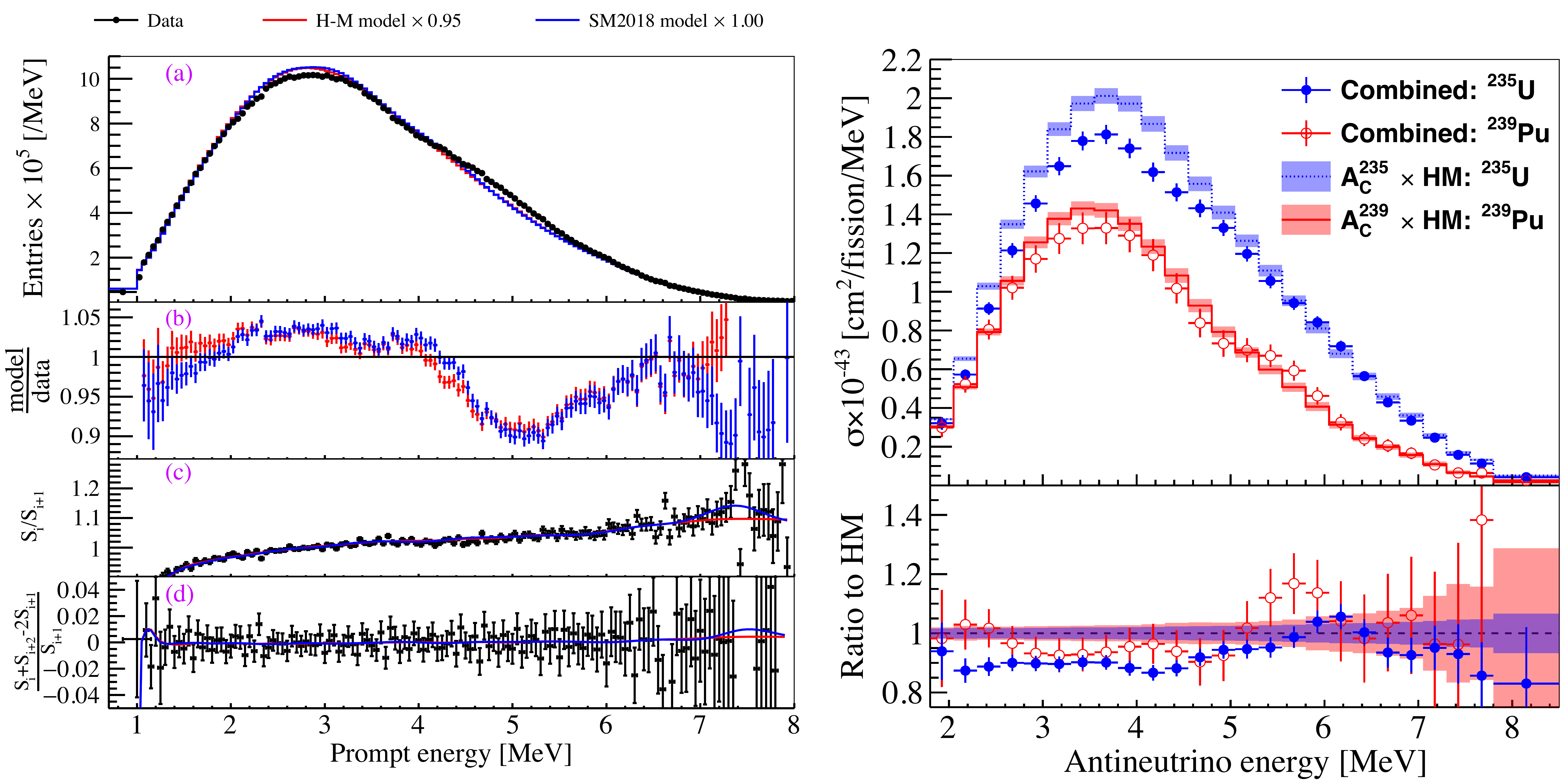}
    \caption{Left: (a) Prompt energy reactor antineutrino spectrum measured with 1958 days at Daya Bay, next to the prediction from the HM model~\cite{Mueller:2011nm,Huber:2011wv} scaled by a factor of 0.95, and the SM 2018 summation model~\cite{bib:fallot2}. (b) Model/data ratios. (c) Continuity of the spectrum measured as $\frac{S_i}{S_{i+1}}$, where $S_i$ is the number of events in the $i$-th bin. The expectation from the two models is also shown, given Daya Bay's energy resolution. (d) Same as (c), but with the continuity of the spectrum evaluated as $\frac{S_i+S_{i+2}-2S_{i+1}}{S_{i+1}}$. Image from Ref.~\cite{DayaBay:2021dqj}. Right: unfolded $^{235}$U and $^{239}$Pu antineutrino energy spectra from a joint analysis between Daya Bay and PROSPECT. Both spectra are compared to the expectation from the HM model, with the ratios shown in the bottom panel. Image from Ref.~\cite{DayaBay:2021owf}.}
    \label{fig:speccomp}
\end{figure}

Soon after the reactor antineutrino anomaly was uncovered, discrepancies were also found between the measured and expected reactor antineutrino spectra. As an example, the top panel on the left side of Fig.~\ref{fig:speccomp} shows the time-integrated spectrum observed at Daya Bay compared with a rescaled version of the HM model (to compensate for the reactor antineutrino anomaly) and the SM2018 summation model. The most notable discrepancy is an excess in the data around 5~MeV, which has also been consistently seen by other LEU experiments~\cite{DoubleChooz:2019qbj,RENO:2018dro,NEOS:2016wee}. It is noteworthy that this feature is present when comparing to both the conversion and summation models, as shown in Fig.~\ref{fig:speccomp}, as well as in HEU experiments, such as PROSPECT and STEREO. In fact, the data so far are consistent with this feature existing in both isotopic spectra, as illustrated on the right side of Fig.~\ref{fig:speccomp} which shows the measured $^{235}$U and $^{239}$Pu spectra from a joint analysis between Daya Bay and PROSPECT next to their prediction from the HM model. Most notably, the feature is not seen in the recent measurements done at the Kurchatov Institute. All this information suggests that an issue with an input common to all isotopes and both prediction methods, such as the assumed theoretical shape of the beta spectra, is at play~\cite{Sonzogni:2017wxy}. 

% first by LEU and then by HEU. 
% Not solved by Kurchatov 

\subsection{Future Improvements} \label{sec:futureimprovements}
\subsubsection{IBD Neutrino Measurements}

Efforts are already underway to improve existing IBD measurements of isotopic flux and spectra with ongoing experiments. The final data set of the Daya Bay experiment, consisting of more than six million reactor antineutrino events, is still being analyzed. The additional statistics are expected to improve the flux evolution measurement. The time-integrated flux measurement, which is already systematics dominated, could still be improved through the use of more efficient and robust inverse beta-decay selection criteria. The collaboration plans to release the full data set after all the final results have been released, allowing the data to be re-examined, tested for new models, and used as a benchmark for other experiments, phenomenologists, and nuclear databases~\cite{DYB_LOI}. 

STEREO's latest published results were based on 179 days of reactor-on data and 235 days of reactor-off data near the compact core of the ILL reactor. A new data set with roughly twice the statistics has already been collected and is under analysis~\cite{Labit:2021zma}.  

The NEOS experiment is located at a distance of about 24~m from the 2.8~GW~commercial LEU core at the YoungKwang Hanbit nuclear power plant~\cite{NEOS:2016wee}. After 180~days of reactor-on data, a second phase measurement, referred to as NEOS-II, collected $\sim$100~days of reactor-off data and $\sim$400~days of reactor-on data covering an entire fuel cycle~\cite{Oh:2020inx}. While the statistics are not as large as Daya Bay's, the proximity to the reactor core allows NEOS-II to observe a broader range of reactor fuel content, providing complementary information to that of Daya Bay and RENO and potentially extracting the isotopic yields and spectra with similar or even superior precision. Preliminary results have been released~\cite{neos2neutrino2022} but have not yet been published at the time of writing. Ref.~\cite{Gebre:2017vmm} includes a discussion on the plausible gains in isotopic IBD yield measurement precision achievable in a single-core LEU experiment. 
%%% Preliminary results have been released but are not yet published. 
%%% Note to self: SNO+ submitted an LOI that may be worth mentioning here. 

\begin{figure}
    \centering
    \includegraphics[width=0.51\textwidth]{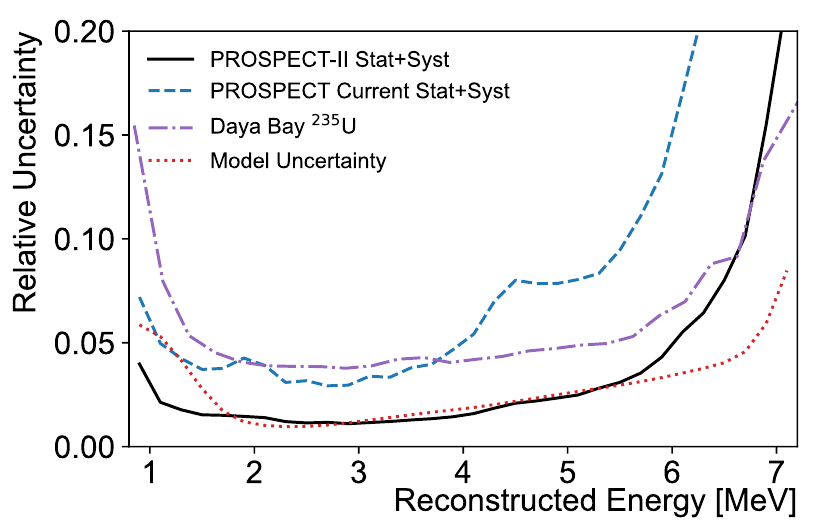}
    \includegraphics[width=0.48\textwidth]{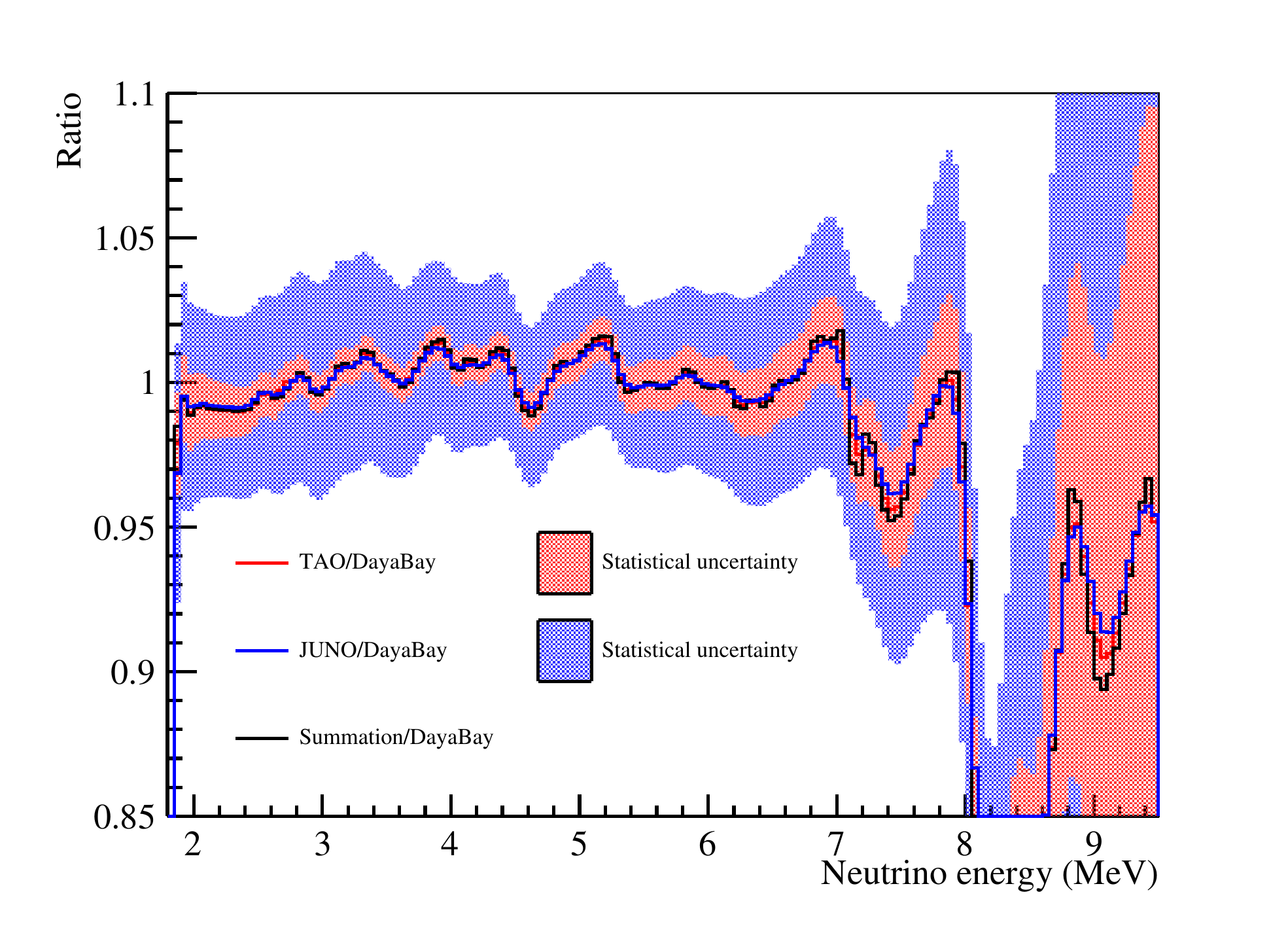}
    \caption{Left: PROSPECT-II $^{235}$U spectrum measurement uncertainties after two years of data-taking, from~\cite{Andriamirado:2021qjc}.  Right: Comparison of projected JUNO-TAO and JUNO measurements and uncertainties with Daya Bay measurements, assuming that the true LEU reactor spectrum measured by JUNO-TAO and JUNO is given by Ref.~\cite{bib:fallot2}; JUNO-TAO's sensitivity to fine structure in the LEU reactor antineutrino spectrum is clearly illustrated. From Ref.~\cite{JUNO:2020ijm}.}
    \label{fig:prostao}
\end{figure}

Several experimental efforts on the horizon also hold the promise of further advancing our knowledge of reactor antineutrino emission. A follow-up to the PROSPECT experiment, PROSPECT-II, is being proposed to be deployed at 7-9 m from the High Flux Isotope Reactor at Oak Ridge National Lab~\cite{Andriamirado:2022psq}. With five times more data than PROSPECT's first run, and an increased signal-to-background ratio, PROSPECT-II endeavors to measure the $^{235}$U spectrum to a $\sim$2.5\% precision with 2 years of data. This is a precision rivaling that of the prediction models, as shown in Fig.~\ref{fig:prostao}. A deployment of PROSPECT-II near an LEU reactor is also foreseen, which through correlated flux measurements between core types would further enhance knowledge of individual isotopic contributions. 

The JUNO experiment, currently under construction in China, plans to deploy a satellite detector called JUNO-TAO at a baseline of about 30~m from one of the 4.6~GW$_\mathrm{th}$ cores of the Taishan nuclear power plant~\cite{TAO_LOI,JUNO:2020ijm}. This detector will measure the reactor antineutrino spectrum with an unprecedented energy resolution $<2\%$ at 1~MeV, revealing some of the fine structure caused by signatures of individual fission products for the first time, as illustrated on the right side of Fig.~\ref{fig:prostao}. Accordingly, this measurement will serve as an important reference to other experiments and nuclear databases. By measuring several reactor fuel cycles, JUNO-TAO will also be able to improve the flux evolution measurements done by Daya Bay and RENO. When combined with a high-precision HEU experiment such as PROSPECT-II, these data should allow major improvements in our knowledge of the isotopic fluxes and spectra~\cite{reactorwp}. 

Finally, the deployment of mobile detectors such as CHANDLER~\cite{chandlerloi} and ROADSTER~\cite{roadstrloi} could also result in unique contributions to the global knowledge of reactor antineutrino emission. These detectors, both of which use segmented designs with plastic scintillator and whose primary purpose is demonstrating the feasibility of mobile reactor neutrino detection at the surface for nuclear safeguards, will allow the collection of large samples of systematic-correlated data at different reactors and different stages of the reactor fuel cycle. Further information on the synergy between neutrino physics and applications can be found in the report of the NF07 topical group~\cite{nf07report}. 

\subsubsection{Other Measurements}

Direct reactor antineutrino detection has been dominated by the IBD reaction. However, as mentioned in Sec.~\ref{sec:nrintro}, a new generation of experiments aiming to detect reactor antineutrinos via the threshold-free CE$\nu$NS reaction, first observed in 2017 with neutrinos from stopped-pion decay~\cite{COHERENT:2017ipa}, is being deployed. While no detection of reactor antineutrinos has been made with this reaction yet, some of the technologies being developed offer the promise of making direct measurements of the reactor antineutrino flux below the 1.8~MeV IBD threshold, enabling tests of reactor antineutrino models at those energies for the first time.  
This is particularly important given that summation predictions at these low energies have been shown to be very sensitive to uncertainties in the nuclear data, especially concerning isomeric transition corrections in fission products~\cite{Ang:2021svv}. 
%% The primary purpose of these experiments is to observe reactor antineutrinos via this process, which has not been yet accomplished, and to develop the technology to be able to for nuclear safeguards, search for physics beyond the SM, or study nuclear structure. It is very hard to project when and how well these experiments will be able to probe reactor antineutrino fluxes. 
%A recent evaluation of the uncertainties in the $0-2$~MeV spectrum detected with large uncertainties emanating from nuclear data, particularly isomeric transitions. 
%\textcolor{blue}{[Would be happy to include more quantitative information about CEvNS' contributions to the knowledge of reactor antineutrino fluxes if provided by the community.]}

Additional nuclear physics measurements will be needed to satisfactorily resolve the data/model discrepancies of Section~\ref{sec:datavsmodel}. Recommendations have been issued at the completion of several recent workshops focused on the interplay between reactor antineutrino measurements and nuclear data, most importantly the {\it Technical Meeting on Nuclear Data for Anti-neutrino Spectra and Their Applications}, organized by the International Atomic Energy Agency~\cite{bib:IAEA}, and the {\it Workshop on Nuclear Data for Reactor Antineutrino Measurements}~\cite{wondram2021}. The recommendations from these two workshops have a high degree of overlap, indicating the strong community consensus concerning the path forward. One of the primary recommendations concerns the beta spectrum measurements carried over three decades ago at the ILL reactor, which constitute a single point of failure for conversion models and have been recently put into doubt by recent measurements, including the direct reassessment of aggregate beta spectra $^{239}$Pu/$^{235}$U ratio at Kurchatov Institute mentioned in Section~\ref{sec:datavsmodel}. This issue could be authoritatively resolved with high-precision aggregate beta spectrum measurements using modern neutron facilities and measurement techniques for all fission isotopes. Such measurements could be achievable at a number of US-based neutron facilities. For instance, the high thermal neutron fluxes of Oak Ridge National Laboratory or the National Institute of Standards \& Technology would be ideally suited to the $^{235}$U \& $^{239,241}$Pu measurements, while the fast neutron flux available in the Triangle Universities Nuclear Laboratory would be well-suited for the $^{238}$U measurement. Total uncertainties of 1\% or less in the 2 to 8 MeV electron energy range should be within reach~\cite{wondram2021}. 

Other high-priority recommendations include performing direct measurements of beta spectrum shapes with a precision of 2\% or better for selected forbidden transitions, particularly from high-Q isotopes and/or some of the largest fission product contributors to the total antineutrino spectrum. This would directly quantify the effect of sub-dominant corrections to the spectrum shape, which is a key input to both the summation and conversion calculations. Likewise, continued advancement in the correction of pandemonium-affected nuclear data via additional total absorption spectroscopy measurements is needed for accurate summation predictions of reactor antineutrino spectra at the highest energy range ($\sim$7-12 MeV). Finally, the establishment of {standardized evaluations of reactor antineutrino data sets and predictions in a centralized, publicly accessible domain}, including the development of an open software framework for reactor antineutrino flux predictions, would greatly facilitate rapid and reliable progress in the field. The full list of recommendations can be obtained in Refs.~\cite{wondram2021,bib:IAEA}.

\section{Other Neutrino Sources}
\label{sec:novel}

Nearly all of the artificial neutrino sources currently in operation fall into the categories described in Sections~\ref{sec:convb}-~\ref{sec:reac}.  However, many new potential sources are being considered for the future and some (e.g. neutrinos from the LHC) have already begun operation.  In this section, we describe these new sources.  

\subsection{IsoDAR}
\label{subsec:isodar}
\begin{figure}
\centering
\includegraphics[width=5.5in]{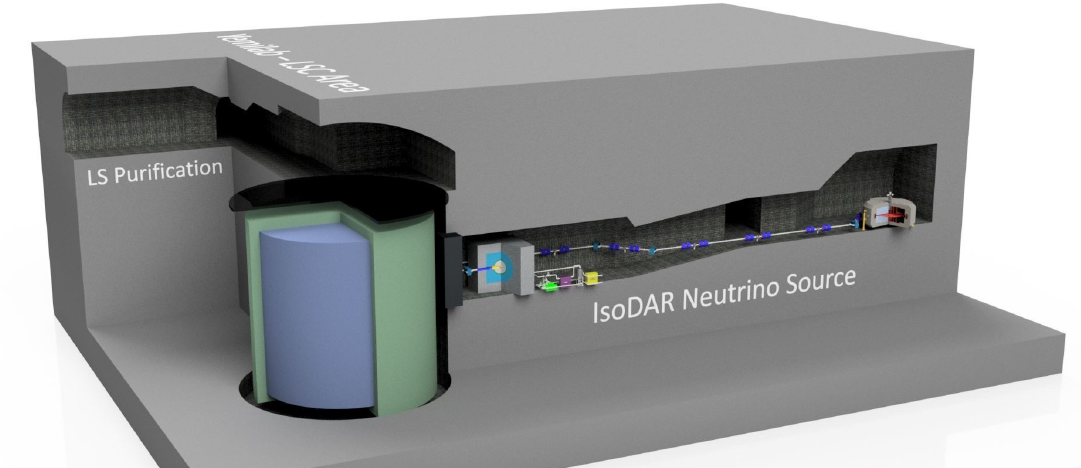}
\caption{The IsoDAR@Yemilab antineutrino source and detector configuration.}
\label{isodar}
\end{figure}

The IsoDAR experiment will rely on a compact cyclotron-based antineutrino source paired to a planned or existing liquid scintillator detector for studying new physics involving neutrino production, propagation, and interactions~\cite{isodar_prl,Alonso:2021kyu}. The current plan is for targeting 600~kW, 60~MeV protons on target about 17~m away from the center of the 2.26~kton target volume Liquid Scintillator Counter (LSC) detector 1~km underground at Yemilab in South Korea~\cite{Alonso:2022mup}. The IsoDAR cavern excavation was completed in 1/2022 and the LSC hall construction is well underway~\cite{Alonso:2021kyu}; a schematic of this configuration is shown in Figure~\ref{isodar}.

In the context of artificial neutrino sources, the IsoDAR concept is completely unique. The idea is to utilize a high power cyclotron (see, e.g., Ref.~\cite{Winklehner:2021qzp}) and resulting 60~MeV proton-on-beryllium target interaction to produce an intense source of neutrons. The neutrons enter an isotopically pure $^{7}$Li sleeve surrounding the target, slow down, and then capture on $^{7}$Li to make $^{8}$Li. The resulting high-$Q$ $\beta^-$ decay ($\rightarrow~^8\mathrm{Be}+e^-+\bar{\nu}_e$; $\tau_{1/2}=839$~ms) produces an intense antineutrino source (mean antineutrino energy of 6.4~MeV and an endpoint of $\sim$15~MeV; $1.15 \times 10^{23}$ $\bar{\nu}_e$ in 4~years of livetime~\cite{Alonso:2021kyu}). Coupled with a large nearby detector, this experiment's expected sample of 1.7$\times 10^6$ inverse beta decay (IBD; $\bar{\nu}_e+p \rightarrow e^+ + n$) and 7000~elastic scattering (ES; $\bar{\nu}_e + e^- \rightarrow \bar{\nu}_e + e^-$) events, will provide high sensitivity to new physics associated with short-baseline oscillations, including involving multiple sterile neutrinos and/or sterile neutrino decay, neutrino wavepacket effects, the weak mixing angle, non-standard neutrino interactions, light boson production, and mirror neutrons~\cite{Hostert:2022ntu,Alonso:2021kyu}.

The power of IsoDAR comes from a combination of the high event statistics, extremely well known flux shape (featuring a $\textit{single}$ isotope source) and interaction cross sections (IBD and ES), and expected high resolution detector. In particular, IsoDAR's physics reach is expected to improve upon existing measurements at a level approaching an order of magnitude in both short-baseline and weak-mixing-angle/NSI. In addition, along with particle physics, the IsoDAR accelerator technology is of significant interest for the accelerator and medical physics communities~\cite{medical}.

\subsection{Intense radioactive electron capture source}
Electron capture sources~\cite{ElectronCaptureLOI} produce neutrinos via the electron capture process, in which an isotope absorbs an atomic electron and emits a neutrino.  The process produces mono-energetic electron neutrinos in the 0.5-1.5 MeV energy range depending on the isotope and process.  Two electron capture sources that have been used to date: $^{51}$Cr and $^{37}$Ar.  The GALLEX experiment~\cite{GALLEX} used $^{51}$Cr as a calibration source, while the SAGE experiment~\cite{sage} used both $^{51}$Cr and $^{37}$Ar as calibration sources.  

The Baksan Experiment on Sterile Neutrino (BEST) experiment~\cite{BEST} (based on the SAGE experiment) is a short-baseline neutrino experiment that took data using a $^{51}$Cr source and recently reported first results~\cite{BEST2}.  Looking for electron neutrino disappearance using capture on Gallium, it found a deficit of electron neutrinos relative to expectation.  The results are consistent with previous Gallium anomaly measurements~\cite{sage,sage2,gallex2}.  Combining SAGE, GALLEX and BEST measurements, the observed neutrino capture rate is $0.8\pm0.05$ relative to the predicted rate.  BEST made measurements at two different baselines but found no significant difference in deficit between the two, indicating that, if the deficit is due to a neutrino oscillation, the oscillation length is similar to or smaller than the meter-scale volume of the BEST detectors.  Future measurements at either a smaller baseline or with a higher energy source would be needed to understand the length scale of the oscillation.     

The SOX experiment~\cite{SOX} proposed to use the BOREXINO detector to search for short-baseline oscillations using both $^{51}$Cr neutrino and $^{144}$Ce–$^{144}$Pr sources.  The project was cancelled in 2018 due to difficults with source development~\cite{SOX2} 

Future sources are expected to be primarily $^{51}$Cr, as $^{37}$Ar requires a a large fast-neutron reactor, which is not currently available.    $^{51}$Cr is created by exposing $^{50}$Cr to a high thermal neutron flux in a nuclear reactor core.  Approximately 90\% of $^{51}$Cr decays produce a 750 keV neutrino, with the remaining 10\% producing a 430 keV neutrino.  The BEST experiment has also considered using a $^{65}$Zn source~\cite{BEST3}, which is created by irradiating enriched $^{64}$Zn in a nuclear reactor thermal neutron flux.  $^{65}$Zn produces roughly equal parts 1.35 MeV and 235 keV neutrinos.    

Although antineutrinos at this energy scale are available from nuclear reactors, electron capture sources produce neutrinos (rather than antineutrinos) and have the advantage that they can be transported to a neutrino detector facility and operated underground with very short and variable baselines.  Calls have  been made for further development and use of these sources, including a proposal for the US to develop a $^{51}$Cr source that could be paired with the LZ or XENONnT detectors and expand the new physics search capabilities of those experiments~\cite{ElectronCaptureLOI}.

\subsection{Neutrinos from the LHC}

The FASER$\nu$ collaboration~\cite{faser_proposal,faser_physics_case} recently reported the first observation of neutrinos from a collider~\cite{fasernu}.  This measurement is likely to be the first of many measurements from the Large Hadron Collider (LHC).  The FASER$\nu$ detector is oriented to receive flux from the ATLAS interaction point and sits just downstream of the main FASER detector (designed to search for new lightly coupled particles).  The initial observation came from a 2018 pilot run, but the experiment plans to take data during LHC Run-3 (2022-2024), collecting samples of $\nu_e$, $\nu_\mu$, and $\nu_\tau$ neutrinos, with the ability to distinguish between all three flavors.  The FASER$\nu$ experiment's physics goals include cross section measurements of all neutrino flavors at previously unmeasured neutrino energies, searches for BSM physics, and neutrino flux measurements as novel constraints to LHC event generators.  Another neutrino detector, SND@LHC, also plans to operate near the ATLAS interaction point (but on the opposite side to FASER$\nu$) during LHC Run-3, with similar physics goals to Faser$\nu$.  Neutrino fluxes at SND@LHC and FASER$\nu$ peak in the 400-800 GeV region depending on neutrino flavor.   

For future measurements of neutrinos from the LHC, a new Forward Physics Facility~\cite{FPF1,FPF2} has been proposed that would provide a new cavern with space for upgraded versions of FASER, FASER$\nu$, SND@LHC, as well as a potential Liquid Argon neutrino detector and other potential non-Neutrino detectors.  If this new facility goes forward, it would add substantial statistics to the FASER and SND@LHC samples, including potentially thousands of $\nu_\tau$ events.  This is would be a dramatic improvement over the $\sim20$ $\nu_\tau$ events that have been observed so far from artificial sources.  

\subsection{Muon Decay Sources}

Muon decay from stored muon beams produces a colimated beam containing equal numbers of muon neutrinos and electron neutrinos (or muon antineutrinos and electron antineutrinos depending on the charge of the muon).  Muon-based sources offer advantages over traditional hadron-focused neutrino beams in that they contain equal parts muon neutrinos and electron antineutrinos and have a well known energy spectrum.  Several facilities have been proposed to produce and make use of neutrinos from stored muons.  Because they share common challenges associated with producing, accelerating, and storing muon beams, there are many overlaps in R\&D efforts towards these facilities. Addressing these challenges were not identified as a priority in the 2014 P5 report~\cite{P52014}, but was set as a high priority in the 2020 update of the European Strategy for Particle Physics~\cite{EuropeanStrategyGroup:2020pow}.  Continued development of these concepts would greatly benefit from an endorsement in the upcoming P5 process.  

\subsubsection{nuSTORM}

\begin{figure}
\centering
\includegraphics[width=0.9\linewidth]{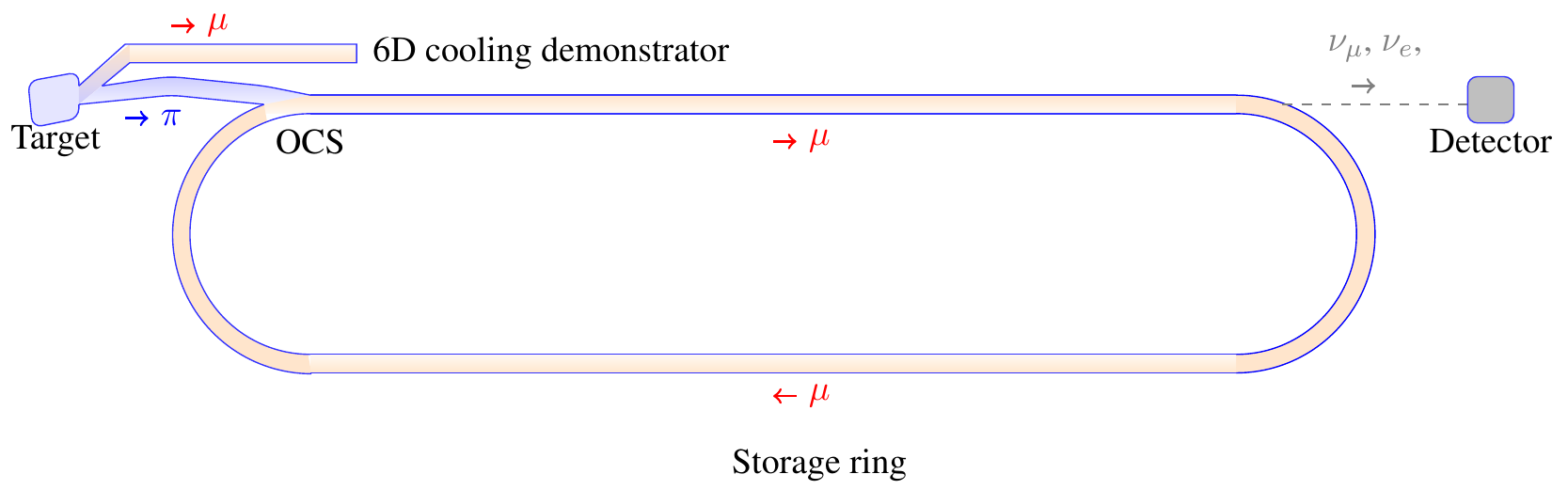}
\caption{Schematic of the nuSTORM Facility.  Pions from collisions of a primary proton beam with the target would be focused and directed towards nuSTORM or a muon cooling demonstrator.  A third potential transfer line to ENUBET is not shown.  Figure reproduced from Ref.~\cite{NuStorm}.}
\label{nustorm}
\end{figure}

The Neutrinos from Stored Muons facility (nuSTORM)~\cite{NuStorm} facility would consist of a $\mu^\pm$ storage ring capable of storing 1-6 GeV muons and producing beams of muon and electron neutrinos spanning the 0.3 - 5.5 GeV energy range.  Its primary physics goal is measurement of electron and muon neutrino cross sections at the few percent level, which would be beneficial to both DUNE and Hyper-Kamiokande.  It would also provide measurements of various nuclear effects, searches for sterile neutrinos, and serve as a test facility for the development of a future neutrino factory and/or muon collider.  Both Fermilab and CERN would be capable of hosting nuSTORM.  A facility design is being developed together with the ENUBET facility described in section~\ref{Sec:beam_instrumentation}.    Although it is a relatively low-intensity source that would facilitate short-baseline measurements, nuSTORM would be a testbed for technology development for a neutrino factory and/or muon collider.  

\begin{figure}
\centering
\includegraphics[width=0.95\linewidth]{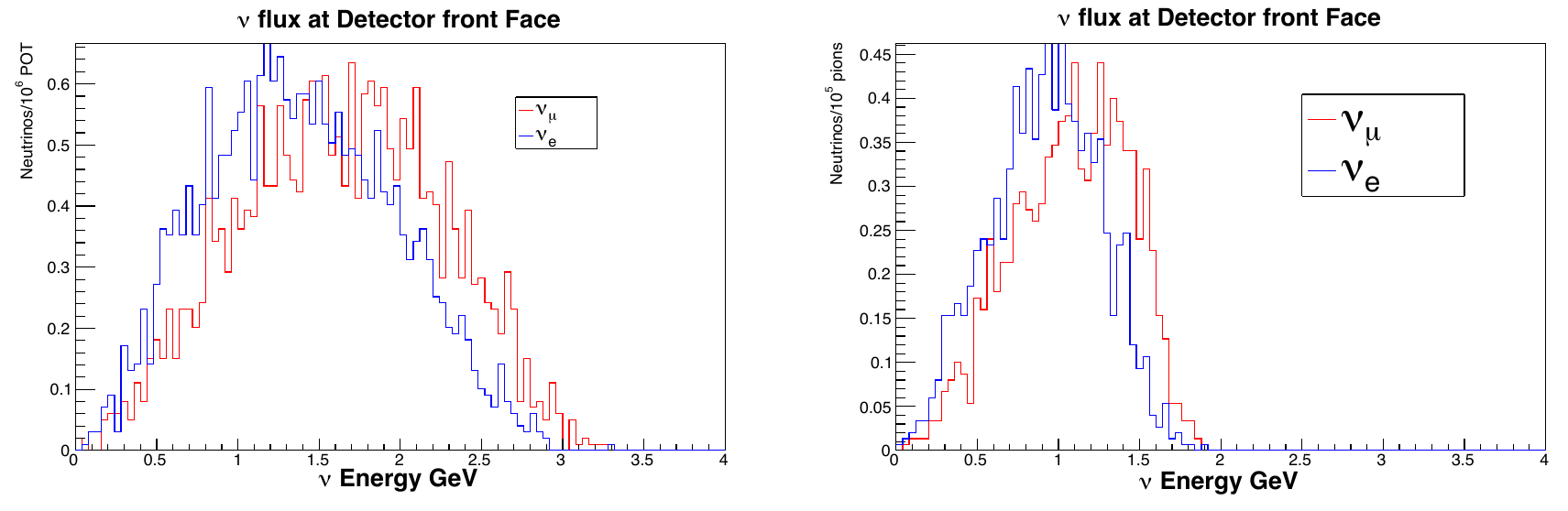}
\caption{Neutrino Fluxes from nuSTORM from 5 GeV (left) and 3 GeV (right) pions.  Reproduced from Ref.~\cite{NuStorm}.  }
\label{nustorm_flux}
\end{figure}

\subsubsection{Neutrino Factories}
\begin{figure}[ht]
\centering
\includegraphics[width=0.6\linewidth,trim={0 2.8cm 0 0.8cm},clip]{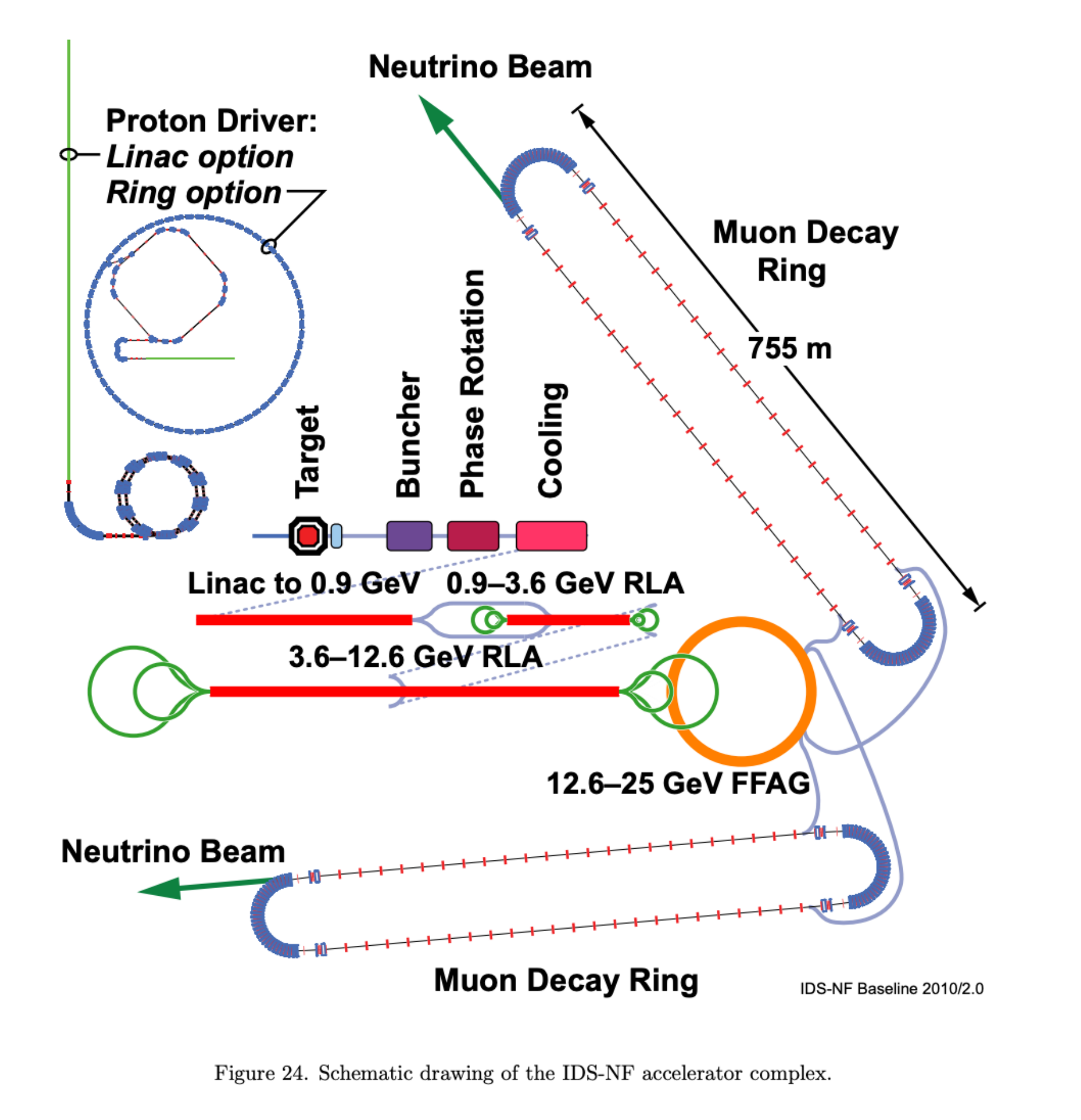}
\caption{Principle elements of the NuMAX neutrino factory concept and a potential muon collider.  Figure reproduced from~\cite{Bogacz:2022xsj}.   }
\label{fig:nufactory1}
\end{figure}

\begin{figure}[ht]
\centering
\includegraphics[width=0.7\linewidth,trim={0 7cm 0 7cm},clip]{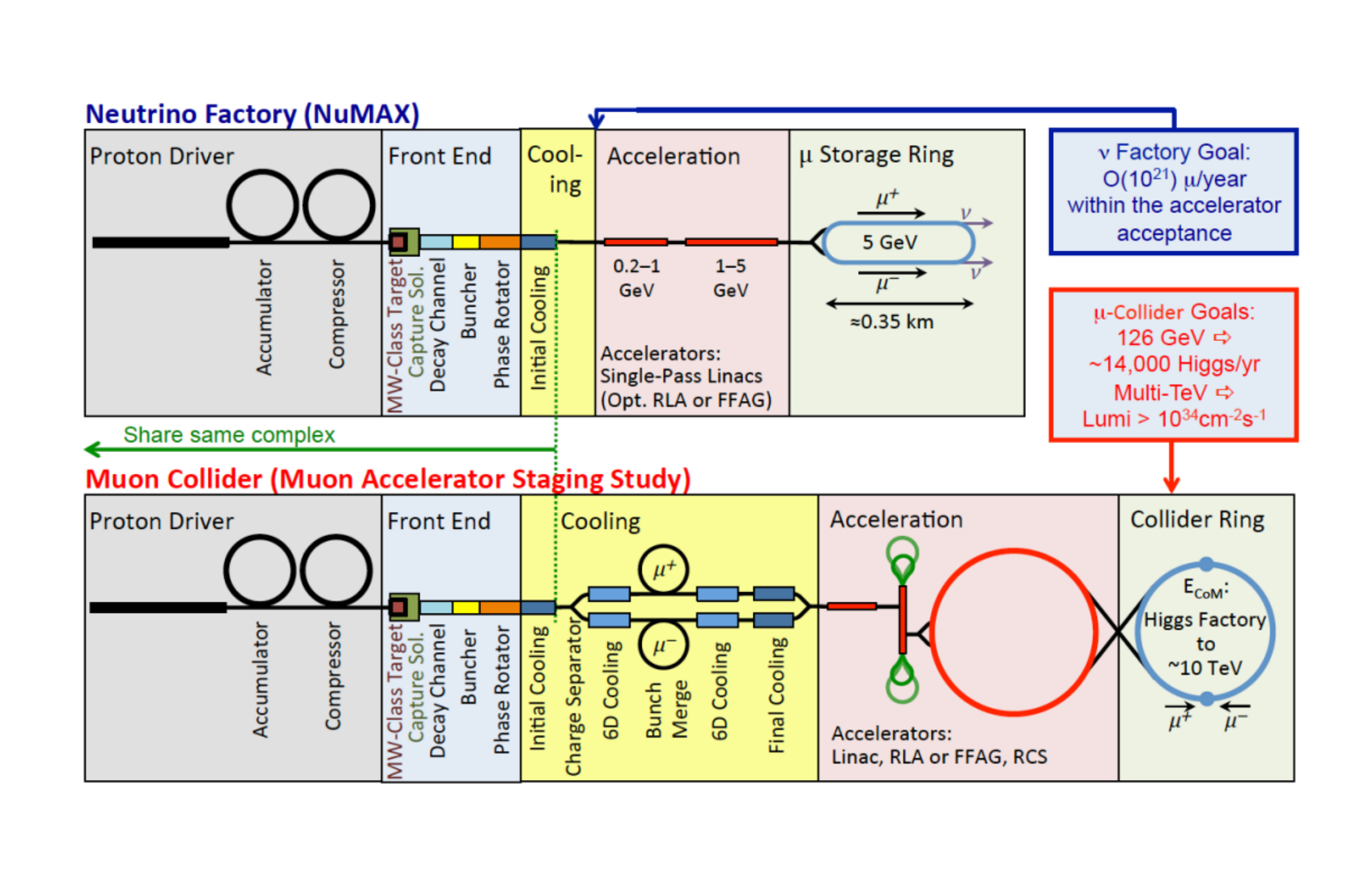}
\caption{Schematic drawing of the International Design Study for the Neutrino Factory.  Figure reproduced from ~\cite{IDS-NF:2011swj}.    }
\label{fig:nufactory2}
\end{figure}

The term ``neutrino factory'' typically refers to muon storage rings optimized for the production of long-baseline neutrino beams.  Since the first neutrino factory proposal in 1997~\cite{Geer:1997iz}, several concepts for neutrino factories have been developed, including the International Design Study for the Neutrino Factory~\cite{IDS-NF:2011swj} (see Fig.~\ref{fig:nufactory1}) and the NuMAX long-baseline neutrino factory concept~\cite{Delahaye:2018yfq} (see Fig.~\ref{fig:nufactory2}.  More recently, the physics case for a neutrino factory was described in a Snowmass 2022 White Paper~\cite{Bogacz:2022xsj}.  

A neutrino factory would expand on capabilities of the currently planned long-baseline program by providing high-statistics measurements of $\nu_\mu$, $\nu_e$, $\bar{\nu}_\mu$, and $\bar{\nu}_e$ oscillations.  It would have greater sensitivity to $\delta_{CP}$ than DUNE or Hyper-Kamiokande and facilitate a wide variety of searches for BSM at both the near and far detector.  A key feature of a neutrino factory is the ability to produce a high energy neutrino beam without compromising the intensity of the neutrino beam.  As such, it would provide high precision measurements of oscillations to tau neutrinos, opening up new tests of the three-flavor mixing paradigm that are not currently possible.  

Several far detector technology options have been considered for neutrino factories, including a liquid argon TPC, magnetized iron spectrometer, or water cherenkov detector.

\subsubsection{Muon Collider}

A muon collider, at either the TeV scale or as a Higgs Factory~\cite{Neuffer:2013wrd} ($E_{CM}\sim 126~\mathrm{GeV}$), is a future possibility being considered by the collider community. A muon collider would also be an excellent source of neutrinos.  In fact, they are such intense sources of neutrinos that neutrino radiological concerns must be taken into account in their design~\cite{Alexahin:2022tav}.  This neutrino radiation will come in the form of high energy neutrinos, hundreds (tens) of GeV in the case of a TeV-scale (Higgs-factor) collider.  This energy scale makes it difficult or impossible to study standard 3-flavor oscillations, but other physics such as BSM searches and observation of $\nu_\tau$ appearance may be possible.  Neutrino physics at the collider interaction point would also be possible~\cite{tau-neutrino-lepton-collider}, including $\nu+\mu$ and $\nu+\nu$ interactions~\cite{DeGouvea2021}.

%\cleardoublepage

%\input{sections/conclusions}
%\cleardoublepage

\section{Acknowledgements}
\label{sec:acknowledgements}

The authors would like to thank everyone who participated in the NF09 activities that led to this report, including speakers and attendees at our workshop, the conveners of the AF02 working group, those who provided comments on drafts of this report, and authors of relevant Letters of Interest and White Papers.  

This material is based upon work supported by the U.S. Department of Energy, Office of Science, Office of High Energy Physics, under Award Numbers DE-SC0007859, DE-SC0009920, DE-SC0010005, DE-SC0022172 and DE-SC0022529, as well as by the National Science Foundation under Grant PHY-2111546. 

%%%%%%%%%%%%%%%%%%%%%%%%%%%%%%%%%%%%%%%%%%%%%%%%%%%%%%%%%%%

%\clearpage

    % this is added just after end of document

% end stuff from init
%\cleardoublepage
%\renewcommand{\bibname}{References}
\renewcommand{\refname}{References}

\printglossary

\bibliographystyle{utphys}
% To understand the style chosen, see:
% https://arxiv.org/hypertex/bibstyles/ (very bottom -- additions) and
% https://www.sharelatex.com/learn/Bibtex_bibliography_styles
% July 2017, AH and BV (and AM)

%\begin{multicols}{2}[\printbibheading]
\bibliography{common/references}

\providecommand{\href}[2]{#2}\begingroup\raggedright\begin{thebibliography}{100}

\bibitem{Danby:1962nd}
G.~Danby, J.~M. Gaillard, K.~A. Goulianos, L.~M. Lederman, N.~B. Mistry,
  M.~Schwartz, and J.~Steinberger
  \href{http://dx.doi.org/10.1103/PhysRevLett.9.36}{{\em Phys. Rev. Lett.}
  {\bfseries 9} (1962) 36--44}.

\bibitem{LSND:1996jxj}
{\bfseries LSND} Collaboration, C.~Athanassopoulos {\em et~al.}
  \href{http://dx.doi.org/10.1016/S0168-9002(96)01155-2}{{\em Nucl. Instrum.
  Meth. A} {\bfseries 388} (1997) 149--172},
  \href{http://arxiv.org/abs/nucl-ex/9605002}{{\ttfamily
  arXiv:nucl-ex/9605002}}.

\bibitem{MicroBooNE:2015bmn}
{\bfseries MicroBooNE, LAr1-ND, ICARUS-WA104} Collaboration, M.~Antonello {\em
  et~al.} \href{http://arxiv.org/abs/1503.01520}{{\ttfamily arXiv:1503.01520
  [physics.ins-det]}}.

\bibitem{BNB_TDR}
{\bfseries BooNE} Collaboration, I.~Stancu {\em et~al.}, ``{Technical Design
  Report for the MiniBooNE Neutrino Beam},'' tech. rep., 2001.
\newblock \url{http://www-boone.fnal.gov/publicpages/target_tdr.ps.gz}.

\bibitem{Crane:1995ky}
{\bfseries NuMI Beam Group} Collaboration, D.~A. Crane, W.~Freeman, M.~C.
  Goodman, D.~Johnson, A.~Malensek, J.~G. Morfin, S.~O'Day, and J.~Thomas.

\bibitem{T2K:2011qtm}
{\bfseries T2K} Collaboration, K.~Abe {\em et~al.}
  \href{http://dx.doi.org/10.1016/j.nima.2011.06.067}{{\em Nucl. Instrum. Meth.
  A} {\bfseries 659} (2011) 106--135},
  \href{http://arxiv.org/abs/1106.1238}{{\ttfamily arXiv:1106.1238
  [physics.ins-det]}}.

\bibitem{NOvA:2007rmc}
{\bfseries NOvA} Collaboration, D.~S. Ayres {\em et~al.},
  \href{http://dx.doi.org/10.2172/935497}{``{The NOvA Technical Design
  Report},''} tech. rep., 2007.

\bibitem{Lebedev:2017vnu}
M.~Ball {\em et~al.}, \href{http://dx.doi.org/10.2172/1346823}{``{The PIP-II
  Conceptual Design Report},''} tech. rep., 2017.

\bibitem{DUNE:2016evb}
{\bfseries DUNE} Collaboration, J.~Strait {\em et~al.}
  \href{http://arxiv.org/abs/1601.05823}{{\ttfamily arXiv:1601.05823
  [physics.ins-det]}}.

\bibitem{NOvALOI}
P.~Shanahan, P.~Vahle, {\em et~al.}, ``{The NOvA Physics Program through
  2025},'' {Snowmass Letter of Interest}, 2020.
\newblock
  \url{https://www.snowmass21.org/docs/files/summaries/NF/SNOWMASS21-NF1_NF3_Patricia_Vahle-145.pdf}.

\bibitem{Super-Kamiokande:2002weg}
{\bfseries Super-Kamiokande} Collaboration, Y.~Fukuda {\em et~al.}
  \href{http://dx.doi.org/10.1016/S0168-9002(03)00425-X}{{\em Nucl. Instrum.
  Meth. A} {\bfseries 501} (2003) 418--462}.

\bibitem{T2KLOI}
K.~Mahn {\em et~al.}, ``{T2K Experiment: future plans and capabilities},''
  {Snowmass Letter of Interest}, 2020.
\newblock
  \url{https://www.snowmass21.org/docs/files/summaries/NF/SNOWMASS21-NF1_NF3__NF06_T2KCollab-130.pdf}.

\bibitem{Hyper-Kamiokande:2018ofw}
{\bfseries Hyper-Kamiokande} Collaboration, K.~Abe {\em et~al.}
  \href{http://arxiv.org/abs/1805.04163}{{\ttfamily arXiv:1805.04163
  [physics.ins-det]}}.

\bibitem{Hyper-Kamiokande:2022smq}
{\bfseries Hyper-Kamiokande} Collaboration, J.~Bian {\em et~al.} in {\em {2022
  Snowmass Summer Study}}.
\newblock 3, 2022.
\newblock \href{http://arxiv.org/abs/2203.02029}{{\ttfamily arXiv:2203.02029
  [hep-ex]}}.

\bibitem{JapanLOI}
S.~Asai {\em et~al.}, ``{Update of the Japanese Strategy for Particle
  Physics},'' {Snowmass Letter of Interest}, 2020.
\newblock
  \url{https://www.snowmass21.org/docs/files/summaries/EF/SNOWMASS21-EF0_EF0-NF0_NF0-RF0_RF0-CF0_CF0-TF0_TF0-AF0_AF0-IF0_IF0-CompF0_CompF0-UF0_UF0-CommF0_CommF0-254.pdf}.

\bibitem{DUNE:2022aul}
{\bfseries DUNE} Collaboration, A.~A. Abud {\em et~al.} {\em {2022 Snowmass
  Summer Study}} , \href{http://arxiv.org/abs/2203.06100}{{\ttfamily
  arXiv:2203.06100 [hep-ex]}}.

\bibitem{DUNE:2020jqi}
{\bfseries DUNE} Collaboration, B.~Abi {\em et~al.}
  \href{http://dx.doi.org/10.1140/epjc/s10052-020-08456-z}{{\em Eur. Phys. J.
  C} {\bfseries 80} no.~10, (2020) 978},
  \href{http://arxiv.org/abs/2006.16043}{{\ttfamily arXiv:2006.16043
  [hep-ex]}}.

\bibitem{HEPAPSubcommittee:2014bsm}
{\bfseries HEPAP Subcommittee} Collaboration, S.~Ritz {\em et~al.}

\bibitem{Pellico:2022dju}
W.~Pellico, C.~Bhat, J.~Eldred, C.~Johnstone, J.~Johnstone, K.~Seiya, C.-Y.
  Tan, M.~Toups, P.~deNiverville, and R.~Van De~Water
  \href{http://arxiv.org/abs/2203.07339}{{\ttfamily arXiv:2203.07339
  [physics.acc-ph]}}.

\bibitem{Eldred:2022vxi}
J.~Eldred, S.~Nagaitsev, V.~Shiltsev, A.~Valishev, R.~Zwaska, and M.~Syphers
  {\em {2022 Snowmass Summer Study Proceedings}} ,
  \href{http://arxiv.org/abs/2203.08276}{{\ttfamily arXiv:2203.08276
  [physics.acc-ph]}}.

\bibitem{Nagaitsev:2021xzy}
S.~Nagaitsev and V.~Lebedev {\em {2022 Snowmass Summer Study Proceedings}} ,
  \href{http://arxiv.org/abs/2111.06932}{{\ttfamily arXiv:2111.06932
  [physics.acc-ph]}}.

\bibitem{Ainsworth:2021ahm}
R.~Ainsworth {\em et~al.} {\em {2022 Snowmass Summer Study Proceedings}} ,
  \href{http://arxiv.org/abs/2106.02133}{{\ttfamily arXiv:2106.02133
  [physics.acc-ph]}}.

\bibitem{Belomestnykh:2022kal}
S.~Belomestnykh, M.~Checchin, D.~Johnson, D.~Neuffer, H.~Padamsee, S.~Posen,
  E.~Pozdeyev, V.~Pronskikh, N.~Solyak, and V.~Yakovlev in {\em {2022 Snowmass
  Summer Study}}.
\newblock \href{http://arxiv.org/abs/2203.05052}{{\ttfamily arXiv:2203.05052
  [physics.acc-ph]}}.

\bibitem{Arrington:2022pon}
J.~Arrington {\em et~al.} {\em {2022 Snowmass Summer Study Proceedings}} ,
  \href{http://arxiv.org/abs/2203.03925}{{\ttfamily arXiv:2203.03925
  [hep-ph]}}.

\bibitem{HurhCSS}
P.~Hurh, ``{LBNF} beamline upgrades for 2.4{MW},'' 2022.
\newblock \url{https://indico.fnal.gov/event/22303/contributions/246067/}. Talk
  at Snowmass Community Summer Study, Seattle, Washington.

\bibitem{PellemoineCSS}
F.~Pellemoine, ``{R} and {D} for target materials,'' 2022.
\newblock \url{https://indico.fnal.gov/event/22303/contributions/246068/}. Talk
  at Snowmass Community Summer Study, Seattle, Washington.

\bibitem{radiate}
C.~J. Densham {\em et~al.}, ``{{R}\&{D} for {MW} Pion Production Targets for
  Next Generation Long Baseline Neutrino Facilities},'' {Snowmass Letter of
  Interest}, 2020.
\newblock
  \url{https://www.snowmass21.org/docs/files/summaries/AF/SNOWMASS21-AF7_AF7_Densham-190.pdf}.

\bibitem{Pellemoine:2022wmo}
F.~Pellemoine {\em et~al.} \href{http://arxiv.org/abs/2203.08239}{{\ttfamily
  arXiv:2203.08239 [physics.acc-ph]}}.

\bibitem{Ammigan:2022ogd}
K.~Ammigan {\em et~al.} \href{http://arxiv.org/abs/2203.08357}{{\ttfamily
  arXiv:2203.08357 [physics.acc-ph]}}.

\bibitem{Garoby:2017vew}
R.~Garoby {\em et~al.} \href{http://dx.doi.org/10.1088/1402-4896/aa9bff}{{\em
  Phys. Scripta} {\bfseries 93} no.~1, (2018) 014001}.

\bibitem{ESSnuLOI}
M.~D. Ekelof, T. {\em et~al.}, ``{The ESS neutrino Super Beam Design Study
  (ESSnuSB) and the High Intensity Frontier Initiative (HIFI)},'' {Snowmass
  Letter of Interest}, 2020.
\newblock
  \url{https://www.snowmass21.org/docs/files/summaries/NF/SNOWMASS21-NF1_NF10-AF2_AF0_Tord_Ekelof_and_Marcos_Dracos-062.pdf}.

\bibitem{Alekou:2022mav}
A.~Alekou {\em et~al.} \href{http://arxiv.org/abs/2203.08803}{{\ttfamily
  arXiv:2203.08803 [physics.acc-ph]}}.

\bibitem{Alekou:2022emd}
A.~Alekou {\em et~al.} \href{http://arxiv.org/abs/2206.01208}{{\ttfamily
  arXiv:2206.01208 [hep-ex]}}.

\bibitem{Ekelof:2019hqi}
T.~Ekelof and M.~Dracos \href{http://dx.doi.org/10.22323/1.337.0029}{{\em PoS}
  {\bfseries NOW2018} (2019) 029},
  \href{http://arxiv.org/abs/1903.08437}{{\ttfamily arXiv:1903.08437
  [physics.ins-det]}}.

\bibitem{Akindinov:2019flp}
A.~V. Akindinov {\em et~al.}
  \href{http://dx.doi.org/10.1140/epjc/s10052-019-7259-5}{{\em Eur. Phys. J. C}
  {\bfseries 79} no.~9, (2019) 758},
  \href{http://arxiv.org/abs/1902.06083}{{\ttfamily arXiv:1902.06083
  [physics.ins-det]}}.

\bibitem{StroboscopicLOI}
M.~Wetstein {\em et~al.}, ``{The use of precision timing in LBNF/DUNE},''
  {Snowmass Letter of Interest}, 2020.
\newblock
  \url{https://www.snowmass21.org/docs/files/summaries/NF/SNOWMASS21-NF1_NF10-AF2_AF0-116.pdf}.

\bibitem{Angelico:2019gyi}
E.~Angelico, J.~Eisch, A.~Elagin, H.~Frisch, S.~Nagaitsev, and M.~Wetstein
  \href{http://dx.doi.org/10.1103/PhysRevD.100.032008}{{\em Phys. Rev. D}
  {\bfseries 100} no.~3, (2019) 032008},
  \href{http://arxiv.org/abs/1904.01611}{{\ttfamily arXiv:1904.01611
  [physics.acc-ph]}}.

\bibitem{Angelico:2020gyn}
E.~Angelico, A.~Elagin, H.~J. Frisch, and M.~Wetstein
  \href{http://arxiv.org/abs/2004.00580}{{\ttfamily arXiv:2004.00580
  [physics.ins-det]}}.

\bibitem{TheiaLOI}
M.~Wilking {\em et~al.}, ``{Long-Baseline Neutrinos at THEIA},'' {Snowmass
  Letter of Interest}, 2020.
\newblock
  \url{https://www.snowmass21.org/docs/files/summaries/NF/SNOWMASS21-NF1_NF10-178.pdf}.

\bibitem{tauphysics}
A.~Aurisano {\em et~al.}, ``{Tau Neutrino Physics},'' {Snowmass Letter of
  Interest}, 2020.
\newblock
  \url{https://www.snowmass21.org/docs/files/summaries/NF/SNOWMASS21-NF1_NF5-TF11_TF0_Kevin_J_Kelly-126.pdf}.

\bibitem{Abraham:2022jse}
R.~M. Abraham {\em et~al.} \href{http://arxiv.org/abs/2203.05591}{{\ttfamily
  arXiv:2203.05591 [hep-ph]}}.

\bibitem{DUNE:2020ypp}
{\bfseries DUNE} Collaboration, B.~Abi {\em et~al.}
  \href{http://arxiv.org/abs/2002.03005}{{\ttfamily arXiv:2002.03005
  [hep-ex]}}.

\bibitem{Yonehara:2022lmk}
K.~Yonehara \href{http://arxiv.org/abs/2203.06024}{{\ttfamily arXiv:2203.06024
  [physics.acc-ph]}}.

\bibitem{Brizzolari:2022seb}
C.~Brizzolari {\em et~al.} \href{http://dx.doi.org/10.22323/1.398.0205}{{\em
  PoS} {\bfseries EPS-HEP2021} (2022) 205}.

\bibitem{Meregaglia:2016vxf}
A.~Meregaglia \href{http://dx.doi.org/10.1088/1748-0221/11/12/C12040}{{\em
  JINST} {\bfseries 11} no.~12, (2016) C12040}.

\bibitem{SPY:1998czp}
{\bfseries SPY} Collaboration, G.~Ambrosini {\em et~al.}
  \href{http://dx.doi.org/10.1016/S0370-2693(98)00237-8}{{\em Phys. Lett. B}
  {\bfseries 425} (1998) 208--214}.

\bibitem{SPY:1998jku}
{\bfseries SPY} Collaboration, G.~Ambrosini {\em et~al.}
  \href{http://dx.doi.org/10.1016/S0370-2693(97)01572-4}{{\em Phys. Lett. B}
  {\bfseries 420} (1998) 225}.

\bibitem{NA56SPY:1999zez}
{\bfseries NA56/SPY} Collaboration, G.~Ambrosini {\em et~al.}
  \href{http://dx.doi.org/10.1007/s100520050601}{{\em Eur. Phys. J. C}
  {\bfseries 10} (1999) 605--627}.

\bibitem{HARP:2005clh}
{\bfseries HARP} Collaboration, M.~G. Catanesi {\em et~al.}
  \href{http://dx.doi.org/10.1016/j.nuclphysb.2005.10.016}{{\em Nucl. Phys. B}
  {\bfseries 732} (2006) 1--45},
  \href{http://arxiv.org/abs/hep-ex/0510039}{{\ttfamily arXiv:hep-ex/0510039}}.

\bibitem{HARP:2007dqt}
{\bfseries HARP} Collaboration, M.~G. Catanesi {\em et~al.}
  \href{http://dx.doi.org/10.1140/epjc/s10052-007-0382-8}{{\em Eur. Phys. J. C}
  {\bfseries 52} (2007) 29--53},
  \href{http://arxiv.org/abs/hep-ex/0702024}{{\ttfamily arXiv:hep-ex/0702024}}.

\bibitem{HARP:2008sqs}
{\bfseries HARP} Collaboration, M.~G. Catanesi {\em et~al.}
  \href{http://dx.doi.org/10.1016/j.astropartphys.2008.02.002}{{\em Astropart.
  Phys.} {\bfseries 29} (2008) 257--281},
  \href{http://arxiv.org/abs/0802.0657}{{\ttfamily arXiv:0802.0657
  [astro-ph]}}.

\bibitem{E910:2007puw}
{\bfseries E910} Collaboration, I.~Chemakin {\em et~al.}
  \href{http://dx.doi.org/10.1103/PhysRevC.77.049903}{{\em Phys. Rev. C}
  {\bfseries 77} (2008) 015209},
  \href{http://arxiv.org/abs/0707.2375}{{\ttfamily arXiv:0707.2375 [nucl-ex]}}.
  [Erratum: Phys.Rev.C 77, 049903 (2008)].

\bibitem{NA49:2006oyk}
{\bfseries NA49} Collaboration, C.~Alt {\em et~al.}
  \href{http://dx.doi.org/10.1140/epjc/s10052-006-0165-7}{{\em Eur. Phys. J. C}
  {\bfseries 49} (2007) 897--917},
  \href{http://arxiv.org/abs/hep-ex/0606028}{{\ttfamily arXiv:hep-ex/0606028}}.

\bibitem{NA49:2012jna}
{\bfseries NA49} Collaboration, B.~Baatar {\em et~al.}
  \href{http://dx.doi.org/10.1140/epjc/s10052-013-2364-3}{{\em Eur. Phys. J. C}
  {\bfseries 73} no.~4, (2013) 2364},
  \href{http://arxiv.org/abs/1207.6520}{{\ttfamily arXiv:1207.6520 [hep-ex]}}.

\bibitem{MIPP:2010vnr}
{\bfseries MIPP} Collaboration, T.~S. Nigmanov {\em et~al.}
  \href{http://dx.doi.org/10.1103/PhysRevD.83.012002}{{\em Phys. Rev. D}
  {\bfseries 83} (2011) 012002},
  \href{http://arxiv.org/abs/1010.6291}{{\ttfamily arXiv:1010.6291 [hep-ex]}}.

\bibitem{MIPP:2014shj}
{\bfseries MIPP} Collaboration, J.~M. Paley {\em et~al.}
  \href{http://dx.doi.org/10.1103/PhysRevD.90.032001}{{\em Phys. Rev. D}
  {\bfseries 90} no.~3, (2014) 032001},
  \href{http://arxiv.org/abs/1404.5882}{{\ttfamily arXiv:1404.5882 [hep-ex]}}.

\bibitem{NA61SHINE:2015bad}
{\bfseries NA61/SHINE} Collaboration, N.~Abgrall {\em et~al.}
  \href{http://dx.doi.org/10.1140/epjc/s10052-016-3898-y}{{\em Eur. Phys. J. C}
  {\bfseries 76} no.~2, (2016) 84},
  \href{http://arxiv.org/abs/1510.02703}{{\ttfamily arXiv:1510.02703
  [hep-ex]}}.

\bibitem{T2K:2012bge}
{\bfseries T2K} Collaboration, K.~Abe {\em et~al.}
  \href{http://dx.doi.org/10.1103/PhysRevD.87.012001}{{\em Phys. Rev. D}
  {\bfseries 87} no.~1, (2013) 012001},
  \href{http://arxiv.org/abs/1211.0469}{{\ttfamily arXiv:1211.0469 [hep-ex]}}.
  [Addendum: Phys.Rev.D 87, 019902 (2013)].

\bibitem{NA61SHINE:2018rhe}
{\bfseries NA61/SHINE} Collaboration, N.~Abgrall {\em et~al.}
  \href{http://dx.doi.org/10.1140/epjc/s10052-019-6583-0}{{\em Eur. Phys. J. C}
  {\bfseries 79} no.~2, (2019) 100},
  \href{http://arxiv.org/abs/1808.04927}{{\ttfamily arXiv:1808.04927
  [hep-ex]}}.

\bibitem{NA61LOI}
E.~D. Zimmerman {\em et~al.}, ``{The future NA61/SHINE program on hadron
  production},'' {Snowmass Letter of Interest}, 2020.
\newblock
  \url{https://www.snowmass21.org/docs/files/summaries/NF/SNOWMASS21-NF9_NF4_Zimmerman,Eric-069.pdf}.

\bibitem{EMPHATIC:2019xmc}
{\bfseries EMPHATIC} Collaboration, T.~Akaishi {\em et~al.}
  \href{http://arxiv.org/abs/1912.08841}{{\ttfamily arXiv:1912.08841
  [hep-ex]}}.

\bibitem{EMPHATICLOI}
J.~Paley {\em et~al.}, ``{The EMPHATIC Table-Top Spectrometer: Enabling Hadron
  Scattering and Production Measurements for Improved Beam Simulations},''
  {Snowmass Letter of Interest}, 2020.
\newblock
  \url{https://www.snowmass21.org/docs/files/summaries/NF/SNOWMASS21-NF9_NF0-EF0_EF0-RF0_RF0-IF0_IF0_EMPHATIC_Paley-173.pdf}.

\bibitem{DsTau:2019wjb}
{\bfseries DsTau} Collaboration, S.~Aoki {\em et~al.}
  \href{http://dx.doi.org/10.1007/JHEP01(2020)033}{{\em JHEP} {\bfseries 01}
  (2020) 033}, \href{http://arxiv.org/abs/1906.03487}{{\ttfamily
  arXiv:1906.03487 [hep-ex]}}.

\bibitem{nuPRISM:2014mzw}
{\bfseries nuPRISM} Collaboration, S.~Bhadra {\em et~al.}
  \href{http://arxiv.org/abs/1412.3086}{{\ttfamily arXiv:1412.3086
  [physics.ins-det]}}.

\bibitem{DUNE:2021tad}
{\bfseries DUNE} Collaboration, A.~Abed~Abud {\em et~al.}
  \href{http://dx.doi.org/10.3390/instruments5040031}{{\em Instruments}
  {\bfseries 5} no.~4, (2021) 31},
  \href{http://arxiv.org/abs/2103.13910}{{\ttfamily arXiv:2103.13910
  [physics.ins-det]}}.

\bibitem{SANDLOI}
R.~Petti {\em et~al.}, ``{Precision Measurements with (Anti)Neutrinos at
  LBNF},'' {Snowmass Letter of Interest}, 2020.
\newblock
  \url{https://www.snowmass21.org/docs/files/summaries/NF/SNOWMASS21-NF5_NF6-EF6_EF4-RF1_RF6-122.pdf}.

\bibitem{DUNE:2022yni}
{\bfseries DUNE} Collaboration, A.~A. Abud {\em et~al.} {\em {2022 Snowmass
  Summer Study}} , \href{http://arxiv.org/abs/2203.06281}{{\ttfamily
  arXiv:2203.06281 [hep-ex]}}.

\bibitem{JSNS2:2021hyk}
{\bfseries JSNS2} Collaboration, S.~Ajimura {\em et~al.}
  \href{http://dx.doi.org/10.1016/j.nima.2021.165742}{{\em Nucl. Instrum. Meth.
  A} {\bfseries 1014} (2021) 165742},
  \href{http://arxiv.org/abs/2104.13169}{{\ttfamily arXiv:2104.13169
  [physics.ins-det]}}.

\bibitem{Ajimura:2017fld}
S.~Ajimura {\em et~al.} \href{http://arxiv.org/abs/1705.08629}{{\ttfamily
  arXiv:1705.08629 [physics.ins-det]}}.

\bibitem{CCM:2021leg}
{\bfseries CCM} Collaboration, A.~A. Aguilar-Arevalo {\em et~al.}
  \href{http://arxiv.org/abs/2105.14020}{{\ttfamily arXiv:2105.14020
  [hep-ex]}}.

\bibitem{COHERENT:2017ipa}
{\bfseries COHERENT} Collaboration, D.~Akimov {\em et~al.}
  \href{http://dx.doi.org/10.1126/science.aao0990}{{\em Science} {\bfseries
  357} no.~6356, (2017) 1123--1126},
  \href{http://arxiv.org/abs/1708.01294}{{\ttfamily arXiv:1708.01294
  [nucl-ex]}}.

\bibitem{Kelly:2021jgj}
K.~J. Kelly, P.~A.~N. Machado, A.~Marchionni, and Y.~F. Perez-Gonzalez
  \href{http://dx.doi.org/10.1007/JHEP08(2021)087}{{\em JHEP} {\bfseries 08}
  (2021) 087}, \href{http://arxiv.org/abs/2103.00009}{{\ttfamily
  arXiv:2103.00009 [hep-ph]}}.

\bibitem{Toups:2022yxs}
M.~Toups {\em et~al.} in {\em {2022 Snowmass Summer Study}}.
\newblock 3, 2022.
\newblock \href{http://arxiv.org/abs/2203.08079}{{\ttfamily arXiv:2203.08079
  [hep-ex]}}.

\bibitem{Zettlemoyer}
J.~Zettlemoyer. Private communication (2022).

\bibitem{Kahn:2014sra}
Y.~Kahn, G.~Krnjaic, J.~Thaler, and M.~Toups
  \href{http://dx.doi.org/10.1103/PhysRevD.91.055006}{{\em Phys. Rev. D}
  {\bfseries 91} no.~5, (2015) 055006},
  \href{http://arxiv.org/abs/1411.1055}{{\ttfamily arXiv:1411.1055 [hep-ph]}}.

\bibitem{Jordan:2018gcd}
J.~R. Jordan, Y.~Kahn, G.~Krnjaic, M.~Moschella, and J.~Spitz
  \href{http://dx.doi.org/10.1103/PhysRevD.98.075020}{{\em Phys. Rev. D}
  {\bfseries 98} no.~7, (2018) 075020},
  \href{http://arxiv.org/abs/1806.05185}{{\ttfamily arXiv:1806.05185
  [hep-ph]}}.

\bibitem{Akimov:2022oyb}
D.~Akimov {\em et~al.} in {\em {2022 Snowmass Summer Study}}.
\newblock 4, 2022.
\newblock \href{http://arxiv.org/abs/2204.04575}{{\ttfamily arXiv:2204.04575
  [hep-ex]}}.

\bibitem{Lindner:2016wff}
M.~Lindner, W.~Rodejohann, and X.-J. Xu
  \href{http://dx.doi.org/10.1007/JHEP03(2017)097}{{\em JHEP} {\bfseries 03}
  (2017) 097}, \href{http://arxiv.org/abs/1612.04150}{{\ttfamily
  arXiv:1612.04150 [hep-ph]}}.

\bibitem{LSND:1996ubh}
{\bfseries LSND} Collaboration, C.~Athanassopoulos {\em et~al.}
  \href{http://dx.doi.org/10.1103/PhysRevLett.77.3082}{{\em Phys. Rev. Lett.}
  {\bfseries 77} (1996) 3082--3085},
  \href{http://arxiv.org/abs/nucl-ex/9605003}{{\ttfamily
  arXiv:nucl-ex/9605003}}.

\bibitem{Hino:2021uwz}
Y.~Hino {\em et~al.} \href{http://arxiv.org/abs/2111.07482}{{\ttfamily
  arXiv:2111.07482 [hep-ex]}}.

\bibitem{Ajimura:2020qni}
S.~Ajimura {\em et~al.} \href{http://arxiv.org/abs/2012.10807}{{\ttfamily
  arXiv:2012.10807 [hep-ex]}}.

\bibitem{Baxter:2019mcx}
D.~Baxter {\em et~al.} \href{http://dx.doi.org/10.1007/JHEP02(2020)123}{{\em
  JHEP} {\bfseries 02} (2020) 123},
  \href{http://arxiv.org/abs/1911.00762}{{\ttfamily arXiv:1911.00762
  [physics.ins-det]}}.

\bibitem{Anderson:2012pn}
A.~J. Anderson, J.~M. Conrad, E.~Figueroa-Feliciano, C.~Ignarra, G.~Karagiorgi,
  K.~Scholberg, M.~H. Shaevitz, and J.~Spitz
  \href{http://dx.doi.org/10.1103/PhysRevD.86.013004}{{\em Phys. Rev. D}
  {\bfseries 86} (2012) 013004},
  \href{http://arxiv.org/abs/1201.3805}{{\ttfamily arXiv:1201.3805 [hep-ph]}}.

\bibitem{VandeWater:2022qot}
R.~G. Van~de Water {\em et~al.} in {\em {2022 Snowmass Summer Study}}.
\newblock 4, 2022.
\newblock \href{http://arxiv.org/abs/2204.01860}{{\ttfamily arXiv:2204.01860
  [physics.ins-det]}}.

\bibitem{O1GeVBeamDumpLOI}
M.~Toups {\em et~al.}, ``{Fixed-Target Searches for New Physics with O(1 GeV)
  Proton Beams at Fermi National Accelerator Laboratory},'' {Snowmass Letter of
  Interest}, 2020.
\newblock
  \url{https://www.snowmass21.org/docs/files/summaries/RF/SNOWMASS21-RF6_RF0-NF2_NF3-AF2_AF5-099.pdf}.

\bibitem{O10GeVBeamDumpLOI}
M.~Toups {\em et~al.}, ``{Fixed-Target Searches for New Physics with O(10 GeV)
  Proton Beams at Fermi National Accelerator Laboratory},'' {Snowmass Letter of
  Interest}, 2020.
\newblock
  \url{https://www.snowmass21.org/docs/files/summaries/RF/SNOWMASS21-RF6_RF0-NF3_NF0-AF5_AF0-084.pdf}.

\bibitem{Spitz:2012gp}
J.~Spitz \href{http://dx.doi.org/10.1103/PhysRevD.85.093020}{{\em Phys. Rev. D}
  {\bfseries 85} (2012) 093020},
  \href{http://arxiv.org/abs/1203.6050}{{\ttfamily arXiv:1203.6050 [hep-ph]}}.

\bibitem{Spitz:2014hwa}
J.~Spitz \href{http://dx.doi.org/10.1103/PhysRevD.89.073007}{{\em Phys. Rev. D}
  {\bfseries 89} no.~7, (2014) 073007},
  \href{http://arxiv.org/abs/1402.2284}{{\ttfamily arXiv:1402.2284
  [physics.ins-det]}}.

\bibitem{Axani:2015dha}
S.~Axani, G.~Collin, J.~Conrad, M.~Shaevitz, J.~Spitz, and T.~Wongjirad
  \href{http://dx.doi.org/10.1103/PhysRevD.92.092010}{{\em Phys. Rev. D}
  {\bfseries 92} no.~9, (2015) 092010},
  \href{http://arxiv.org/abs/1506.05811}{{\ttfamily arXiv:1506.05811
  [physics.ins-det]}}.

\bibitem{MiniBooNE:2018dus}
{\bfseries MiniBooNE} Collaboration, A.~A. Aguilar-Arevalo {\em et~al.}
  \href{http://dx.doi.org/10.1103/PhysRevLett.120.141802}{{\em Phys. Rev.
  Lett.} {\bfseries 120} no.~14, (2018) 141802},
  \href{http://arxiv.org/abs/1801.03848}{{\ttfamily arXiv:1801.03848
  [hep-ex]}}.

\bibitem{SHIP1}
{\bfseries SHIP} Collaboration, C.~Ahdida {\em et~al.}
  \href{http://arxiv.org/abs/2112.01487}{{\ttfamily arXiv:2112.01487
  [physics.ins-det]}}.

\bibitem{Cowan:1956rrn}
C.~L. Cowan, F.~Reines, F.~B. Harrison, H.~W. Kruse, and A.~D. McGuire
  \href{http://dx.doi.org/10.1126/science.124.3212.103}{{\em Science}
  {\bfseries 124} (1956) 103--104}.

\bibitem{wnassoc}
{World Nuclear Association}, ``{Plans for New Reactors Worldwide}.''
  \url{https://world-nuclear.org/information-library/current-and-future-generation/plans-for-new-reactors-worldwide.aspx}.
\newblock Accessed: 2022-05-10.

\bibitem{reactorwp}
O.~Akindele {\em et~al.} \href{http://arxiv.org/abs/2203.07214}{{\ttfamily
  arXiv:2203.07214 [hep-ph]}}.

\bibitem{Cogswell:2016aog}
B.~K. Cogswell and P.~Huber
  \href{http://dx.doi.org/10.1080/08929882.2016.1184531}{{\em Science \& Global
  Security} {\bfseries 24} no.~2, (2016) 114--130}.

\bibitem{wondram2021}
C.~Romano, N.~Bowden, A.~Conant, B.~Goldblum, P.~Huber, J.~Link, B.~Littlejohn,
  H.~Mumm, J.~Ochoa-Ricoux, S.~Prasad, C.~Riddle, A.~Sonzogni, and
  W.~Wieselquist. \url{{http://dx.doi.org/10.2172/1842423}}.

\bibitem{nf07report}
{N.~S. Bowden, J.~M. Link, and W.~Wang}, ``{NF07 Topical Group Report},''
  {Snowmass Topical Group Report}, 2022.
\newblock
  \url{https://drive.google.com/file/d/1N8tlcxV-maLQ-NXHNe-6sHCbGAFQrUjY/view}.

\bibitem{bib:IAEA}
``'antineutrino spectra and their applications'.'' Iaea report indc(nds)-0786,
  2019.
\newblock {https://www-nds.iaea.org/publications/indc/indc-nds-0786.pdf}.

\bibitem{BROWN20181}
D.~Brown {\em et~al.}
  \href{http://dx.doi.org/https://doi.org/10.1016/j.nds.2018.02.001}{{\em
  Nuclear Data Sheets} {\bfseries 148} (2018) 1--142}.
  \url{https://www.sciencedirect.com/science/article/pii/S0090375218300206}.
  Special Issue on Nuclear Reaction Data.

\bibitem{bib:JEFF33}
A.~Plompen {\em et~al.} {\em Eur. Phys. J. A} {\bfseries 56} (2020) 181.

\bibitem{bib:ENSDF}
J.~Tuli {\em Nucl. Inst. Meth. A} {\bfseries 506} (1996) 506.

\bibitem{Huber:2016fkt}
P.~Huber \href{http://dx.doi.org/10.1016/j.nuclphysb.2016.04.012}{{\em Nucl.
  Phys.} {\bfseries B908} (2016) 268--278},
\href{http://arxiv.org/abs/1602.01499}{{\ttfamily arXiv:1602.01499 [hep-ph]}}.
%%CITATION = ARXIV:1602.01499;%%.

\bibitem{PhysRevLett.105.202501}
A.~Algora {\em et~al.}
  \href{http://dx.doi.org/10.1103/PhysRevLett.105.202501}{{\em Phys. Rev.
  Lett.} {\bfseries 105} (Nov, 2010) 202501}.
  \url{https://link.aps.org/doi/10.1103/PhysRevLett.105.202501}.

\bibitem{tas_lots}
A.~Fija\l{}kowska {\em et~al.}
  \href{http://dx.doi.org/10.1103/PhysRevLett.119.052503}{{\em Phys. Rev.
  Lett.} {\bfseries 119} (Aug, 2017) 052503}.
  \url{https://link.aps.org/doi/10.1103/PhysRevLett.119.052503}.

\bibitem{tas_few}
B.~Rasco {\em et~al.}
  \href{http://dx.doi.org/10.1103/PhysRevLett.117.092501}{{\em Phys. Rev.
  Lett.} {\bfseries 117} no.~9, (2016) 092501}.

\bibitem{tas_br}
E.~Valencia {\em et~al.}
  \href{http://dx.doi.org/10.1103/PhysRevC.95.024320}{{\em Phys. Rev. C}
  {\bfseries 95} no.~2, (2017) 024320},
  \href{http://arxiv.org/abs/1609.06128}{{\ttfamily arXiv:1609.06128
  [nucl-ex]}}.

\bibitem{tas_nb}
V.~Guadilla {\em et~al.}
  \href{http://dx.doi.org/10.1103/PhysRevC.100.024311}{{\em Phys. Rev. C}
  {\bfseries 100} no.~2, (2019) 024311},
  \href{http://arxiv.org/abs/1904.07036}{{\ttfamily arXiv:1904.07036
  [nucl-ex]}}.

\bibitem{tas_nb2}
V.~Guadilla {\em et~al.}
  \href{http://dx.doi.org/10.1103/PhysRevLett.122.042502}{{\em Phys. Rev.
  Lett.} {\bfseries 122} no.~4, (2019) 042502}.

\bibitem{tas_rb}
{\bfseries IGISOL} Collaboration, A.~Zakari-Issoufou {\em et~al.}
  \href{http://dx.doi.org/10.1103/PhysRevLett.115.102503}{{\em Phys. Rev.
  Lett.} {\bfseries 115} no.~10, (2015) 102503},
  \href{http://arxiv.org/abs/1504.05812}{{\ttfamily arXiv:1504.05812
  [nucl-ex]}}.

\bibitem{tas_rbbr}
S.~Rice {\em et~al.} \href{http://dx.doi.org/10.1103/PhysRevC.96.014320}{{\em
  Phys. Rev. C} {\bfseries 96} no.~1, (2017) 014320},
  \href{http://arxiv.org/abs/1704.01915}{{\ttfamily arXiv:1704.01915
  [nucl-ex]}}.

\bibitem{tas_rbi}
V.~Guadilla {\em et~al.}
  \href{http://dx.doi.org/10.1103/PhysRevC.100.044305}{{\em Phys. Rev. C}
  {\bfseries 100} no.~4, (2019) 044305},
  \href{http://arxiv.org/abs/1907.02748}{{\ttfamily arXiv:1907.02748
  [nucl-ex]}}.

\bibitem{Matthews2021}
E.~F. Matthews, L.~A. Bernstein, and W.~Younes
  \href{http://dx.doi.org/https://doi.org/10.1016/j.adt.2021.101441}{{\em
  Atomic Data and Nuclear Data Tables} {\bfseries 140C} (2021) 101441}.
  \url{https://www.sciencedirect.com/science/article/pii/S0092640X21000279}.

\bibitem{SCHRECKENBACH1985325}
K.~Schreckenbach, G.~Colvin, W.~Gelletly, and F.~{Von Feilitzsch}
  \href{http://dx.doi.org/https://doi.org/10.1016/0370-2693(85)91337-1}{{\em
  Physics Letters B} {\bfseries 160} no.~4, (1985) 325--330}.
  \url{https://www.sciencedirect.com/science/article/pii/0370269385913371}.

\bibitem{VONFEILITZSCH1982162}
F.~{von Feilitzsch}, A.~Hahn, and K.~Schreckenbach
  \href{http://dx.doi.org/https://doi.org/10.1016/0370-2693(82)90622-0}{{\em
  Physics Letters B} {\bfseries 118} no.~1, (1982) 162--166}.
  \url{https://www.sciencedirect.com/science/article/pii/0370269382906220}.

\bibitem{HAHN1989365}
A.~Hahn, K.~Schreckenbach, W.~Gelletly, F.~{von Feilitzsch}, G.~Colvin, and
  B.~Krusche
  \href{http://dx.doi.org/https://doi.org/10.1016/0370-2693(89)91598-0}{{\em
  Physics Letters B} {\bfseries 218} no.~3, (1989) 365--368}.
  \url{https://www.sciencedirect.com/science/article/pii/0370269389915980}.

\bibitem{hayes_first}
A.~Hayes, J.~Friar, G.~Garvey, G.~Jungman, and G.~Jonkmans
  \href{http://dx.doi.org/10.1103/PhysRevLett.112.202501}{{\em Phys. Rev.
  Lett.} {\bfseries 112} (2014) 202501},
  \href{http://arxiv.org/abs/1309.4146}{{\ttfamily arXiv:1309.4146 [nucl-th]}}.

\bibitem{Hayen:2019eop}
L.~Hayen, J.~Kostensalo, N.~Severijns, and J.~Suhonen
  \href{http://dx.doi.org/10.1103/PhysRevC.100.054323}{{\em Phys. Rev. C}
  {\bfseries 100} no.~5, (2019) 054323},
  \href{http://arxiv.org/abs/1908.08302}{{\ttfamily arXiv:1908.08302
  [nucl-th]}}.

\bibitem{Wang:2016rqh}
X.~B. Wang, J.~L. Friar, and A.~C. Hayes
  \href{http://dx.doi.org/10.1103/PhysRevC.94.034314}{{\em Phys. Rev. C}
  {\bfseries 94} no.~3, (2016) 034314},
  \href{http://arxiv.org/abs/1607.02149}{{\ttfamily arXiv:1607.02149
  [nucl-th]}}.

\bibitem{Haag:2013raa}
N.~Haag, A.~G\"utlein, M.~Hofmann, L.~Oberauer, W.~Potzel, K.~Schreckenbach,
  and F.~M. Wagner \href{http://dx.doi.org/10.1103/PhysRevLett.112.122501}{{\em
  Phys. Rev. Lett.} {\bfseries 112} no.~12, (2014) 122501},
  \href{http://arxiv.org/abs/1312.5601}{{\ttfamily arXiv:1312.5601 [nucl-ex]}}.

\bibitem{Mueller:2011nm}
T.~A. Mueller {\em et~al.}
  \href{http://dx.doi.org/10.1103/PhysRevC.83.054615}{{\em Phys. Rev. C}
  {\bfseries 83} (2011) 054615},
  \href{http://arxiv.org/abs/1101.2663}{{\ttfamily arXiv:1101.2663 [hep-ex]}}.

\bibitem{Huber:2011wv}
P.~Huber \href{http://dx.doi.org/10.1103/PhysRevC.85.029901}{{\em Phys. Rev. C}
  {\bfseries 84} (2011) 024617},
  \href{http://arxiv.org/abs/1106.0687}{{\ttfamily arXiv:1106.0687 [hep-ph]}}.
  [Erratum: Phys.Rev.C 85, 029901 (2012)].

\bibitem{bib:fallot2}
M.~Estienne {\em et~al.}
  \href{http://dx.doi.org/10.1103/PhysRevLett.123.022502}{{\em Phys. Rev.
  Lett.} {\bfseries 123} no.~2, (2019) 022502},
  \href{http://arxiv.org/abs/1904.09358}{{\ttfamily arXiv:1904.09358
  [nucl-ex]}}.

\bibitem{STEREO:2020fvd}
{\bfseries STEREO} Collaboration, H.~Almaz\'an {\em et~al.}
  \href{http://dx.doi.org/10.1103/PhysRevLett.125.201801}{{\em Phys. Rev.
  Lett.} {\bfseries 125} no.~20, (2020) 201801},
  \href{http://arxiv.org/abs/2004.04075}{{\ttfamily arXiv:2004.04075
  [hep-ex]}}.

\bibitem{DayaBay:2016ssb}
{\bfseries Daya Bay} Collaboration, F.~P. An {\em et~al.}
  \href{http://dx.doi.org/10.1088/1674-1137/41/1/013002}{{\em Chin. Phys. C}
  {\bfseries 41} no.~1, (2017) 013002},
  \href{http://arxiv.org/abs/1607.05378}{{\ttfamily arXiv:1607.05378
  [hep-ex]}}.

\bibitem{DayaBay:2021dqj}
{\bfseries Daya Bay} Collaboration, F.~P. An {\em et~al.}
  \href{http://dx.doi.org/10.1088/1674-1137/abfc38}{{\em Chin. Phys. C}
  {\bfseries 45} no.~7, (2021) 073001},
  \href{http://arxiv.org/abs/2102.04614}{{\ttfamily arXiv:2102.04614
  [hep-ex]}}.

\bibitem{RENO:2020dxd}
{\bfseries RENO} Collaboration, S.~G. Yoon {\em et~al.}
  \href{http://dx.doi.org/10.1103/PhysRevD.104.L111301}{{\em Phys. Rev. D}
  {\bfseries 104} no.~11, (2021) L111301},
  \href{http://arxiv.org/abs/2010.14989}{{\ttfamily arXiv:2010.14989
  [hep-ex]}}.

\bibitem{STEREO:2020hup}
{\bfseries STEREO} Collaboration, H.~Almaz\'an {\em et~al.}
  \href{http://dx.doi.org/10.1088/1361-6471/abd37a}{{\em J. Phys. G} {\bfseries
  48} no.~7, (2021) 075107}, \href{http://arxiv.org/abs/2010.01876}{{\ttfamily
  arXiv:2010.01876 [hep-ex]}}.

\bibitem{DayaBay:2017jkb}
{\bfseries Daya Bay} Collaboration, F.~P. An {\em et~al.}
  \href{http://dx.doi.org/10.1103/PhysRevLett.118.251801}{{\em Phys. Rev.
  Lett.} {\bfseries 118} no.~25, (2017) 251801},
  \href{http://arxiv.org/abs/1704.01082}{{\ttfamily arXiv:1704.01082
  [hep-ex]}}.

\bibitem{RENO:2018pwo}
{\bfseries RENO} Collaboration, G.~Bak {\em et~al.}
  \href{http://dx.doi.org/10.1103/PhysRevLett.122.232501}{{\em Phys. Rev.
  Lett.} {\bfseries 122} no.~23, (2019) 232501},
  \href{http://arxiv.org/abs/1806.00574}{{\ttfamily arXiv:1806.00574
  [hep-ex]}}.

\bibitem{Giunti:2017nww}
C.~Giunti \href{http://dx.doi.org/10.1103/PhysRevD.96.033005}{{\em Phys. Rev.
  D} {\bfseries 96} no.~3, (2017) 033005},
  \href{http://arxiv.org/abs/1704.02276}{{\ttfamily arXiv:1704.02276
  [hep-ph]}}.

\bibitem{Gebre:2017vmm}
Y.~Gebre, B.~R. Littlejohn, and P.~T. Surukuchi
  \href{http://dx.doi.org/10.1103/PhysRevD.97.013003}{{\em Phys. Rev. D}
  {\bfseries 97} no.~1, (2018) 013003},
  \href{http://arxiv.org/abs/1709.10051}{{\ttfamily arXiv:1709.10051
  [hep-ph]}}.

\bibitem{Giunti:2019qlt}
C.~Giunti, Y.~F. Li, B.~R. Littlejohn, and P.~T. Surukuchi
  \href{http://dx.doi.org/10.1103/PhysRevD.99.073005}{{\em Phys. Rev. D}
  {\bfseries 99} no.~7, (2019) 073005},
  \href{http://arxiv.org/abs/1901.01807}{{\ttfamily arXiv:1901.01807
  [hep-ph]}}.

\bibitem{Stereo:2021wfd}
{\bfseries Stereo, Prospect} Collaboration, H.~Almaz\'an {\em et~al.}
  \href{http://dx.doi.org/10.1103/PhysRevLett.128.081802}{{\em Phys. Rev.
  Lett.} {\bfseries 128} no.~8, (2022) 081802},
  \href{http://arxiv.org/abs/2107.03371}{{\ttfamily arXiv:2107.03371
  [nucl-ex]}}.

\bibitem{DayaBay:2021owf}
{\bfseries Daya Bay, PROSPECT} Collaboration, F.~P. An {\em et~al.}
  \href{http://dx.doi.org/10.1103/PhysRevLett.128.081801}{{\em Phys. Rev.
  Lett.} {\bfseries 128} no.~8, (2022) 081801},
  \href{http://arxiv.org/abs/2106.12251}{{\ttfamily arXiv:2106.12251
  [nucl-ex]}}.

\bibitem{Kopeikin:2021rnb}
V.~I. Kopeikin, Y.~N. Panin, and A.~A. Sabelnikov
  \href{http://dx.doi.org/10.1134/S1063778821010129}{{\em Phys. Atom. Nucl.}
  {\bfseries 84} no.~1, (2021) 1--10}.

\bibitem{Kopeikin:2021ugh}
V.~Kopeikin, M.~Skorokhvatov, and O.~Titov
  \href{http://dx.doi.org/10.1103/PhysRevD.104.L071301}{{\em Phys. Rev. D}
  {\bfseries 104} no.~7, (2021) L071301},
  \href{http://arxiv.org/abs/2103.01684}{{\ttfamily arXiv:2103.01684
  [nucl-ex]}}.

\bibitem{huber_berryman}
J.~M. Berryman and P.~Huber
  \href{http://dx.doi.org/10.1007/JHEP01(2021)167}{{\em JHEP} {\bfseries 01}
  (2021) 167}, \href{http://arxiv.org/abs/2005.01756}{{\ttfamily
  arXiv:2005.01756 [hep-ph]}}.

\bibitem{Giunti:2021kab}
C.~Giunti, Y.~F. Li, C.~A. Ternes, and Z.~Xin
  \href{http://arxiv.org/abs/2110.06820}{{\ttfamily arXiv:2110.06820
  [hep-ph]}}.

\bibitem{DoubleChooz:2019qbj}
{\bfseries Double Chooz} Collaboration, H.~de~Kerret {\em et~al.}
  \href{http://dx.doi.org/10.1038/s41567-020-0831-y}{{\em Nature Phys.}
  {\bfseries 16} no.~5, (2020) 558--564},
  \href{http://arxiv.org/abs/1901.09445}{{\ttfamily arXiv:1901.09445
  [hep-ex]}}.

\bibitem{RENO:2018dro}
{\bfseries RENO} Collaboration, G.~Bak {\em et~al.}
  \href{http://dx.doi.org/10.1103/PhysRevLett.121.201801}{{\em Phys. Rev.
  Lett.} {\bfseries 121} no.~20, (2018) 201801},
  \href{http://arxiv.org/abs/1806.00248}{{\ttfamily arXiv:1806.00248
  [hep-ex]}}.

\bibitem{NEOS:2016wee}
{\bfseries NEOS} Collaboration, Y.~J. Ko {\em et~al.}
  \href{http://dx.doi.org/10.1103/PhysRevLett.118.121802}{{\em Phys. Rev.
  Lett.} {\bfseries 118} no.~12, (2017) 121802},
  \href{http://arxiv.org/abs/1610.05134}{{\ttfamily arXiv:1610.05134
  [hep-ex]}}.

\bibitem{Sonzogni:2017wxy}
A.~A. Sonzogni, E.~A. McCutchan, and A.~C. Hayes
  \href{http://dx.doi.org/10.1103/PhysRevLett.119.112501}{{\em Phys. Rev.
  Lett.} {\bfseries 119} no.~11, (2017) 112501}.

\bibitem{DYB_LOI}
B.~Littlejohn, K.~B. Luk, and J.~P. Ochoa-Ricoux. {\href{Legacy of the Daya Bay
  Reactor Neutrino Experiment
  https://www.snowmass21.org/docs/files/summaries/NF/SNOWMASS21-NF1_NF2_Daya_Bay-086.pdf},
  Snowmass 2021 Letter of Interest}.

\bibitem{Labit:2021zma}
{\bfseries STEREO} Collaboration, L.-R. Labit
  \href{http://dx.doi.org/10.1088/1742-6596/2156/1/012156}{{\em J. Phys. Conf.
  Ser.} {\bfseries 2156} no.~2, (2021) 012156}.

\bibitem{Oh:2020inx}
Y.~Oh \href{http://dx.doi.org/10.1088/1742-6596/1468/1/012185}{{\em J. Phys.
  Conf. Ser.} {\bfseries 1468} no.~1, (2020) 012185}.

\bibitem{neos2neutrino2022}
J.~Kim, ``{NEOS-II New Results},'' zenodo.
\newblock \url{https://doi.org/10.5281/zenodo.6680618}.

\bibitem{Andriamirado:2022psq}
M.~Andriamirado {\em et~al.} in {\em {2022 Snowmass Summer Study}}.
\newblock 2, 2022.
\newblock \href{http://arxiv.org/abs/2202.12343}{{\ttfamily arXiv:2202.12343
  [hep-ex]}}.

\bibitem{Andriamirado:2021qjc}
M.~Andriamirado {\em et~al.} \href{http://arxiv.org/abs/2107.03934}{{\ttfamily
  arXiv:2107.03934 [hep-ex]}}.

\bibitem{JUNO:2020ijm}
{\bfseries JUNO} Collaboration, A.~Abusleme {\em et~al.}
  \href{http://arxiv.org/abs/2005.08745}{{\ttfamily arXiv:2005.08745
  [physics.ins-det]}}.

\bibitem{TAO_LOI}
G.~Cao, J.~P. Ochoa-Ricoux, W.~Wang, L.~Wen, M.~Wurm, and L.~Zhan.
  {\href{https://www.snowmass21.org/docs/files/summaries/NF/SNOWMASS21-NF9_NF7_Pedro_Ochoa-035.pdf}{The
  JUNO-TAO Experiment}, Snowmass 2021 Letter of Interest}.

\bibitem{chandlerloi}
C.~Awe {\em et~al.}, ``{CHANDLER: A Technology for Surface-level Reactor
  Neutrino Detection},'' {Snowmass Letter of Interest}, 2020.
\newblock
  \url{https://www.snowmass21.org/docs/files/summaries/NF/SNOWMASS21-NF2_NF7_Jon_Link-075.pdf}.

\bibitem{roadstrloi}
O.~A. Akindele {\em et~al.}, ``{ROADSTR: a Mobile Antineutrino Detector
  Platform for enabling Multi-Reactor Spectrum, Oscillation, and Application
  Measurements},'' {Snowmass Letter of Interest}, 2020.
\newblock
  \url{https://www.snowmass21.org/docs/files/summaries/NF/SNOWMASS21-NF9_NF7_ROADSTR_Mobile_Antineutrino-184.pdf}.

\bibitem{Ang:2021svv}
W.~E. Ang, S.~Lee, and S.~Prasad
  \href{http://arxiv.org/abs/2112.12250}{{\ttfamily arXiv:2112.12250
  [hep-ex]}}.

\bibitem{isodar_prl}
A.~Bungau {\em et~al.}
  \href{http://dx.doi.org/10.1103/PhysRevLett.109.141802}{{\em Phys. Rev.
  Lett.} {\bfseries 109} (Oct, 2012) 141802}.
  \url{https://link.aps.org/doi/10.1103/PhysRevLett.109.141802}.

\bibitem{Alonso:2021kyu}
J.~Alonso {\em et~al.}
  \href{http://dx.doi.org/10.1103/PhysRevD.105.052009}{{\em Phys. Rev. D}
  {\bfseries 105} no.~5, (2022) 052009},
  \href{http://arxiv.org/abs/2111.09480}{{\ttfamily arXiv:2111.09480
  [hep-ex]}}.

\bibitem{Alonso:2022mup}
J.~R. Alonso {\em et~al.} \href{http://arxiv.org/abs/2201.10040}{{\ttfamily
  arXiv:2201.10040 [physics.ins-det]}}.

\bibitem{Winklehner:2021qzp}
D.~Winklehner, A.~Adelmann, J.~M. Conrad, S.~Mayani, S.~Muralikrishnan,
  D.~Schoen, and M.~Yampolskaya
  \href{http://arxiv.org/abs/2103.09352}{{\ttfamily arXiv:2103.09352
  [physics.acc-ph]}}.

\bibitem{Hostert:2022ntu}
M.~Hostert, D.~McKeen, M.~Pospelov, and N.~Raj
  \href{http://arxiv.org/abs/2201.02603}{{\ttfamily arXiv:2201.02603
  [hep-ph]}}.

\bibitem{medical}
J.~R. Alonso, R.~Barlow, J.~M. Conrad, and L.~H. Waites
  \href{http://dx.doi.org/10.1038/s42254-019-0095-6}{{\em Nature Reviews
  Physics} {\bfseries 1} no.~9, (2019) 533--535}.
  \url{https://doi.org/10.1038/s42254-019-0095-6}.

\bibitem{ElectronCaptureLOI}
P.~Coloma {\em et~al.}, ``{Physics with Electron Capture Neutrino Sources},''
  {Snowmass Letter of Interest}, 2020.
\newblock
  \url{https://www.snowmass21.org/docs/files/summaries/NF/SNOWMASS21-NF2_NF9-CF1_CF0_Jon_Link-038.pdf}.

\bibitem{GALLEX}
{\bfseries GALLEX} Collaboration, W.~Hampel {\em et~al.}
  \href{http://dx.doi.org/10.1016/S0370-2693(97)01562-1}{{\em Phys. Lett. B}
  {\bfseries 420} (1998) 114--126}.

\bibitem{sage}
J.~N. Abdurashitov, V.~N. Gavrin, S.~V. Girin, V.~V. Gorbachev, T.~V.
  Ibragimova, A.~V. Kalikhov, N.~G. Khairnasov, T.~V. Knodel, V.~N. Kornoukhov,
  I.~N. Mirmov, A.~A. Shikhin, E.~P. Veretenkin, V.~M. Vermul, V.~E. Yants,
  G.~T. Zatsepin, Y.~S. Khomyakov, A.~V. Zvonarev, T.~J. Bowles, J.~S. Nico,
  W.~A. Teasdale, D.~L. Wark, M.~L. Cherry, V.~N. Karaulov, V.~L. Levitin,
  V.~I. Maev, P.~I. Nazarenko, V.~S. Shkol’nik, N.~V. Skorikov, B.~T.
  Cleveland, T.~Daily, R.~Davis, K.~Lande, C.~K. Lee, P.~S. Wildenhain, S.~R.
  Elliott, and J.~F. Wilkerson
  \href{http://dx.doi.org/10.1103/physrevc.59.2246}{{\em Physical Review C}
  {\bfseries 59} no.~4, (Apr, 1999) 2246–2263}.
  \url{http://dx.doi.org/10.1103/PhysRevC.59.2246}.

\bibitem{BEST}
V.~Gavrin, B.~Cleveland, S.~Danshin, S.~R. Elliott, V.~Gorbachev,
  T.~Ibragimova, A.~Kalikhov, T.~Knodel, Y.~Kozlova, Y.~Malyshkin, V.~Matveev,
  I.~Mirmov, J.~Nico, R.~G.~H. Robertson, A.~Shikhin, D.~Sinclair,
  E.~Veretenkin, and J.~Wilkerson
  \href{http://dx.doi.org/10.1134/S1063779615020100}{{\em Physics of Particles
  and Nuclei} {\bfseries 46} no.~2, (3, 2015) }.

\bibitem{BEST2}
V.~V. Barinov, B.~T. Cleveland, S.~N. Danshin, H.~Ejiri, S.~R. Elliott,
  D.~Frekers, V.~N. Gavrin, V.~V. Gorbachev, D.~S. Gorbunov, W.~C. Haxton,
  T.~V. Ibragimova, I.~Kim, Y.~P. Kozlova, L.~V. Kravchuk, V.~V. Kuzminov,
  B.~K. Lubsandorzhiev, Y.~M. Malyshkin, R.~Massarczyk, V.~A. Matveev, I.~N.
  Mirmov, J.~S. Nico, A.~L. Petelin, R.~G.~H. Robertson, D.~Sinclair, A.~A.
  Shikhin, V.~A. Tarasov, G.~V. Trubnikov, E.~P. Veretenkin, J.~F. Wilkerson,
  and A.~I. Zvir, ``A search for electron neutrino transitions to sterile
  states in the best experiment,'' 2022.
\newblock \url{https://arxiv.org/abs/2201.07364}.

\bibitem{sage2}
J.~N. Abdurashitov {\em et~al.}
  \href{http://dx.doi.org/10.1134/S1063778806110032}{{\em J. Phys. Conf. Ser.}
  {\bfseries 39} (2006) 284--286}.

\bibitem{gallex2}
F.~Kaether, W.~Hampel, G.~Heusser, J.~Kiko, and T.~Kirsten
  \href{http://dx.doi.org/10.1016/j.physletb.2010.01.030}{{\em Phys. Lett. B}
  {\bfseries 685} (2010) 47--54},
  \href{http://arxiv.org/abs/1001.2731}{{\ttfamily arXiv:1001.2731 [hep-ex]}}.

\bibitem{SOX}
K.~Altenm\"uller {\em et~al.}
  \href{http://dx.doi.org/10.1088/1748-0221/13/09/P09008}{{\em JINST}
  {\bfseries 13} no.~09, (2018) P09008}.

\bibitem{SOX2}
``Infn sox press release.''
  \url{https://web.archive.org/web/20180309012856/http://home.infn.it/en/media-outreach/press-releases/2789-infn-e-cea-annullano-il-progetto-sox-per-l-impossibilita-di-realizzare-la-sorgente-con-le-caratteristiche-necessarie-all-esperimento-2}.
\newblock Accessed: 2022-05-29.

\bibitem{BEST3}
V.~N. Gavrin, V.~V. Gorbachev, T.~V. Ibragimova, V.~N. Kornoukhov, A.~A.
  Dzhanelidze, S.~B. Zlokazov, N.~A. Kotelnikov, A.~L. Izhutov, S.~V. Mainskov,
  V.~V. Pimenov, V.~P. Borisenko, K.~B. Kiselev, and M.~P. Tsevelev, ``On the
  gallium experiment best-2 with a $^{65}$zn source to search for neutrino
  oscillations on a short baseline,'' 2018.
\newblock \url{https://arxiv.org/abs/1807.02977}.

\bibitem{faser_proposal}
{\bfseries FASER} Collaboration, H.~Abreu {\em et~al.}
  \href{http://arxiv.org/abs/2001.03073}{{\ttfamily arXiv:2001.03073
  [physics.ins-det]}}.

\bibitem{faser_physics_case}
{\bfseries FASER} Collaboration, H.~Abreu {\em et~al.}
  \href{http://dx.doi.org/10.1140/epjc/s10052-020-7631-5}{{\em Eur. Phys. J. C}
  {\bfseries 80} no.~1, (2020) 61},
  \href{http://arxiv.org/abs/1908.02310}{{\ttfamily arXiv:1908.02310
  [hep-ex]}}.

\bibitem{fasernu}
{\bfseries FASER} Collaboration, H.~Abreu {\em et~al.}
  \href{http://dx.doi.org/10.1103/PhysRevD.104.L091101}{{\em Phys. Rev. D}
  {\bfseries 104} no.~9, (2021) L091101},
  \href{http://arxiv.org/abs/2105.06197}{{\ttfamily arXiv:2105.06197
  [hep-ex]}}.

\bibitem{FPF1}
L.~A. Anchordoqui {\em et~al.}
  \href{http://arxiv.org/abs/2109.10905}{{\ttfamily arXiv:2109.10905
  [hep-ph]}}.

\bibitem{FPF2}
J.~L. Feng {\em et~al.} in {\em {2022 Snowmass Summer Study}}.
\newblock 3, 2022.
\newblock \href{http://arxiv.org/abs/2203.05090}{{\ttfamily arXiv:2203.05090
  [hep-ex]}}.

\bibitem{P52014}
{\bfseries HEPAP Subcommittee} Collaboration, S.~Ritz {\em et~al.}

\bibitem{EuropeanStrategyGroup:2020pow}
{\bfseries European Strategy~Group} Collaboration.
\newblock CERN Council, Geneva, 2020.

\bibitem{NuStorm}
L.~A. Ruso, T.~Alves, S.~Boyd, A.~Bross, P.~R. Hobson, P.~Kyberd, J.~B.
  Lagrange, K.~Long, X.~G. Lu, J.~Pasternak, M.~Pfaff, and C.~Rogers,
  ``Neutrinos from stored muons (nustorm),'' 2022.
\newblock \url{https://arxiv.org/abs/2203.07545}.

\bibitem{Geer:1997iz}
S.~Geer \href{http://dx.doi.org/10.1103/PhysRevD.57.6989}{{\em Phys. Rev. D}
  {\bfseries 57} (1998) 6989--6997},
  \href{http://arxiv.org/abs/hep-ph/9712290}{{\ttfamily arXiv:hep-ph/9712290}}.
  [Erratum: Phys.Rev.D 59, 039903 (1999)].

\bibitem{IDS-NF:2011swj}
{\bfseries IDS-NF} Collaboration, S.~Choubey {\em et~al.}
  \href{http://arxiv.org/abs/1112.2853}{{\ttfamily arXiv:1112.2853 [hep-ex]}}.

\bibitem{Delahaye:2018yfq}
J.~P. Delahaye, C.~M. Ankenbrandt, S.~A. Bogacz, P.~Huber, H.~G. Kirk,
  D.~Neuffer, M.~A. Palmer, R.~Ryne, and P.~V. Snopok
  \href{http://dx.doi.org/10.1088/1748-0221/13/06/T06003}{{\em JINST}
  {\bfseries 13} (2018) T06003},
  \href{http://arxiv.org/abs/1803.07431}{{\ttfamily arXiv:1803.07431
  [physics.acc-ph]}}.

\bibitem{Bogacz:2022xsj}
A.~Bogacz {\em et~al.} in {\em {2022 Snowmass Summer Study}}.
\newblock 3, 2022.
\newblock \href{http://arxiv.org/abs/2203.08094}{{\ttfamily arXiv:2203.08094
  [hep-ph]}}.

\bibitem{Neuffer:2013wrd}
D.~Neuffer, M.~Palmer, Y.~Alexahin, C.~Ankenbrandt, and J.~P. Delahaye in {\em
  {4th International Particle Accelerator Conference}}.
\newblock 5, 2013.
\newblock \href{http://arxiv.org/abs/1502.02042}{{\ttfamily arXiv:1502.02042
  [physics.acc-ph]}}.

\bibitem{Alexahin:2022tav}
Y.~I. Alexahin {\em et~al.} in {\em {2022 Snowmass Summer Study}}.
\newblock 3, 2022.
\newblock \href{http://arxiv.org/abs/2203.10431}{{\ttfamily arXiv:2203.10431
  [physics.acc-ph]}}.

\bibitem{tau-neutrino-lepton-collider}
G.~Dallavalle {\em et~al.} in {\em {2022 Snowmass Summer Study}}.
\newblock
  \url{https://www.snowmass21.org/docs/files/summaries/AF/SNOWMASS21-AF4_AF0-EF3_EF0-NF6_NF10-081.pdf}.

\bibitem{DeGouvea2021}
A.~de~Gouvea Snowmass Muon Collider Forum, 2021.

\end{thebibliography}\endgroup
%\printbibliography
%\end{multicols}

\end{document}